\documentclass[10pt-default,11pt]{article}
\usepackage{amsmath}
\usepackage{amsfonts}
\usepackage{amssymb}
\usepackage{geometry}

\input{tcilatex}
\begin{document}

{\Large How the modified Bertrand theorem explains regularities}

{\Large of the periodic table I. From conformal invariance to }

{\Large Hopf mapping\bigskip \bigskip }

\textbf{Arkady Kholodenko}$^{a}\footnote{%
Corresponding author.
\par
E-mail address: \ string@clemson.edu}$ \textbf{and Louis Kauffman}$%
^{b}\bigskip $

$^{a}$\ 375 H.L. Hunter Laboratories, Clemson \ University, Clemson, SC
29634-0973, USA.

$^{b}$ Department of Mathematics, Statistics and Computer Science,
University of \ Illinois

\ \ at Chicago, 851 S.Morgan Street, Chicago, Il,60607-7045, USA \ \ \ \ \ \
\ \ \ \ \ \ \ \ \ \ \ \ \ \ \ \ \ \ \ \ \ \ \ \ \ \ \ \ \ \ \ \ \ \ \ \ \ \
\ \ \ \ \ \ \ \ \ \ \ \ \ \ \ \ \ \ \ \ \ \ \ \ \ \ \ \ \ \ \ \ \ \ \ \ \ \
\ \ \ \ \ \ \ \ \ \ \ \ \ \ \ \ \ 

\bigskip

\ \ \ \ \ \ \ \ \ \ \ \ \ \ \ \ \ \ \ \ \ \ \ \ \ \ \ \ \ \ \ \ \ \ \ \ \ \
\ \ \ \ \ \ \ \ \ \ \ Abstract\bigskip \textbf{\ }

Bertrand theorem permits closed orbits in 3d Euclidean space only for 2
types of central

potentials. These are of Kepler- Coulomb and harmonic oscillator type.
Volker Perlick \ 

recently extended Bertrand theorem. He designed new static spherically
symmetric

(Bertrand) spacetimes obeying Einstein's equations and supporting closed
orbits. In this

work we demonstrate that the topology and geometry of these spacetimes
permits to

solve quantum many-body problem for any atom of periodic system exactly. The

computations of spectrum for any atom of periodic system becomes analogous
to that

for hydrogen atom. Initially the exact solution of the Schr\"{o}dinger
equation for any

multielectron atom was obtained by Tietz in 1956. However, neither himself
nor

others fully comprehended what \ actually was obtained. We recalculated
Tietz results

by applying the methodology consistent with new (different from that
developed by Fock

in 1936) way of solving Schr\"{o}dinger's equation for hydrogen atom. In the
light of this

new result it had become possible to demonstrate rigorously that the Tietz-
type Schr\"{o}dinger's

equation is in fact describing the quantum motion in Bertrand spacetime. As
a bonus,

we obtained the analytical proof of the Madelung rule defined in the text.

\textit{Keywords}:\textbf{\ \bigskip\ }

Maxwell's fish-eye potential$,$

conformally invariant classical and quantumequations$,$

superintegrability$,$quantum Bertrand spacetimes\bigskip \bigskip \medskip

MSC 2010: 81Q05, 81Q99, 83C15, 83C20\bigskip

\ \ \ \ \ \ \ \ \ \ \ \ \ \ \ \ \ \ \ \ \ \ \ \ \ \ \ \ \ \ \ \ \ \ 

\textbf{1. Background and summary \bigskip\ \ \ \ }

Although quantum mechanical description of multielectron atoms and molecules
is considered to be a well established domain of research, recently
published book [$1$] indicates that there are still many topics left for
development. As is well known, the quantum mechanical description of
multielectron atom (with atomic number Z and infinitely heavy nucleus)
begins with writing down the stationary Schr\"{o}dinger equation 
\begin{equation}
\hat{H}\Psi (\mathbf{r}_{1},\mathbf{r}_{2},...,\mathbf{r}_{Z})=E\Psi (%
\mathbf{r}_{1},\mathbf{r}_{2},...,\mathbf{r}_{Z})  \tag{1.1a}
\end{equation}%
with the Hamiltonian%
\begin{equation}
\hat{H}=-\dsum\limits_{i=1}^{Z}\frac{\hslash ^{2}}{2m}\nabla
_{i}^{2}-\dsum\limits_{i=1}^{Z}\frac{Ze^{2}}{r_{i}}+\frac{1}{2}\dsum\limits 
_{\substack{ i,j=1  \\ i\neq j}}^{Z}\frac{e^{2}}{r_{ij}}.  \tag{1.1b}
\end{equation}%
Following Bohr's \textsl{Aufbauprinzip} the atom with atomic number Z is
made up of electrons added in succession to the bare atomic nucleus. At the
initial stages of this process electrons are assumed to occupy the
one-electron levels of lowest energy. \ This process is described in terms
of the one electron eigenvalue problem 
\begin{equation}
\hat{H}_{i}\psi _{\square _{i}}(r_{i})=[-\frac{\hslash ^{2}}{2m}\nabla
_{i}^{2}+V_{eff}(\mathbf{r}_{i})]\psi _{\square _{i}}(\mathbf{r}%
_{i})=\varepsilon _{nl}(i)\psi _{\square _{i}}(\mathbf{r}_{i}),i=1\div Z, 
\tag{1.2}
\end{equation}%
where $V_{eff}(\mathbf{r}_{i})$ is made of the combined nuclear potential - $%
\frac{Ze^{2}}{r_{i}}$ for the i-th electron and the centrally symmetric
Hartree-Fock type potential $\mathcal{F}$(\textbf{r}$_{i})$ \ coming from
the presence of the rest of atomic electrons. The fact that $\mathcal{F}$(%
\textbf{r}$_{i})$ is centrally symmetric \ was discussed, for example, in
the \ book by Bethe and Jackiw [$2$]. Later on in the text and, especially,
in the Appendix G, we provide much more details on this topic. The symbol $%
\square _{i}$ indicates the i-th entry into the set\ made out of
hydrogen-like quantum numbers characterizing individual electrons. Recall
that the concept of an\textsl{\ orbital} is determined by the major quantum
number $n$ having its origin in studies of hydrogen atom. The number of
electrons allowed to sit on a given orbital is determined by the \textsl{%
Pauli exclusion principle. }Thus, with increasing Z the electrons are
occupying successive orbitals according to Bohr's Aufbau scheme until the 
\textsl{final ground state electron configuration} is reached. \ Since
electrons are indistinguishable, the hydrogen-like quantum numbers $n,l,m$
and $m_{s}$ cannot be associated with a particular electron. Therefore, the
symbol $\ \square _{i\text{ }}$ should be understood as representing a
specific set of quantum numbers otherwise used for description of individual
(that is not collectivized) electrons.

The problem with just described \textsl{Aufbauprinzip }lies in the
assumption that the guiding principle in designing the \ final ground state
electron configuration is made out of two components: a) knowledge of
hydrogen atom-like wave functions supplying the quantum boxes/numbers $%
\square _{i}$ and, b) the Pauli principle which is mathematically restated
in the form of the fully antisymmetric wavefunction $\Psi (\mathbf{r}_{1},%
\mathbf{r}_{2},...,\mathbf{r}_{Z})$. Should these requirements be sufficient
then, we would be able to replace $V_{eff}(\mathbf{r}_{i})$ by -$\frac{Ze^{2}%
}{r_{i}}$ \ so that the filling of electronic levels would occur according
to the Fock $n$-rule\medskip

\textbf{Fock n-rule: }\textsl{With increasing Z the nl orbitals are filled
in order of increasing n.\medskip }

This rule leads to the problems already for the lithium as explained in [1],
page 330. As result, the ($n,l$) rule was proposed instead.\medskip

\textbf{The hydrogenic (}\textit{n},\textit{l})\textbf{\ rule: }\textsl{With
increasing Z, the orbitals are filled in order of increasing n while for a
fixed n the orbitals are filled in order of increasing \textit{l}.\medskip }

After $Z=18$ the ($n,l$) rule breaks down though. Therefore, it was
subsequently replaced by the ($n+l,n$) rule suggested by Madelung- the
person who reformulated Schr\"{o}dinger's equation in hydrodynamic form
[3].\medskip

\textbf{The Madelung (}\textit{n+l},\textit{n}\textbf{) rule: }\textsl{With
increasing Z, the orbitals are filled in order of increasing n+}$\QTR{sl}{l}%
=N.$ \textsl{For fixed }$N$\textsl{, the orbitals are filled in order of
increasing n.\medskip }

All the above rules are empirical. As such, they require theoretical
explanation. This fact brings us to the\medskip

\textbf{L\"{o}wdin's challenge problem}: Find a way to derive the Madelung
rule ab initio. \textsl{\footnote{%
This problem was posed by Per-Olov L\"{o}wdin [4]. Additional details and
references are given in [5].}}.\medskip

The essence of Mendeleev's periodic system of elements lies exactly in the
discovered periodicity of properties of chemical elements. Although there
are 100's of ways this periodicity is exhibited\footnote{%
Results of www searches indicate that this process is still ongoing.}, the
commonly accepted periodic table \ of elements consists of \ seven periods:
2-8-8-18-18-32-32. Notice that all period lengths occur in pairs (period
doubling), except for the very first period of size 2. To determine whether
this exception is intrinsic or not, analysis of work by Charles Janet on \
periodic table done in 1930 (6 years before work by Madelung!) is the most
helpful, [1], pages 336-340. \ Although initially Janet developed his
version of periodic table without guidance of quantum mechanics, eventually
he did make a connection with Bohr's results. Janet's periodic table has 8
periods. The periods in Janet's table are characterized (without exception)
by the constant value of $N=n+l.$ It is in excellent agreement \ with the
Madelung rule suggesting elevation of $N=n+l$ to the rank of new quantum
number. By organizing the elements in periods \ of constant $n+l$ and groups
of constant $l,m_{l}$ and $m_{s}$, the period doubling emerges naturally and
leads to the sequence of periods: 2-2-8-8-18-18-32-32.\bigskip\ 

A concise and convincing explanation of the period doubling and its
connection with the Madelung rule is given in [6]. The Madelung rule had
attracted the attention of Demkov and Ostrovsky in 1971 resulting in their
publication [7$]$ of major importance. Nevertheless, subsequently Kitagawara
and Barut [8,9] found apparent flaws in the logic of Demkov-Ostrovsky
calculations. The authors of \ the book [1] also expressed their objections
to the Demkov-Ostrovsky cycle of works. On page 381 of [1] we found the
following statement: "Demkov and Ostrovsky developed an atomic physics model
that incorporates the Madelung rule, but by replacing the quantization of
level energies with quantization of coupling constants at zero
energy."\medskip\ \ Furthermore, Demkov and Ostrovsky, while obtained the
correct results, had been unable to provide their rigorous justification
since their effective potential $V_{eff}(\mathbf{r}_{i})$ was correctly
guessed. The authors of [1$]$ conclude that, even though the Demkov
-Ostrovsky results do reproduce the Madelung rule correctly, the way these
results were obtained \ cannot be considered as solution of the L\"{o}wdin
challenge.\medskip\ \bigskip

In this work we demonstrate that the objections raised in [$1,8,9$] are
coming from the lack of knowledge of needed mathematical apparatus. Thus,
one of the tasks of this work is to describe this apparatus. By doing so, a
number of problems of major practical importance is solved and all of the
objections raised in [$1,8,9$ ] are removed. Some additional objections to
be described shortly below are also removed. In their seminal work [$7]$
Demkov-Ostrovsky (D-O) realized that the key to success of solving problems
of periodic table lies is Eq.(1.2) where $V_{eff}(\mathbf{r}_{i})$ should be
correctly chosen. The Bertrand theorem of classical mechanics [$10$] imposes
seemingly insurmountable restrictions on selection of $V_{eff}(\mathbf{r}%
_{i})$ since for spherically symmetric potentials only the Coulombic -$\frac{%
Ze^{2}}{r_{i}}$ and the harmonic oscillator $kr^{2}$ potentials allow
dynamically closed orbits. D-O believed that, in spite of the
indistinguishability of electrons, the discrete spectrum of multielectron
atoms should be associated with closed orbits. \ From the point of view of
theory of chaos\footnote{%
Begining from the motion of electrons in helium, all atomic multielectron
systems are believed to be chaotic,} at the semiclassical level\ of
description the role of closed orbits recently was discussed, e.g. in [$11]$
and, much earlier, in the seminal book by Gutzwiller [$12$]\footnote{%
See also the book by Cvitanovic [$13$].}.\bigskip

Hoping to bypass limitations of the Bertrand theorem, D-O employed the
optical-mechanical analogy in their calculations. It permitted them to use
the Maxwell fish-eye potential (and, later on, its conformally deformed
modifications) instead of the Coulombic potential. They demonstrated in [7$]$
the equivalence at the classical level between the Hamilton-Jacobi equations
employing these two, seemingly different, potentials. D-O hoped that use of
the fish-eye potential (or its conformal modification) might allow to bypass
limitations of the Bertrand theorem. This fact attracted attention of John
Wheeler who in [$14,16$] and, with his student, in [$15$], studied
classically and semiclassically the motion in the fish-eye and conformally
deformed fish-eye potentials used by D-O in [$7$]. The dynamics of electrons
in such conformally deformed potentials happens to involve orbits \ which
are closed, planar and have self-intersections$.$ In his paper [17$]$
Ostrovski argued that the self-intersections of orbits do not contradict the
Bertrand theorem. This statement by Ostrovski happens to be wrong as
demonstrated in section 5. In it, we mention works by other authors who
obtained the same results prior to the discovery of curved Bertrand
spacetimes. None of these authors was concerned with the limitations of
Bertrand theorem. Very recently, in 2017, the same self-intersecting
patterns were obtained for dynamical trajectories in Bertrand spacetimes,
e.g. see page 3362 of [$18$]. The results were obtained without any
reference to atomic physics.

According to work by Little [19$]$ there are only \ 3 distinct regular
homotopy classes of oriented closed curves on $S^{2}.$These are: a) those
for curves without self-intersections, b) those for curves with \ just 1
self--intersection and, c) those with 2 self-intersections. The b)-type
homotopy class of curves were \ obtained by Wheeler and \ by other authors
as discussed in section 5. Thus, from the work by Little it follows that the
effects of curvature and presence of self-intersection in dynamical
trajectories are connected with each other. The Kepler-Coulomb dynamics in
flat 3d Euclidean space does not allow self-intersections. The
self-intersections are allowed if the Bertrand theorem is extended to
motions on curved (Bertrand) manifolds [$18$].The sphere $S^{2}$ is
conformally flat\footnote{%
This concept was defined in our work [3$]$ and is going to be also explained
in the bulk of this work.}.The results of the modified Bertrand theorem are
valid exactly for the conformally flat manifolds. Details are given in
section 5.These are corollaries of results obtained in [$20$]. Thus,
flatness of self-intersecting patterns obtained in [$18$] is in fact
conformal flatness.

To connect Bertrand spacetimes with atomic physics we begin with D-O
statement made in [7] : "The Maxwell's fish-eye problem is \textbf{closely
related to} the Coulomb problem". \ Being aware of the book by Luneburg [$21$%
]\footnote{%
Ref.4 in [7$].$} D-O nevertheless \ underestimated the nature of connection
between the Coulombic and optical (fish-eye) problems described in [$21$].
This fact is discussed in detail in section 2. The assumption of only "close
relationship" caused D-O to replace Eq.(1.2) by

\begin{equation}
\lbrack -\frac{\hslash ^{2}}{2m}\nabla _{i}^{2}+V_{eff}(\mathbf{r}_{i})]\psi
(\mathbf{r}_{i})=0.  \tag{1.3}
\end{equation}%
Eq.(1.3) is looking differently from Eq.(1.2). Eq.(1.2) is the eigenvalue
problem while Eq.(1.3) is the Sturmian problem. For such type of problem to
make sense \ the parameters entering into $V_{eff}(\mathbf{r}_{i})$ must be
quantized. Such quantization of parameters is making Sturmian and eigenvalue
problems equivalent. This fact is nontrivial. It is considered in detail in
sections 2 and 4 and, in Appendices A,B, E and F. Without further
explanations this equivalence represents the major reason for using \ the
fish-eye-type potentials in treatments of atomic systems. In fact, in many
ways (discussed in section 3 and Appendices C, D and F) to use Eq.(1.3) is
more advantageous than Eq.(1.2).\footnote{%
Mathematically, there is alternative proof of this fact which is interesting
on its own. It will be discussed in a separate publication.}

In [1], page 377, \ we read that Eq.(1.3) "does not describe the bound
states of the atom". That this is not the case could be concluded already by
D-O themselves should they read the corresponding \ places in books by
Lunenburg [$21$] and Caratheodory [$22$]. They do quote these works though
in [$7$]. \ This misunderstanding of importance of Eq.(1.3) and its relation
with Eq.(1.2) resulted in critique and neglect of D-O works, e.g. read [1],
page 377. \ In this paper we are going to demonstrate that Maxwell's
fish-eye and related to the fish-eye classical and quantum Coulombic
(hydrogen atom) problems are not \textbf{closely related.} Instead, \textbf{%
they are} \textbf{isomorphic.} This is illustrated in section 4. By
overlooking the Coulombic-fish-eye isomoprphism at the quantum level D-O
argued, nevertheless, that Schr\"{o}dinger's Eq.(1.3) with Coulombic and
fish-eye-type potentials \textbf{both} possess the O(4,2) (or SO(4,2))
dynamical symmetry known for the hydrogen atom and later established for the
rest of atoms of the periodic table [$1$], [$3$]. Selection of the
fish-eye-like potentials by D-O \ was guided in part by their desire to
describe the \textsl{atoms} \textsl{other than} \textsl{hydrogen. }In
addition, they believed that: a) $V_{eff}(\mathbf{r}_{i})$ in Eq.(1.3) \ can
be represented by the conformally deformed fish-eye potential so that b) use
of such potential removes restrictions posed by the Bertrand theorem in flat
space. Ostrovsky concludes his paper [17] with the following remark: "It 
\footnote{%
...that is the group-theoretical (our insert from previous discussion)
consideration...} leaves a very interesting question unresolved, the
question of why the interaction of a number of electrons with each other and
with an atomic nucleus leads to an effective one-electron potential having
some approximate hidden symmetry. The method of solution of this question
can hardly be envisaged at the present time." Thus, with complete formal
success of quantum description of atoms of the whole periodic system [7$]$
culminating in formal proof of the Madelung rule, Ostrovsky admits that with
all his results published to date, the L\"{o}wdin's challenge problem still
remains out of reach. This is so because the deformed fish -eye potential
used in D-O calculations had no visible connection with $V_{eff}(\mathbf{r}%
_{i})$ coming from the Hartree-Fock calculations. \ In addition \ to their
inability of solving the L\"{o}wdin challenge, D-O also failed to solve the
Bertrand challenge problem\footnote{%
We are posing the Bertrand challenge problem at this point in addition to
the L\"{o}wdin challenge problem known for some time}: What makes the
deformed fish-eye potential used by D-O as substitute of $V_{eff}(\mathbf{r}%
_{i})$ so good that it removes the restrictions of the classical Bertrand
theorem?\bigskip\ 

In their cycle of works on proving the Madelung rule D-O used the fish-eye ($%
\gamma =1)$ and conformally deformed ($\gamma \neq 1)$ fish -eye potentials%
\begin{equation}
V(x,y,z)=-\left( \frac{a}{r}\right) ^{2}\left[ \frac{n_{0}}{\left(
r/a\right) ^{-\gamma }+\left( r/a\right) ^{\gamma }}\right] ^{2},  \tag{1.4}
\end{equation}%
$r^{2}=x^{2}+y^{2}+z^{2},a=const,\gamma $ is a rational number, as an
alternative to the $V_{eff}(\mathbf{r}_{i})$ Hartree-Fock-type potentials
routinely used in atomic physics literature. Such a replacement required
them to switch from Eq.(1.2) to Eq.(1.3) for reasons explained in detail
below, in sections 2 and 3 and Appendices A and B. Since Eq.(1.3) seemingly
allows only to look for egenfunctions with zero eigenvalue, both D-O and the
rest of researchers in the field considered this limitation as a serious
deficiency. In this work we demonstrate that this restriction is harmless
and, in fact, very helpful. \ We remind to our readers that use of the
potential, Eq.(1.4), was made by D-O for the purpose of taking care of
limitations of the classical Bertrand theorem. No other authors were
concerned with these limitations. Apparently, this fact allowed D-O to
neglect works by other authors on the same or related subjects and to draw
attention of others [1] only to their own works. This happens to be a
fundamental drawback. Details are given in section 5 and Appendices F and G.

Below, we provide a summary of the results just sketched. D-O realized that
when used in Eq.(1.3) the constant $n_{0}$ acquires discrete values as it
happens in all Sturmian type problems (Appendix F). Furthermore, for $\gamma
=1/2$ the solution of Eq.(1.3) provides results compatible with the Madelung
rule. The apparent limitation, $E=0,$ along with (in their expositions) \ no
apparent relationship between the potential $V(x,y,z)$ and $V_{eff}(\mathbf{r%
})$ caused Ostrovski to acknowledge (in [17]) that all D-O results to date
do not solve the L\"{o}wdin challenge problem. Thus, we are left with the
following (unproven yet) facts:

a) use of the potential, Eq.(1.4), \ removes the limitations of the
classical Bertrand theorem;

b) the choice $\gamma =1/2$ in Eq.(1.4) apparently explains the Madelung
rule;

c) finding of the spectral results beyond $E=0$ requires(according to D-O)
uses of sophisticated perturbational methods;

d) the choice $\gamma =1/2$ in Eq.(1.4) is detached from known Hartree-Fock
results;

e) the case $\gamma =1$ corresponds to the standard Maxwell's fish-eye
potential. \ Classical dynamics in such

\ \ \ \ potential is isomorphic to that in Kepler-Coulomb potential.
However, because of the apparent $E=0$ limitation,

\ \ \ neither D-O nor other researchers \ reproduced known eigenvalue
spectrum for the hydrogen atom.

Subsequently, other authors studied \ Eq.(1.3) with D-O potential, Eq.(1.4),
in 2 dimensions where use of conformal transformations leaves Eq.(1.3)
form-invariant. This fact provides many advantages. Still, no attempts to
reproduce known 2 dimensional results for hydrogen atom were made.

In our work we use the observation, in section 4, that results on \textbf{R}$%
^{2}$ can be lifted to $S^{2}$ and then lifted further to $S^{3}$ via
inverse Hopf mapping. These lifting procedures will be discussed in detail
in our next paper, Part II. Using stereographic projection: from $S^{3}$ to 
\textbf{R}$^{3}$, it is possible then to reobtain D-O results done in 
\textbf{R}$^{3}$. Even though the connection between the Hartree-Fock and
the D-O potential, Eq.(1.4), will be discussed in detail from geometrical
and topological perspective in Part II, already available results allow us
to discuss rigorously such a connection in Part I , that is this paper.
Details are given in Appendix G and section 5. The noticed connection is
inseparably linked with the validity of the Madelung rule as explained in
Appendix G and section 5.

Going back to a) we direct our readers attention to work [23] by Volker
Perlick. In it the results of classical Bertrand theorem [10] valid in
Euclidean 3 space had been generalized to static spherically symmetric
spacetimes of general relativity. By design, the motion in such spacetimes
takes place on closed orbits. In section 5 we demonstrate that the
potential, Eq.(1.4), indeed removes limitations of the classical Bertrand
theorem. However, it is used not in flat Euclidean \textbf{R}$^{3}$ but in
curved Bertrand spacetime.

The choice $\gamma =1/2$ listed in b) and d) is indeed connected directly
with the results of Hartree-Fock calculations and with the Madelung rule. In
atomic physics literature the potential, Eq.(1.4), $\gamma =1/2,$ is known
as the Tietz potential\footnote{%
It is bearing the name of its creator.}. Its origin and many properties are
discussed in the book by Flugge [24]. Its remarkable numerical coincidence
with the corresponding Hartree-Fock potentials were discussed in many
places, e.g. read [25], p.664, fig.10. Tietz, the author who invented the
Tietz potential, was initially driven by the desire to simplify the Thomas
-Fermi (T-F) claculations. \ Much more analytically cumbersome T-F type
potentials were used by Latter [26] in his numerical study of the Schr\"{o}%
dinger equation spectra of \ low lying excitations for all atoms of periodic
system. Numerical results of Latter had been subsequently analyzed by March.
On page 76 of [27] without explicit mention of the Madelung rule he
described results by Latter in terms of the Madelung rule. After discovery
of his potential Tietz used it in the stationary Schr\"{o}dinger equation,
Eq.(1.2), in which $V_{eff}(\mathbf{r}_{i})$ was replaced by the Tietz
potential, that is by Eq.(1.4) with $\gamma =1/2$ [28]$.$ Tietz used
Eq.(1.2) in which $E\neq 0$ instead of D-O version of this equation,
Eq.(1.3). In the light of results of Appendix F, this ha[ppens to be
permissible. Tietz also recognized that Eq.(1.2) with his potential can be
solved exactly. His first attempt to do so was initially made in 1956 [$28$%
]. The rest of his attempts is summarized in [$29$]. The exact solution
methods in these works differ in style from that used by D-O. Besides, Tietz
used these exact solutions \ only to check them against the Hartree-Fock
results for the Mercury ($Z=80$). In doing so he got a good agreement with
the Hartree-Fock results. In this work we prove that the Tietz potential
used in [$28$], [29] is describing the classical and quantum motion in
Bertand spacetimes. Unlike Tietz and, in accord with D-O, we used Eq.(1.3)
for solving the corresponding eigenvalue problem. Our method of solving this
equation differs somewhat from that used by D-O though since their method is
unable to reproduce the hydrogen atom spectrum (the problem e)). Our way of
solving Eq.(1.3) with $\gamma =1/2$ is consistent with our new way
(different from that developed by Fock in 1936 (discussed in [3])) of
obtaining the spectrum of hydrogen atom. We were able to reproduce this
spectrum correctly also in 2 dimensions. These facts allowed us to bypass
entirely the item c) present in D-O works. They also allow us to obtain the
low lying spectrum of any atom of periodic system consistent with the
Madelung rule. In another paper of ours [$30$], we discuss some exceptions
to the Madelung rule\footnote{%
https://en.wikipedia.org/wiki/Aufbau\_principle} and theoretical means of
explaining these exceptions. More details are still forthcoming.

\bigskip

\textbf{2. \ Known and unknown properties of Maxwell's fish-eye lenses}

\ \ \ \ \textbf{\ and potentials associated with them}. \textbf{Their role
in optics and}

\ \ \ \ \textbf{\ quantum mechanics \ }

\ \ 

Maxwell's fish eye lens was invented by James Clerk Maxwell. It is nicely
described in [6$],$[21],[22]. Its invention was motivated by the desire to
make perfect (absolute) optical instrument. In such an instrument it is
expected that \textsl{\ all} rays leaving point $\mathbf{x}$ meet again at
the (\textsl{conjugated or focal}) point $\mathbf{x}^{\prime }.$
Mathematically \ this can be expressed with help of the Fermat principle. It
is described as follows. Introduce the \textsl{optical path} length $S(%
\mathbf{x},\mathbf{x}^{\prime })$ via the length functional%
\begin{equation}
S(\mathbf{x},\mathbf{x}^{\prime })=\dint\limits_{0}^{t}dl,  \tag{2.1}
\end{equation}%
where the conformally flat metric $dl^{2}$ is given by $dl^{2}=n^{2}(\mathbf{%
x}(t))\left( \dfrac{d\mathbf{x}}{dt}\right) ^{2}dt^{2}$ \ with $n(\mathbf{x}%
(t))$ being position-dependent refractive index of the medium.\bigskip

\textbf{Fermat principle}: Light rays through spatial points $\mathbf{x(}0%
\mathbf{)}$ and $\mathbf{x}^{\prime }\mathbf{(}t)$ move on optically
shortest paths . \bigskip

Based on this definition, it follows that minimization of the functional,
Eq.(2.1), \ results in the Euler-Lagrange equations for geodesics in the
optical medium in analogy with equations for the point-like particle moving
in the gravitational field [$31$],[32]. We shall demonstrate below that such
\ noticed known optical-mechanical-gravitational correspondence is capable
of producing new \ results. Being armed with the above information, now we
are in the position to describe the origins of Maxwell's fish-eye lens
(additional information is given in Appendix A). \ For this purpose, \ let
us consider the $n-$dimensional sphere $S^{n}$ in the flat Euclidean space $%
\mathbf{R}^{n+1}$ described \ by 
\begin{equation}
S^{n}=\{(z_{1},...,z_{n};y)\mid \dsum\limits_{i=1}^{n}z_{i}^{2}+y^{2}=1\}. 
\tag{2.2}
\end{equation}%
The stereographic projection: from $\mathbf{R}^{n}$ to $S^{n},$ is given by 
\begin{equation}
z_{i}=\frac{x_{i}}{1+\frac{1}{4}\mathbf{x}^{2}},y=\frac{1-\frac{1}{4}\mathbf{%
x}^{2}}{1+\frac{1}{4}\mathbf{x}^{2}},i=1\div n,\text{ }x_{i}\in \mathbf{R}%
^{n},\mathbf{x}^{2}=x_{i}x^{i}.  \tag{2.3}
\end{equation}%
Here and below the summation over repeated indices is assumed. From the
mathematical literature \ [33], [34] it is known that the stereographic
projection is a conformal mapping since it is angle-preserving. This means
that using Eq.(2.3) it is possible to rewrite the metric $ds^{2}$ on the
sphere $S^{n}$ in terms of the metric on \textbf{R}$^{n},$ that is 
\begin{equation}
ds_{S^{n}}^{2}=\rho (\{x_{i}\})\left( d\mathbf{x}\right) ^{2},  \tag{2.4a}
\end{equation}%
where 
\begin{equation}
\rho (\{x_{i}\})=\left( \frac{1}{1+\frac{1}{4}\mathbf{x}^{2}}\right) ^{2}. 
\tag{2.4b}
\end{equation}%
An example of such relation, e.g. between the metric of the 2-sphere $S^{2}$
and the metric of the 2-plane \textbf{R}$^{2}$ is described in detail in
[34],page 88. \ Since \textbf{R}$^{2}+\{\infty \}=S^{2}$. The same is true
for the spheres of any dimension. The geodesics on $S^{n}$ are the great
circles. All great circles passing through some point $P$ on $S^{n}$
intersect again in the conjugate point $P^{\prime },$ that is in the
diametrally opposite point $-P$ (the antipodal map [33]). This circumstance
can be used for design of perfect optical \ instruments. \ Specifically, let
us adopt \ Eq.(2.4a) to 4 and 3 dimensions respectively. Then, in view of
Eq.(2.1), the conformal factor $\rho (\{x_{i}\})$ can be interpreted as the
square of refractive index $n^{2}(\mathbf{x}(t))$. This explains the origin
of Maxwell's fish-eye refractive index. Written \ with account of
dimensionality arguments, the refractive index is given by 
\begin{equation}
n(\mathbf{x})=\frac{n_{0}}{1+\mathbf{x}^{2}/a^{2}}.  \tag{2.5}
\end{equation}%
The light rays in 3 dimensional Maxwell's fish-eye medium are images of the
great circles, \ that is of \textsl{the geodesics,} in $S^{3}.$ The
antipodal map $P\rightarrow -P,P\in S^{3},$ corresponds to the negative
inversion map $x_{i}^{\prime }=-\frac{x_{i}}{\left\vert \mathbf{x}%
\right\vert ^{2}},x_{i}\in \mathbf{R}^{3}$ from $\mathbf{R}^{3}$ to $\mathbf{%
R}^{3}.$ The antipodal points on $S^{3}$ correspond to points $\mathbf{x}%
_{1} $ and $\mathbf{x}_{2}$ in $\mathbf{R}^{3}$ which lie on a common line
through the origin, and satisfy $\mathbf{x}_{1}\cdot \mathbf{x}_{2}=-1.$ The
great circles on \ $S^{3}$ are just \ those circles (or lines through the
origin in $\mathbf{R}^{3}$) \ which map into themselves under the negative
inversion operation [33$].$

These facts, we hope, remind our readers about the classical (by Moser [$%
35,36])$ and quantum mechanical (Fock [37$])$ treatments of hydrogen atom
respectively. They were discussed to some extent in our previous work [3].
The connection between the description of optical (fish-eye refractive
index) instruments and the classical/quantum mechanical description \ of
hydrogen atom was discovered by Luneburg [21]. \ Some important \ additional
details (not of quantum mechanical nature though) were independently
developed by Caratheodory [2$2]$. \ They are discussed in Appendix A.
Contrary to the D-O claim [7] that the Maxwell's fish-eye and the Coulomb
(that is hydrogen atom) problems are "closely related" , Luneburg found the
exact mapping between these problems. To avoid further literature
interpretation ambiguities, we present Luneburg's arguments below.\bigskip

To shorten our exposition to the absolute minimum, we extensively use the
results of our previous work [3]. It also discusses results of Luneburg's
book [21$]$, albeit with different purpose. Thus, we begin with Eq.(4.11) of
[3]\ for the wavefront. We write it in Luneburg's notations (e.g. see
Eq.(28.13) of [21$]),$ 
\begin{equation}
\psi _{x}^{2}+\psi _{y}^{2}+\psi _{z}^{2}=n^{2}(x,y,z)  \tag{2.6}
\end{equation}%
and select for the refractive index 
\begin{equation}
n^{2}(x,y,z)=C+\frac{1}{\sqrt{x^{2}+y^{2}+z^{2}}}.  \tag{2.7}
\end{equation}%
If this equation is to be used quantum mechanically, Eq.(2.6) must be
rewritten in the form of Eq.(4.21) of [21], that is in the form 
\begin{equation}
\psi _{x}^{2}+\psi _{y}^{2}+\psi _{z}^{2}=2m(E-V).  \tag{2.8a}
\end{equation}%
Then, following Scrh\"{o}dinger, the fully reversible substitution: $\psi
\rightleftarrows \hbar \ln \varphi $ \ done in Eq.(2.8a) is converting it
into Eq.(4.22) of [3]. That is into 
\begin{equation}
\varphi _{x}^{2}+\varphi _{y}^{2}+\varphi _{z}^{2}=\frac{2m}{\hbar ^{2}}%
(E-V)\varphi ^{2}.  \tag{2.8b}
\end{equation}%
Eq.s (2.8a) and (2.8b) are \textsl{equivalent classical} Hamilton Jacobi
(H-J) equations \textsl{even though} Eq.(2.8b) \textsl{has} $\hbar $ \textsl{%
in it}! The rationale of rewriting Eq.(2.8a) into Eq.(2.8b) lies in the fact
that the stationary Schr\"{o}dinger equation is obtainable from Eq.(2.8b)
variationally. That is by minimization of the functional%
\begin{equation}
J[\varphi ]=\frac{1}{2}\dint d^{3}x[\left( \mathbf{\nabla }\varphi \right)
^{2}-\frac{2m}{\hbar ^{2}}(E-V)\varphi ^{2}]  \tag{2.9}
\end{equation}%
under subsidiary condition%
\begin{equation}
\dint d^{3}x\varphi ^{2}=1  \tag{2.10}
\end{equation}%
the Schr\"{o}dinger equation 
\begin{equation}
\left( \frac{\partial ^{2}}{\partial x^{2}}+\frac{\partial ^{2}}{\partial
y^{2}}+\frac{\partial ^{2}}{\partial z^{2}}\right) \varphi +\frac{2m}{\hbar
^{2}}(E-V)\varphi =0  \tag{2.11}
\end{equation}%
is obtained. As discussed in [3] the quantum solutions of Eq.(2.11) are also 
\textsl{exact solutions} of the classical Eq.s(2.8a), (2.8b). Because of
this, it is sufficient to look for solutions of quantum Schrodinger's
equation \ by using classical H-J Eq.s (2.8). This \textbf{fundamentally} 
\textbf{nontrivial} fact was exploited by Luneburg and, apparently, was\
overlooked by Demkov and Ostrovsky [7] who wrote\ that "Maxwell's fish-eye
problem \ is \textsl{closely related to the Coulomb }(that is hydrogen atom)
problem". By comparing Eq.s(2.6),(2.7) with Eq.(2.8a) we realize that
Luneburg selected the system of units in which $C=E,2m=1$ and $e^{2}=1.$ \
We hope that his choice should not cause any difficulty in reading.
Therefore, we shall proceed along with Luneburg's choices for the time
being. \ Furthermore, in addition he performs the following nontrivial
Legendre transformation (which is canonical in nature as it will be
explained below) 
\begin{eqnarray}
\xi &=&\psi _{x},\eta =\psi _{y},\zeta =\psi _{z},  \notag \\
x &=&\omega _{\xi },y=\omega _{y},z=\omega _{z},  \notag \\
\psi +\omega &=&x\xi +y\eta +z\zeta  \TCItag{2.12}
\end{eqnarray}%
resulting in 
\begin{equation}
\xi ^{2}+\eta ^{2}+\zeta ^{2}=C+\frac{1}{\sqrt{\omega _{\xi }^{2}+\omega
_{y}^{2}+\omega _{z}^{2}}}  \tag{2.13a}
\end{equation}%
or, equivalently,%
\begin{equation}
\omega _{\xi }^{2}+\omega _{y}^{2}+\omega _{z}^{2}=\left( \frac{1}{-C+\xi
^{2}+\eta ^{2}+\zeta ^{2}}\right) ^{2}.  \tag{2.13b}
\end{equation}%
By comparing Eq.s (2.5)-(2.7) and (2.13) we realize that for negative
energies, $C=-\left\vert E\right\vert .$ That is presence of the constant $C$
accounts for the spectrum of bound states of the Coulomb (hydrogen atom)
problem. The r.h.s. of Eq.(2.13) is describing the fish-eye potential. \
Luneburg's choice $C=-1$ \ apparently caused Demkov and Ostrovsky to
conclude that the Schr\"{o}dinger equation originating from the quantum
version of the H-J Eq.(2.13b) employing the fish-eye potential should be
written with $E=0$ (not to be confused with the choices for $C$, though).
This conclusion is correct and is of fundamental importance to all
discussions which follow. However, it is \ still fundamentally misleading
since the condition $E=0$ for the Schr\"{o}dinger equation with the fish-eye
potential is \textbf{not} related to the energy of the Coulombic problem as
we just have demonstrated! Much more on thie topic is presented in section 4
and Appendix F. Alternative arguments are presented below.

Specifically, we write the Schr\"{o}dinger equation with the fish-eye
potential as 
\begin{equation}
(-\left( \frac{\partial ^{2}}{\partial x^{2}}+\frac{\partial ^{2}}{\partial
y^{2}}+\frac{\partial ^{2}}{\partial z^{2}}\right) -\left( \frac{n_{0}}{%
\left\vert E\right\vert +x^{2}+y^{2}+z^{2}}\right) ^{2})\psi =0  \tag{2.14a}
\end{equation}%
or, accounting for the discreteness of $\left\vert E\right\vert \equiv A$, as%
\begin{equation*}
(-\left( \frac{\partial ^{2}}{\partial \frac{x^{2}}{A}}+\frac{\partial ^{2}}{%
\partial \frac{y^{2}}{A}}+\frac{\partial ^{2}}{\partial \frac{z^{2}}{A}}%
\right) \frac{1}{A}-\left( \frac{n_{0}}{A}\right) ^{2}\left( \frac{1}{1+%
\frac{x^{2}}{A}+\frac{y^{2}}{A}+\frac{z^{2}}{A}}\right) ^{2})\psi =0,
\end{equation*}%
or,equivalently, 
\begin{equation}
(-\left( \frac{\partial ^{2}}{\partial x^{2}}+\frac{\partial ^{2}}{\partial
y^{2}}+\frac{\partial ^{2}}{\partial z^{2}}\right) -\beta _{n}\left( \frac{1%
}{1+x^{2}+y^{2}+z^{2}}\right) ^{2})\psi =0  \tag{2.14b}
\end{equation}%
resulting in the manifestly discrete constant $\beta _{n}\simeq \frac{%
n_{0}^{2}}{A}\footnote{%
The problem of solving the Schr\"{o}dinger equation resulting in discretness
of $\beta _{n}$ is known as Sturmian problem. It is discussed in detail in
section 4 and Appendix F.}.$ To make a comparison with the Schr\"{o}dinger
equation for hydrogen atom, in view of Eq.s (2.11),(2.13), we have: a) to
choose system of units in which $\hslash =1,$b) to replace $n_{0}$ by $%
Ze^{2},$ c) to multiply the Laplacian by factor of 1/2 (kinetic energy
standard form). The combination $\beta _{n}\simeq \frac{n_{0}^{2}}{A}$ is
then being replaced by $\beta _{n}=\frac{\left( Ze^{2}\right) ^{2}}{%
\left\vert E_{n}\right\vert }.$In the case of 2 dimensions the \ discrete
values of $\beta _{n}$ were calculated by Makowski and Gorska [38]. The
reason why this was done in two dimensions is worthy of discussion. \bigskip
We shall do so in sections 3 and 4 and Appendix F.\bigskip \medskip

The extension of idea of designing perfect optical instrument is based on
the assumption that typically such instruments are axially symmetric. Also,
the Euclidean space \textbf{R}$^{3}$ in which the fish-eye potential lives
can be foliated by the two-spheres $S^{2}.$ The stereographic projection,
Eq.s(2.2),(2,3) adopted to $S^{2}$ in notations of Luneburg [21] reads%
\begin{equation}
X=\frac{2x}{1+r^{2}},Y=\frac{2y}{1+r^{2}},Z=\frac{1-r^{2}}{1+r^{2}}%
,r^{2}=x^{2}+y^{2},  \tag{2.15}
\end{equation}%
so that the metric is converted into 
\begin{equation}
ds^{2}=dX^{2}+dY^{2}+dZ^{2}=4\left( d\frac{x}{1+r^{2}}\right) ^{2}+4\left( d%
\frac{y}{1+r^{2}}\right) ^{2}+\left( d\frac{r^{2}-1}{r^{2}+1}\right) ^{2}=%
\frac{4}{\left( 1+r^{2}\right) ^{2}}(dx^{2}+dy^{2}).  \tag{2.16}
\end{equation}%
In view of axial symmetry it is possible to replace Eq.s(2.4a),(2.4b) by
their 2-dimensional analog%
\begin{equation}
ds^{2}=n^{2}(x,y)(dx^{2}+dy^{2}),n^{2}(x,y)=\frac{4}{\left( 1+r^{2}\right)
^{2}}.  \tag{2.17}
\end{equation}%
Let now 
\begin{equation}
x=u(\xi ,\eta ),y=v(\xi ,\eta ).  \tag{2.18}
\end{equation}%
and introduce a complex z-plane variable via 
\begin{equation}
z=f(\xi +i\eta )=u(\xi ,\eta )+iv(\xi ,\eta ).  \tag{2.19}
\end{equation}%
The Cauchy-Riemann equations imply%
\begin{equation}
\left\vert dz\right\vert ^{2}=(dx^{2}+dy^{2})=\left\vert f_{z}(z)\right\vert
^{2}(d\xi ^{2}+d\eta ^{2})\text{ with }\left\vert f_{z}(z)\right\vert
^{2}=u_{\xi }^{2}+v_{\xi }^{2}=u_{\eta }^{2}+v_{\eta }^{2}  \tag{2.20}
\end{equation}%
By comparing Eq.s(2.17),(2.20) we obtain:%
\begin{equation}
n(\xi ,\eta )=\frac{2}{1+r^{2}}\left\vert f_{z}(z)\right\vert =\frac{2}{%
1+\left\vert f(z)\right\vert ^{2}}\left\vert f_{z}(z)\right\vert . 
\tag{2.21}
\end{equation}%
Consider now $f(z)=z^{\gamma }$, $\gamma \geq 1.$ Using Eq.(2.21), we obtain,%
\begin{equation}
n(r)=\frac{2\gamma r^{\gamma -1}}{1+r^{2\gamma }}.  \tag{2.22}
\end{equation}%
For $\gamma =1$ we recover the fish-eye potential reduced to 2-dimensional
space defined in Eq.(2.17). Since $r=\sqrt{x^{2}+y^{2},}$ it is apparently
permissible\footnote{%
Actually, it is not permissible!. The rest of our \ 2 parts work is devoted
to correction of this unintensional error by D-O.} by continuity and
complementarity to extend this result to 3 dimensions as it was done by D-O
and the rest of authors who followed D-O to obtain the extended fish-eye
like \ potential (index of refraction) in 3 dimensions 
\begin{equation}
n(r)=\frac{\beta _{n}\left( r/a\right) ^{\gamma -1}}{1+\left( r/a\right)
^{2\gamma }}=\frac{a}{r}\frac{\beta _{n}}{\left( r/a\right) ^{-\gamma
}+\left( r/a\right) ^{\gamma }}.  \tag{2.23}
\end{equation}%
Here $r=\sqrt{x^{2}+y^{2}+z^{2}.}$ The constant $a$ \ making the ratio
dimensionless is specified in section 4. The $n(r)$ defined by Eq.(2.23)
should not be confused with the analogous result used in 2 dimensional
calculations. For $\gamma =1,$ by design,\ such a potential coincides with
that presented in Eq.(2.5) as required. The D-O potential, used for the
whole periodic system, is obtained by selecting $\gamma =1/2$ in Eq.(2.23)$.$
This choice is motivated by some classical and semiclassical arguments
apparently allowing to by-pass restrictions of the classical Bertrand
theorem. This hypothesis by D-O is incorrect in the form discussed by D-O. \
Instead, it is studied in detail in section 5 from the point of view of
Bertrand spaces discovered by Volker Perlick. There are yet other, less
sophisticated, arguments causing selection of the exponent $\gamma =1/2$.
First of them lies in the fact that \textsl{only for} \textsl{such exponent
it is possible to repeat word-for -word scaling calculations done in}
Eq.s(2.14 a) and (2.14b) and to get the quantization of the coupling
constant $\beta _{n}$ in a manner used for the fish-eye potential$.$ Such a
quantization \textbf{was postulated} by D-O. They entirely overlooked the
consequences of the scaling analysis. Second of them lies in the observation
that, when used in the Schr\"{o}dinger Eq.(2.14), such \ a potential
coincides exactly (up to scaling) with that proposed by Tietz [28],[29] 
\footnote{%
See also a paper by Belokolos [39$]$ who used the WKB method applied to
Schrodinger's equation with Tietz potential for proving the validity of the
Madelung rule.}, as discussed in section where it was also mentioned that
the Tietz potential wonderfully coincides with that obtained with help of
the Hartee-Fock calculations [25].\bigskip Now, we would like to connect
these results with those by Konstantin Caratheodory presented in Appendix A.

Eq.(A.5) of \ this appendix A coincides exactly with Eq.(3.4) of lecture
notes by Dirac [40] on quantum mechanics of the dynamical systems with
constraints. It is also given in Part II, section 4, of the classical book
by Arnol'd [41]. More is presented in the book by Oliver Johns[42] and in [43%
$]$. We follow, in addition, Ref.s[44] and [45$]$ in which modern treatment
of constrained systems is given in the context of gauge fields and
time-dependent processes. Thus, in Eq.(A.5) we shall assume that $t=q_{0}$($%
\tau )$ so that $\dfrac{dt}{d\tau }=\mathring{q}_{0}$ . Using these changes
of notations, we obtain as well: $\mathring{q}_{^{i}}=\QDABOVE{1pt}{dq_{^{i}}%
}{d\tau }=\dot{q}_{^{i}}\dfrac{dt}{d\tau },i=1\div n,$ so that $L^{\prime
}(t^{\prime },q_{i}^{\prime },\dfrac{dq_{i}^{\prime }}{dt})=L^{\ast }(\tau
,q_{i},\mathring{q}_{^{i}}).$ If initially $L(t,q_{i},\dot{q}_{i})=\frac{1}{2%
}g_{ij}\dot{q}_{i}\dot{q}_{j}-V(\{q_{i}\})$ then, finally we obtain: $%
L^{\ast }(q_{0},q_{i},\mathring{q}_{0},\mathring{q}_{i})=(\frac{1}{2}g_{ij}%
\mathring{q}_{i}\mathring{q}_{j}/\left( \mathring{q}_{0}\right)
^{2}-V(\{q_{i}\}))\mathring{q}_{0}.$ The canonical momenta now are given as
follows:%
\begin{equation}
p_{i}=\frac{\partial L^{\ast }}{\partial \mathring{q}_{^{i}}}=g_{ij}\frac{%
\mathring{q}_{j}}{\mathring{q}_{0}}\text{ and }p_{0}=\frac{\partial L^{\ast }%
}{\partial \mathring{q}_{0}}=-\frac{1}{2}g_{ij}\mathring{q}_{i}\mathring{q}%
_{j}/\left( \mathring{q}_{0}\right) ^{2}-V.  \tag{2.24}
\end{equation}%
Since the first of these equations yields%
\begin{equation}
\frac{\mathring{q}_{j}}{\mathring{q}_{0}}=g^{ji}p_{i},  \tag{2.25}
\end{equation}%
its use in the second of Eq.s(2.24) produces 
\begin{equation}
p_{0}+\frac{1}{2}g^{ij}p_{i}p_{j}+V=0\text{ or }p_{0}+H(q_{i},p_{i})=0. 
\tag{2.26}
\end{equation}%
According to known rules of mechanics, the "true" Hamiltonian\ $\mathcal{H}$(%
$q_{\alpha },p_{\beta }),$ $\alpha ,\beta =0\div n,$ is obtainable \ now via
prescription 
\begin{equation}
\mathcal{H}(q_{\alpha },p_{\beta })=p_{i}\mathring{q}_{^{i}}+p_{0}\mathring{q%
}_{0}-L^{\ast }(q_{0},q_{i},\mathring{q}_{0},\mathring{q}_{i})=\mathring{q}%
_{0}(p_{0}+\frac{1}{2}g^{ij}p_{i}p_{j}+V)=0.  \tag{2.27}
\end{equation}%
Evidently, the quantized version of just obtained result can formally be
written as 
\begin{equation}
\mathcal{H}(\hat{q}_{\alpha },\hat{p}_{\beta })\psi (\{\hat{q}_{\alpha
}\})=0,  \tag{2.28}
\end{equation}%
where the hats denote the Hilbert space operators. Eq.(2.14a) belongs to the
class of \ "timeless\textquotedblright\ equations of the type defined by
Eq.(2.28),e.g.read [45]-[47]. This circumstance provides us with the
opportunity to illustrate general results described by Eq.s (2.27),(2.28) in
sufficient detail. These are to be used in the rest of this paper.\bigskip\
Our readers should be warned that the description of timeless equations in
quantum mechanics and gravity in full generality requires many volumes. For
the latest effort, e.g. read [45$].$ In gravitation the story also begins
with Eq.(2.26), e.g. read [46$],$ Eq. (1.11). A condensed summary is given
in [47$]$ by the same author. To keep focus on tasks to be completed in this
paper, we squeeze general description of time problem to the absolute
minimum. We shall study time-related issues using Eq.(2.14b) first because \
at least in 2 dimensions (rigorously)\ the fish-eye potential, Eq.(2.5), can
be replaced by the Lenz potential, Eq.(2.23), without D-O restrictions to
the case of $\gamma =1/2$. This can be achieved because the two dimensional
Laplacian is conformally invariant. The conformal transformations \ leading
to extension of the fish-eye potential discussed in Eq.s(2.15)-(2.22) are
fully compatible with \ those for the two-dimensional Laplacian. \ This \
fact allowed both classical and quantum mechanical problems to be studied in
detail in [38] and [48]. \ In 3 dimensions the situation \ seems intractable
because the Laplacian is no longer conformally invariant. Therefore,\textbf{%
\ by completely ignoring the issue of conformal invariance} \ the fish-eye
potential in Eq.(2.14a) was formally replaced by its 3-dimensional Lenz
extension \ by Ostrovsky [17$]$ and by Barut and Kitagawara [8,9] for the
special case $\gamma =1/2$. For arbitrary $\gamma \geq 1$ it was studied
already in [7$]$ and$,$ later on, in [49$],$ again, by totally neglecting
the issue of conformal invariance. Thus, in this work we are confronted with
the task of relating conformally the Schr\"{o}dinger Eq.(2.14a), where the
fish-eye potential is used legitimately, with the analogous 3-dimensional
Schrodinger's equation with the Lenz fish-eye-like potential whose use in
3-dimensions is lacking justification thus far. Our study begins with
rewriting Eq.s(2.13),(2.14)\textsl{\ classically} first. We obtain, 
\begin{equation}
(A+\left\vert \mathbf{r}\right\vert ^{2})^{2}\left\vert \mathbf{p}%
\right\vert ^{2}=n_{0}^{2}.  \tag{2.29}
\end{equation}%
Here $\left\vert \mathbf{r}\right\vert ^{2}=x^{2}+y^{2}+z^{2}$ and $%
\left\vert \mathbf{p}\right\vert ^{2}=\omega _{\xi }^{2}+\omega
_{y}^{2}+\omega _{z}^{2}.$ Up to rescaling the factor $(A+\left\vert \mathbf{%
r}\right\vert ^{2})^{2}\left\vert \mathbf{p}\right\vert ^{2}$ \ can be
identified with the classical Hamiltonian for the Kepler problem as
discussed in [$50$], Eq.(7). Upon quantization Eq.(2.29) becomes a
Sturmian-type problem. Our task, however, is to bring this result into the
form coinciding with Eq.(2.27) and to study the consequences.

For this purpose following [50$]$ we have to perform canonical
transformations analogous to those displayed in (2.12). This time, following
[50$]$ and [$51],$we would like to arrive at our final result more
systematically. Thus, we begin with Kepler's Hamiltonian 
\begin{equation}
H=\frac{1}{2}\left\vert \mathbf{p}\right\vert ^{2}-\frac{\alpha }{\left\vert 
\mathbf{r}\right\vert }  \tag{2.30}
\end{equation}%
and, using Eq.s (2.26),(2.27), we write 
\begin{equation}
0=(H-E)\frac{\left\vert \mathbf{r}\right\vert }{\alpha }=(\left\vert \mathbf{%
p}\right\vert ^{2}-2E)\frac{\left\vert \mathbf{r}\right\vert }{2\alpha }%
-1\equiv H_{0}.  \tag{2.31}
\end{equation}%
In the above equation, following Moser [35],[36], we used the Levi-Civita
time change%
\begin{equation}
\dfrac{dt}{d\tau }=\mathring{q}_{0}=\frac{\left\vert \mathbf{r}\right\vert }{%
\alpha }  \tag{2.32}
\end{equation}%
whose physical meaning is discussed in detail in Appendix B. Again,
following Moser [35], [36 ], and [51$],$ we introduce the new Hamiltonian
function $\mathcal{F}$ via%
\begin{equation}
\mathcal{F=}\frac{1}{2}(1+H_{0})^{2}=\frac{1}{8\alpha ^{2}}(\left\vert 
\mathbf{p}\right\vert ^{2}-2E)^{2}\left\vert \mathbf{r}\right\vert ^{2}. 
\tag{2.33}
\end{equation}%
But $H_{0}=0!$ Therefore, on isoenergetic surface $H=E$ or, which is the
same, on $\mathcal{F}=\frac{1}{2},$ the trajectories of the Hamiltonian flow
of $H$ traversed in time $t$ coincide with trajectories of the Hamiltonian
flow on $\mathcal{F}$ traversed in time $\tau .$ Details are provided in
Appendix B.

In this subsection, however, we still need to demonstrate that the
Hamiltonian $(A+\left\vert \mathbf{r}\right\vert ^{2})^{2}\left\vert \mathbf{%
p}\right\vert ^{2}$ introduced \ in Eq.(2.29) and the Hamiltonian $\dfrac{1}{%
8\alpha ^{2}}(\left\vert \mathbf{p}\right\vert ^{2}-2E)^{2}\left\vert 
\mathbf{r}\right\vert ^{2}$ introduced in Eq.(2.33) are canonically related.
Already in subsection 2.1. we mentioned that the stereographic projection is
a conformal mapping transformation. It can be proven that for dimensionality
greater than 2 the most general conformal transformation in \textbf{R}$^{n}$
is made out of sphere inversions, euclidean motions and scale
transformations [52]. Following [50$]$ we perform the inversion of
coordinates 
\begin{equation}
\left( x^{\prime }\right) ^{i}=\frac{x^{i}}{\left\vert \mathbf{x}\right\vert
^{2}},  \tag{2.34}
\end{equation}%
so that the corresponding canonical momenta $p_{i}^{\prime }$ =$\left(
\partial x^{j}/\partial \left( x^{\prime }\right) ^{i}\right) p_{j}$\ \
become%
\begin{equation}
p_{i}^{\prime }=\left\vert \mathbf{x}\right\vert ^{2}p_{i}-2x^{i}(\mathbf{x}%
\cdot \mathbf{p}).  \tag{2.35}
\end{equation}%
It can be checked\ explicitly that just described \ transformations are
canonical. Next, one should perform the additional (scale) canonical
transformation%
\begin{equation}
\bar{x}^{i}=\frac{p_{i}^{\prime }}{b},\text{ }\bar{p}_{i}=-b\left( x^{\prime
}\right) ^{i}  \tag{2.36}
\end{equation}%
where $b>0$ is an adjustable parameter . With these conformal
transformations and by adjusting the value of the parameter $b$, the
Hamiltonian $\dfrac{1}{8\alpha ^{2}}(\left\vert \mathbf{p}\right\vert
^{2}-2E)^{2}\left\vert \mathbf{r}\right\vert ^{2},$ $E<0,\left\vert
E\right\vert =A,$ is converted into Hamiltonian, Eq.(2.29), for the fish-eye
potential.\bigskip\ Quantization of the Schr\"{o}dinger equation with such
potential is discussed in detail in section 4.\bigskip\ Additional
independent information is provided in \ Appendix F.\bigskip \medskip

\textbf{3.} \textbf{Known and unknown roles of conformal invariance in
quantum mechanical}

\ \ \textbf{\ problems involving fish-eye potentials\bigskip \medskip }

In two dimensions by using conformal transformations we obtained Eq.(2.21) \
for the refractive index. \ Adapted to two dimensions, Schr\"{o}dinger's
Eq.(2.14a) can be rewritten as 
\begin{equation}
-(\frac{\partial ^{2}}{\partial x^{2}}+\frac{\partial ^{2}}{\partial y^{2}}%
)\psi -V(x,y)\psi =0.  \tag{3.1a}
\end{equation}%
In such a form this equation was studied by Makowski [53] for the fish-eye
and related potentials. By performing conformal transformation on Eq.(3.1a)
\ and using \ Eq.(2.21), Eq.(3.1a) \ can be rewritten in equivalent form%
\begin{equation}
-\left\vert f_{z}(z)\right\vert ^{2}(\frac{\partial ^{2}}{\partial \xi ^{2}}+%
\frac{\partial ^{2}}{\partial \eta ^{2}})\psi -\left\vert
f_{z}(z)\right\vert ^{2}U(\xi ,\eta )\psi =0  \tag{3.1b}
\end{equation}%
allowing the conformal factor $\left\vert f_{z}(z)\right\vert ^{2}$ to drop
out. Such an omission of the conformal factor is possible because the 2
dimensional Laplacian is conformally invariant. In dimensions higher than
two this is no longer true [54$].$ Thus, beginning with the two-dimensional
version of the fish-eye potential and by applying the conformal
transformations to this potential, e.g. to Eq.(2.22), and to the Laplacian,
Makowski [53] studied the Schr\"{o}dinger- type\footnote{%
Actually, the Sturmian-type.} equations for a large class of fish-eye-type
potentials. Other authors, e.g. see [7-9],[17$],[49]$ used the 3-dimensional
version of the generalized fish-eye potential, Eq.(2.23),\footnote{%
For $\gamma =1/2$ in [7-9],[17] and for arbitrary $\gamma >1$ in [49$].$In
view of scaling analysis of Eq.(2.14), arbitrary $\gamma ^{\prime }s$ are
not physically realizable though.} in the corresponding 3-dimensional Schr%
\"{o}dinger-type equations without any concern about the mathematical
correctness of these equations. By ignoring the issue of conformal
invariance, these authors had lost some fundamentally important physics
essential \ not only for solution of the L\"{o}wdin problem in atomic
physics but, more broadly, for \ solution of dynamical problems in non flat
Bertrand spaces. The role of Lie sphere and conformal geometry in finding
the solution of the Schr\"{o}dinger, the Klein-Gordon and other wave
equations \ was discussed in detail in our previous work [3]. Here we are
going to add some new details to the topic of conformal invariance by making
the unexpected connections between results of section 2 and \ our previous
work [3$].$

\bigskip\ We begin this section by connecting the results of Ref.[55$]%
\footnote{%
This paper is containing many typos causing us to rewrite needed equations}$
with those discussed in section 2 and Appendices A and B. Starting with the
Lagrangian 
\begin{equation}
L(t,q_{i},\dot{q}_{i})=\frac{1}{2}g_{ij}\dot{q}^{i}\dot{q}^{j}-V(\{q_{i}\}) 
\tag{3.2}
\end{equation}%
we are defining the conformal transformation 
\begin{equation}
\bar{g}_{ij}(q_{i})=\omega ^{2}(q_{i})g_{ij}(q_{i}),  \tag{3.3}
\end{equation}%
where $\omega ^{2}(q_{i})$ is some positive function of $\ $coordinates $%
\{q_{i}\}.$ Instead of previously used $\dfrac{dt}{d\tau }=\mathring{q}_{0}$
we write now 
\begin{equation}
d\tau =\omega ^{2}(q_{i})dt.  \tag{3.4}
\end{equation}%
In view of this result, the action 
\begin{equation}
S=\dint dtL(t,q_{i},\dot{q}_{i})  \tag{3.5a}
\end{equation}%
should be rewritten now as%
\begin{equation}
S=\dint \frac{d\tau }{\omega ^{2}(x)}[\frac{1}{2}\omega
^{4}(q_{i})g_{ij}q^{\prime i}q^{\prime j}-V(\{q_{i}\}],\text{ \ }%
q_{i}^{\prime }=\frac{dq_{i}}{d\tau },\text{\ }q_{{}}^{\prime
i}=g^{ij}q_{j}^{\prime },  \tag{3.5b}
\end{equation}%
or, equivalently,%
\begin{equation}
S=\dint d\tau \lbrack \frac{1}{2}\bar{g}_{ij}q^{\prime i}q^{\prime j}-\bar{V}%
(\{q_{i}\}]\equiv \dint d\tau \bar{L}(t,q_{i},q_{i}^{\prime }),  \tag{3.5.c}
\end{equation}%
where $\bar{g}_{ij}=\bar{g}_{ij}(q)$ is defined by Eq.(3.3) and $\bar{V}%
(\{q_{i}\})=\dfrac{V(\{q_{i}\})}{\omega ^{2}(q_{i})}.$ In Appendix B we
demonstrate that the Euler-Lagrange equation of motion for $L(t,q_{i},\dot{q}%
_{i})$ and $\bar{L}(t,q_{i},q_{i}^{\prime })$ produce the same trajectories.
It is of interest to reobtain this result now using different method for a
reason which will become \ clear upon reading.

For $L(t,q_{i},\dot{q}_{i})$ the Euler-Lagrange equations are%
\begin{equation}
\ddot{q}^{i}+\Gamma _{jk}^{i}\dot{q}^{j}\dot{q}^{k}+V^{,i}=0  \tag{3.6}
\end{equation}%
while for $\bar{L}(t,q_{i},q_{i}^{\prime })$ these equation acquire the
following form%
\begin{equation}
q^{\prime \prime i}+\hat{\Gamma}_{jk}^{i}q^{\prime j}q^{\prime k}+\frac{%
V^{,i}}{\omega ^{4}(q_{i})}-\frac{2V}{\omega ^{5}(q_{i})}\omega
(q_{i})^{,i}=0.  \tag{3.7}
\end{equation}%
Here, according to [52$],$ we have for the Christoffel symbols $\Gamma
_{jk}^{i}$ the following transformation law: 
\begin{equation}
\hat{\Gamma}_{jk}^{i}=\Gamma _{jk}^{i}+\delta _{j}^{i}(\ln \omega
),_{k}+\delta _{k}^{i}(\ln \omega ),_{j}-g_{jk}(\ln \omega )^{_{,}i}. 
\tag{3.8}
\end{equation}%
Since 
\begin{equation}
q^{\prime i}=\frac{dq^{i}}{dt}\frac{dt}{d\tau }\equiv \dot{q}^{i}\frac{1}{%
\omega ^{2}(q_{i})}  \tag{3.9a}
\end{equation}%
and 
\begin{equation}
q^{\prime \prime i}=\frac{1}{\omega ^{4}(q_{i})}[\ddot{q}^{i}-2\dot{q}^{i}%
\dot{q}^{j}(\ln \omega ),_{j}],  \tag{3.9b}
\end{equation}%
by replacing in Eq.(3.7) $q^{\prime \prime i}$ and $q^{\prime i}$\ by
Eq.s(3.9a) and (3.9b) and using Eq.(3.8) \ we finally obtain after some
algebra 
\begin{equation}
\ddot{q}^{i}+\Gamma _{jk}^{i}\dot{q}^{j}\dot{q}^{k}+V^{,i}-2(\ln \omega
)^{_{,}i}(\frac{1}{2}g_{ij}\dot{q}^{i}\dot{q}^{j}+V(\{q_{i}\})=0.  \tag{3.10}
\end{equation}%
Here 
\begin{equation}
\frac{1}{2}g_{ij}\dot{q}^{i}\dot{q}^{j}+V(\{q_{i}\})=E  \tag{3.11}
\end{equation}%
is an energy. By reversing arguments we also obtain the equations of motion
for primed variables as well as the analog of $E$ rewritten in primed
variables. Thus, we just obtained the \bigskip

\textbf{Theorem 3.1. }\textsl{\ Two dynamical systems are conformally
invariant \ at the level of their respective equations of motion if and only
if they both have vanishing total energies}.\bigskip

\textbf{Corollary 3.2.} From results of Appendix B it follows \ that the
dynamics of Reeb vector fields is conformally invariant. It is describing
the motion on geodesics.\bigskip

\textbf{Corollary 3.3. }Conformal\textbf{\ }invariance is result of time
changing transformations. They make time a canonical variable with
energy/Hamiltonian \ being its conjugate. Use of such transformations brings
the Newtonian mechanics into covariant form used for description of dynamics
of general relativity. More on this is presented in section 5 \ in
connection with description of the Eisenhart lift and Bertrand spaces.

\bigskip

Extension of these results to nonquantized field theories proceeds
analogously. Some representative examples are discussed in [55$].$ The
question arises: How just obtained results can be transferred to quantum
mechanics/quantum field theories? In 2 dimensions Eq.(3.1b) demonstrates
that this is indeed possible in the case of nonrelativistic quantum
mechanics. Therefore, we would like now to demonstrate that since the
Theorem 3.1. holds in higher dimensional spaces as well, there should be an
analog of this theorem in quantum mechanics/quantum field theories in higher
dimensions.\bigskip\ If we would follow standard textbooks on quantum
mechanics \ and replace the classical kinetic energy term by the term
involving the Laplacian (perhaps defined on the curved space) we would not
get the correct answer for conformally related \ quantum analogs of
conformally related classical systems. This is so because above two
dimensions the Laplacian is not going to remain form-invariant \ under the
conformal transformations. \ Instead, the \textsl{conformal} (Yamabe)
Laplacian $\square _{g}$ must be used\footnote{%
For a quick introduction to this topic we refer our readers to our works
[58,59].}. It is defined as follows%
\begin{equation}
\square _{g}=-\Delta _{g}+\alpha (d)R(g),  \tag{3.12a}
\end{equation}%
where $\alpha (d)=\frac{d-2}{4\left( d-1\right) }$ and $R(g)$ is the scalar
curvature of $d$-dimensional Riemannian manifold $\mathcal{M}_{d}$ whose
metric tensor is $g_{ij}$. \ In the case when the Laplacian $\Delta _{g}$ \
is acting on scalar function $\psi ,$ its action is defined as [60$]$:%
\begin{equation}
\Delta _{g}\psi =g^{ij}[\frac{\partial ^{2}\psi }{\partial ^{i}x\partial
^{j}x}-\Gamma _{ij}^{k}\frac{\partial \psi }{\partial x^{k}}].  \tag{3.12b}
\end{equation}%
Under the conformal change of metric $\tilde{g}=g\exp (2f)$ the conformal
Laplacian is transforming as follows 
\begin{equation}
\square _{\tilde{g}}=e^{-\left( \frac{d}{2}+1\right) f}\square _{g}e^{\left( 
\frac{d}{2}-1\right) f}.  \tag{3.13}
\end{equation}%
Ovsyannikov [59] and Ibragimov [60$]$ considered in \ detail the solutions
of partial differential equations of the following type%
\begin{equation}
F[u]=g^{ij}u_{ij}+b^{i}u_{i}+cu=0.  \tag{3.14}
\end{equation}%
Here $u_{i}=\dfrac{\partial u}{\partial x_{i}},u_{ij}=\dfrac{\partial ^{2}u}{%
\partial x_{i}\partial x_{ij}},g^{ij},b^{i}$ and $c$ are the known functions
of $x_{i}$ $,$ and $g^{ij}=g^{ji}.$

Following ideas of Hadamard discussed in our previous work [3$],$ Ibragimov
[60$],$ using work by Ovsyannikov [59], demonstrated that for spaces of
Riemannian signature use of Hadamard-type transformations of the type
discussed in [3] converts Eq.(3.14) into conformally invariant equation 
\begin{equation}
\square _{g}\Psi =0.  \tag{3.15}
\end{equation}%
In spaces of Lorentzian signature it is known as well, e.g. read Appendix B
of [3$]$ and [60$],$ section 12.2, Lemma 1., that the wave equation\footnote{%
Here $\Delta $ is the flat Laplacian} 
\begin{equation}
\square u\equiv \left( \frac{\partial ^{2}}{\partial t^{2}}-\Delta \right)
u=0  \tag{3.16}
\end{equation}%
adapted for spaces of Lorentzian signature having the pp-type metric (the
Penrose limit metric) is conformally invariant and obeys the Huygens
principle. Thus,\ in spaces \ with pp metrics the conformal invariance is
inseparably \ linked with the Huygens principle [60] for such an equation.
In [60$]$ it is demonstrated that equations of the type 
\begin{equation}
\square u+b^{i}u_{i}+cu=0  \tag{3.17}
\end{equation}%
with help of the Hadamard-like transformations are reducible to the standard
wave equation 
\begin{equation}
\square \Phi =0.  \tag{3.18}
\end{equation}%
This property causes Eq.(3.17) \ to belong to the class of equations obeying
Huygens' principle. Summarizing, we obtain the following\medskip

\textbf{Corollary 3.4. }\textsl{In spaces of Riemannian and pseudo
Riemannian signature Eq.s(3.14) and, respectively, (3.17) are conformally
invariant. \ For spaces of \ Lorentzian signature the conformal invariance
is inseparably linked with the validity of Huygens' principle. These
equations are quantum analogs of the conformally invariant dynamical
equations of classical mechanics.\bigskip }

The 3 dimensional analog of the 2-dimensional Schr\"{o}dinger Eq.(3.1a)
reads as follows%
\begin{equation}
-(\frac{\partial ^{2}}{\partial x^{2}}+\frac{\partial ^{2}}{\partial y^{2}}+%
\frac{\partial ^{2}}{\partial z^{2}})\psi +V(x,y,z)\psi =0.  \tag{3.19a}
\end{equation}%
Here, in view of Eq.(2.23), the potential $V(x,y,z)$ is given by 
\begin{equation}
V(x,y,z)=-\left( \frac{a}{r}\right) ^{2}\left[ \frac{n_{0}}{\left(
r/a\right) ^{-\gamma }+\left( r/a\right) ^{\gamma }}\right]
^{2},r^{2}=x^{2}+y^{2}+z^{2},a=const.  \tag{3.19b}
\end{equation}%
The quantum mechanical eigenvalue (Sturmian) problem for Eq.(3.19a) with
such a potential was formally solved in the foundational work Ref.[7$].$
Details were further elaborated in [49] for any $\gamma $. The case $\gamma
=1$ corresponds to the Maxwell's fish -eye potential (and, hence, to the
hydrogen atom) while the $\gamma =1/2$ case \ corresponds \ to the D-O
potential used by these authors for description of electronic structure of
the whole periodic system of elements. Corollary 3.4. guarantees that for $\
any$ $\gamma $ these equations are conformally equivalent to each other. But
in section 2 we found \ that Eq.s(2.14a) and (2.14b) make sence physically
only for $\gamma =1$ or $1/2.$

Given these facts, two questions arise: a) If these Eq.s(3.19) are
conformally equivalent, which one of these should be considered as the seed
from which others are being obtained? b) What makes Maxwell's fish-eye ($%
\gamma =1)$ and D-O ($\gamma =1/2)$ potentials so special if, in fact, they
are conformally equivalent?

To answer a) we have to remind to our readers some general facts about the
conformally related spaces. We begin with\medskip

\textbf{Definition 3.5.} The (pseudo) Riemannian space $\mathcal{M}_{d}$ of
dimension $d$ is \textsl{flat} if in some region around any point $x\in 
\mathcal{M}_{d}$ there is a metric-preferred coordinate system such that 
\begin{equation}
ds^{2}=g_{ij}(x)dx^{i}dx^{j}  \tag{3.20}
\end{equation}%
in which the metric tensor $g_{ij}(x)=e_{i}\delta _{ij},e_{i}=\pm 1.\bigskip 
$

\textbf{Definition \ 3.6.} \ A (pseudo) Riemannian space $\mathcal{M}_{d}$
is called \textsl{conformally flat }if it can be obtained from the flat
space via conformal change of metric tensor according to Eq.(3.3), that is
with help of transformation $\bar{g}_{ij}(x)=\omega ^{2}(x)g_{ij}(x)$ with $%
g_{ij}(x)$ \ defined in Eq.(3.20).\bigskip

\textbf{Definition 3.7. }When\textbf{\ }$\omega ^{2}(x)=const,$ the above
conformal transformation is called \textsl{homothety} (or \textsl{similarity}%
), when $\omega ^{2}(x)=1$ the homothety is\textsl{\ isometry}. The
isometric-type of motion on $\mathcal{M}_{d}$ is described in terms of
Killing's equations. The Killing vectors \textbf{X} are solutions of
Killing's equations%
\begin{equation}
\mathcal{L}_{\mathbf{X}}g_{ij}=0,  \tag{3.21}
\end{equation}%
where the Lie derivative $\mathcal{L}_{\mathbf{X}}$ is defined in
Eq.(B.5).\bigskip

\textbf{Definition 3.8.} The conformal type of motion on $\mathcal{M}_{d}$
is described in terms of the \textsl{conformal Killing equations}%
\begin{equation}
\mathcal{L}_{\mathbf{X}}g_{ij}=2\Phi (x)g_{ij}  \tag{3.22a}
\end{equation}%
in which \textbf{X} is the \textsl{conformal} Killing vector [61$]\footnote{%
For reader's convenience some basic information on Killing's equations is
presented in the Appendix C}.\bigskip $

Consider now two conformally related spaces, e.g. let $\bar{g}%
_{ij}(x)=e^{2\sigma (x)}g_{ij}(x)$. It is of interest to study the
relationship between Eq.(3.22a) and 
\begin{equation}
\mathcal{L}_{\mathbf{X}}\bar{g}_{ij}=2\bar{\Phi}(x)\bar{g}_{ij}.  \tag{3.22b}
\end{equation}%
It can be demonstrated [60$],[62]$ that 
\begin{equation}
\bar{\Phi}(x)=\Phi (x)+2\xi ^{i}\frac{\partial \sigma }{\partial x^{i}}. 
\tag{3.23}
\end{equation}%
Here \ the (Killing) vector $\xi ^{i}$ is defined by Eq.s (C.2),(C.3) in
Appendix C. Eq.(3.23) reminds us about the analogous \ cases \ in gauge
theories associated with the choices of gauges. In particular, the choice $%
\bar{\Phi}(x)=0$ implies that the underlying space with the metric $\bar{g}%
_{ij}$ is completely rigid. It admits only isometric (non bendable/ non
stretchable) motions. Although physically non realistic, such spaces do
exist mathematically [60$]$ and can be constructed based on solutions to the
the equation%
\begin{equation}
\Phi (x)+2\xi ^{i}\frac{\partial \sigma }{\partial x^{i}}=0.  \tag{3.24}
\end{equation}%
These spaces are serving as boundaries of \ conformally flexibe/bendable
spaces or can occur as \ subspaces of these spaces. In Appendix C we \
established that for the Riemannian spaces $\mathcal{M}_{d}$ of
dimensionality $d$ \ the isometric motion is described in terms of $\ $the\
(maximum) $\dfrac{d(d+1)}{2}$ parameters. The maximum is achieved for spaces
of constant curvature.

The question arises: What is the maximal number of parameters required for
description of conformally related spaces? \ 

To answer this question, we have to extend the analysis presented in
Appendix C. In particular, we have to count $d$ Killing vectors $\xi ^{i}$
as well as the factor $\Phi .$ Altogether there are $d+1$ such terms. In
addition, however, we should account for $d^{2}$ terms like $\nabla _{k}\xi
^{i}$ and $d$ terms like $\partial _{k}\Phi $. Their role is nicely
explained in [62$],$ section $1.$ Thus, altogether we obtain: $%
d^{2}+d+d+1=\left( d+1\right) ^{2}$ parameters. The truly nontrivial
conformal motion is described in terms of $\left( d+1\right) ^{2}-\dfrac{%
d(d+1)}{2}=\frac{1}{2}(d+1)(d+2)$ parameters. \ In [60$]$ there is a proof
of the following remarkable \bigskip

\textbf{Theorem 3.9. }\textit{(Pseudo) Riemannian space }$\mathcal{M}_{d}$%
\textit{\ has \textsl{nontrivial }conformal group of motions only if \ }$%
\mathcal{M}_{d}$\textit{\ is conformally equivalent to the conformally flat
space\bigskip \footnote{%
This theorem is compatible with \ thr existence of conformally flat Bertrand
spaces [18$]$ described in section 5 and Appendix E.}}

\textbf{Definition 3.10. }The conformal group is\textsl{\ trivial}\textbf{\ }%
if use of the relation\textbf{\ }$\bar{g}_{ij}(x)=\omega ^{2}(x)g_{ij}(x)$
leads to \textbf{\ }Eq.(3.24) possessing a nonzero solution.\bigskip

\textbf{Corollary 3.11.} According to Theorem C.1. The maximum number of
parameters $\dfrac{d(d+1)}{2}$ is possible \ only for spaces $\mathcal{M}%
_{d} $ of full isometry group which are of constant scalar curvature. But,
according to Corollary C.3., all spaces of constant scalar curvature are
conformally flat. This means that such manifolds have exactly\ $\frac{1}{2}%
(d+1)(d+2)$ conformal parameters involved in nontrivial conformal group
according to Theorem 3.9.\bigskip

This fact can be illustrated with help of Eq.s(3.12a) and (3.13). Suppose
that in Eq.(3.12a) the metric \~{g} is describing the space of constant
scalar curvature. Because such a space is conformally flat, we can write 
\begin{equation}
e^{-\left( \frac{d}{2}+1\right) f}\left( -\Delta _{g=1}\right) e^{\left( 
\frac{d}{2}-1\right) f}=\square _{\tilde{g}},  \tag{3.25}
\end{equation}%
where $g=1$ is actually $g=\delta _{ij}$ is the Euclideanized version of the
metric defined in Eq.(3.20). Thus, we are having two reference manifolds :
a) The manifolds of constant scalar (nonzero) curvature and b) The manifolds
whose scalar curvature is exactly zero. All other conformally flat manifolds
whose scalar curvatures are nonconstant can be related (deformed) either to
a) or to b)-type of manifolds (these are conformally equivalent, of course).
This was the point of departure of Yamabe's work [63]. An excellent modern
exposition of Yamabe ideas is given in [64$].$ For a rapid introduction to
this topic, please, consult our previous works [56$]$ and [57$].$

Based on just described results, we argue below that the Schr\"{o}dinger
equation for the Maxwell fish-eye potential ($\gamma =1)$ should be taken as
a seed.This will provide us an answer to the question a). Foregoing
development makes perfect sense physically since such an equation
corresponds to the \ stationary Schr\"{o}dinger equation \ for the hydrogen
atom. This is demonstrated in detail in section 4 and also below. It means
as well that \textbf{the equation proposed by D-O} ($\gamma =1/2)$ \textbf{%
for the description of spectra of other multielectron atoms is obtainable as}
\textbf{conformal deformation of the Schr\"{o}dinger equation \ for the
hydrogen atom.}\bigskip\ 

Such a deformation is to be described below. It is also discussed in the
Appendix F from the alternative standpoint. The results are fully compatible
with the modified Betrand theorem discussed in section 5.\ \bigskip
Following Ovsyannikov [59] we apply the Hadamard-type transformations
described in some detail in our previous work [3] culminating there in
Theorem 4.1.of that work. Below these transformations are described in much
greater detail with the purpose of transforming Eq.(3.14) into the D-O form,
Eq.(3.19) (firstly with with $\gamma =1$).

In compliance with Hadamard method [3]$,$ we consider now the 3 types of
Hadamard transformations in full generality.\bigskip

$\left( \alpha \right) $ The coordinate transformation $\ $%
\begin{equation}
x^{\prime i}=p^{i}(x),i=1,...,d.  \tag{3.26a}
\end{equation}

It leads to $u=u^{\prime }\circ p$ \ implying 
\begin{equation}
F^{\prime }[u^{\prime }]=g^{\prime ij}u_{ij}^{\prime }+b^{^{\prime
}i}u_{i}^{\prime }+c^{\prime }u^{\prime }=0.  \tag{3.26b}
\end{equation}%
In this equation $u_{i}^{\prime }=\frac{\partial }{\partial x_{i}^{\prime }}%
u^{\prime },$ etc$.,$ $g^{\prime kl}\circ
p=g^{ij}p_{i}^{k}p_{j}^{l},b^{\prime k}\circ
p=g^{ij}p_{ij}^{k}+b^{i}p_{i}^{k},$ $c^{\prime }\circ p=c,$ $p_{i}^{k}=\frac{%
\partial }{\partial x_{i}}p^{k},$ etc$.$

$\left( \beta \right) $ The scale transformation 
\begin{equation}
F^{\prime }[u]=e^{-\theta }F[e^{\theta }u].  \tag{3.27a}
\end{equation}%
It is converting Eq.(3.14) into 
\begin{equation}
F^{\prime }[u]=g^{ij}u_{ij}+b^{\prime i}u_{i}+c^{\prime }u=0.  \tag{3.27b}
\end{equation}%
Here $b^{\prime i}=b^{i}+2g^{ij}\theta _{j}$ and\ $c^{\prime }=c+b^{i}\theta
_{i}+g^{ij}\theta _{ij}+g^{ij}\theta _{i}\theta _{j}.\bigskip $

$\left( \gamma \right) $ The conformal transformation%
\begin{equation}
F^{\prime }[u]=e^{-2\theta }F[u].  \tag{3.28a}
\end{equation}%
It results in 
\begin{equation}
F^{\prime }[u]=\tilde{g}^{ij}u_{ij}+\tilde{b}^{i}u_{i}+\tilde{c}u=0. 
\tag{3.28b}
\end{equation}%
Here $g^{ij}=e^{2\theta }\tilde{g}^{ij},b^{i}=e^{2\theta }\tilde{b}^{i}$ and 
$c=e^{2\theta }\tilde{c}.\bigskip $

Transformations $\left( \alpha \right) $ and $\left( \beta \right) $ \ leave
Eq.(3.14) in the same (pseudo) Riemannian space $\mathcal{M}_{d}$ . The
transformation $\left( \gamma \right) $ \ conformally relates $\mathcal{M}%
_{d}$ to $\mathcal{\tilde{M}}_{d}.$

These statements require proofs. We shall not reproduce these proofs
referring our readers to Ovsyannikov's book [59$].$ Nevertheless, results of
these proofs cause us to introduce \ the scalar $H$ and the tensor $K_{ij}$
to be defined momentarily. For this purpose, following Ovsyannikov [59$],$
we have to define the contravariant vector $a^{i}$ via 
\begin{equation}
a^{i}=b^{i}\pm g^{kl}\Gamma _{kl}^{i},i=1,...,d.  \tag{3.29a}
\end{equation}%
Ovsyannikov uses only sign "+" in his calculations but \ below we shall
argue that, in fact, both signs should be used. \ Since under ($\alpha )-$%
type of transformations Christoffel's symbols transform as 
\begin{equation}
\Gamma _{ij}^{l}p_{l}^{k}=p_{i}^{k}+\Gamma _{ml}^{\prime
k}p_{i}^{m}p_{j}^{l},  \tag{3.30}
\end{equation}%
it can be demonstrated that the contravariant vector $a^{i}$ obeys the
standard rule of transformation for vectors: 
\begin{equation}
a^{\prime k}=b^{\prime k}\pm g^{\prime ml}\Gamma _{ml}^{\prime k}=(b^{i}\pm
g^{kl}\Gamma _{kl}^{i})p_{i}^{k}=a^{i}p_{i}^{k}.  \tag{3.29b}
\end{equation}%
With help of such defined vector $a^{i}$ it becomes possible to introduce a
skew-symmetric tensor $K_{ij}$ via 
\begin{equation}
K_{ij}=a_{i,j}-a_{j,i},  \tag{3.31}
\end{equation}%
where $a_{i,j}=\dfrac{\partial a_{i}}{\partial x^{j}}-a_{l}\Gamma _{ij}^{l}$
\ and the raising and lowering of indices is made with help of the metric
tensor $g_{ij},$ as usual. Accordingly, the scalar $H$ can now be defined as 
\begin{equation}
H=-2c+a_{,i}^{i}+\frac{1}{2}a^{i}a_{i}+2\alpha (d)R(g).  \tag{3.32}
\end{equation}%
Thus defined $H$ and $K_{ij}$ allow us to prove the following \bigskip

\textbf{Theorem 3.12. }$H$\textit{\ and }$K_{ij}$\textit{\ remain unchanged
\ under transformations of (}$\beta )-$\textit{type converting} Eq.(3.14) 
\textit{into} Eq.(3.27b). They are the invariants of \textit{(}$\beta )-$%
\textit{type transformations.}\bigskip

It can be also proven that with respect to transformations $\left( \gamma
\right) $ the tensor $K_{ij}$ \ will remain unchanged. So it is also an
invariant of $\left( \gamma \right) -$type transformations$.$ At the same
time, the scalar $H$ is transformed into $\tilde{H}=e^{-2\theta }H.$
Therefore, it looses its invariance property. The conformal Killing
Eq.s(3.22a) can be rewritten in a more familiar form now as 
\begin{equation}
\nabla _{i}\xi _{j}+\nabla _{j}\xi _{i}=\mu g_{ij},\text{ \ }\mu =2\Phi . 
\tag{3.33}
\end{equation}%
With help of these equations, it also can be demonstrated that%
\begin{equation}
\xi ^{k}H_{,k}+\mu H=0.  \tag{3.34a}
\end{equation}%
\ If $H\neq 0,$ it also can be demonstrated that it is always possible to
select $\theta $ in Eq.(3.28a) so that $H=e^{2\theta }.$ Such a selection
leads to $\tilde{H}=1.$ The transformed equation 
\begin{equation}
\xi ^{k}\tilde{H}_{,k}+\tilde{\mu}\tilde{H}=0  \tag{3.34b}
\end{equation}%
\ is implying that $\tilde{\mu}$ $=0$. \ Therefore, whenever $H=e^{2\theta
}, $ we are back to the Killing Eq.s (C.4) implying that we are dealing with
the isometry subgroup containing $\frac{\left( d+1\right) d}{2}$ generators 
\textsl{inside} \ larger \textsl{nontrivial} conformal group. This fact is
in accord with Yamabe's ideas [63$]$ that "more flexible" spaces $\mathcal{M}%
_{d}$ are conformally deformable into spaces of constant scalar curvature.

To make further progress, from now on we shall deal \ only with the
conformally flat metric, that is \ with the metric of the type $%
g^{ij}(x)=\rho (x)\delta ^{ij}($or with $g_{ij}(x)=\dfrac{1}{\rho (x)}\delta
_{ij}).$ \ To illustrate the value of just described results, we begin with
the simplest but instructive example.\bigskip \bigskip

Instead of considering Eq.(3.14), we shall focus now on Schr\"{o}dinger's
Eq.(3.19a) written in the form%
\begin{equation}
\rho \Delta u+cu=0  \tag{3.35}
\end{equation}%
in which we initially do not require $c(x)$ to be a constant and use the
notation $\Delta =\Delta _{g=1}$ as before$.$ This equation plays a
significal role also in section 5. By looking at Eq.(3.27b) and comparing it
with Eq.(3.14), we realize that in Eq.(3.35) $b^{^{\prime }i}=0$ and $c$ in
this equation should be actually relabelled as $c^{\prime }.$ By looking at
Eq.(3.29b) and by noticing that for the $\left( \beta \right) -$type
transformations $g^{ij}=g^{\prime ij}$ causing $\Gamma _{ml}^{\prime
k}=\Gamma _{kl}^{i}$, the combination $g^{kl}\Gamma _{kl}^{i}=\frac{d-2}{2}%
\frac{\partial }{\partial x^{i}}\ln \rho $ . This result is obtained in the
Appendix D. From Eq.(3.29b) we obtain, 
\begin{equation}
a_{i}^{\prime }=\pm \frac{d-2}{2}\frac{\rho _{i}}{\rho }\text{ or }a^{\prime
i}=\pm \frac{d-2}{2}\delta ^{ij}\rho _{j}.  \tag{3.36}
\end{equation}%
In view of Eq.(3.32) we have to calculate 
\begin{equation}
a_{,i}^{\prime i}=\frac{\partial a^{\prime i}}{\partial x^{i}}+a^{\prime
k}\Gamma _{ki}^{i}=\pm (\frac{d-2}{2}\delta ^{ij}\rho _{ij}-\frac{d(d-2)}{4}%
\delta ^{ij}\frac{\rho _{i}\rho _{j}}{\rho }),  \tag{3.37}
\end{equation}%
where we used Eq.(D.8) of the Appendix D for calculation of $\Gamma
_{ki}^{i}.$ Besides, 
\begin{equation}
\frac{1}{2}a^{\prime i}a_{i}^{\prime }=\frac{\left( d-2\right) ^{2}}{8}%
\delta ^{ij}\frac{\rho _{i}\rho _{j}}{\rho }.  \tag{3.38}
\end{equation}%
In view of Eq.(3.32), by combining Eq.s(3.37) and (3.38) we obtain: 
\begin{equation}
a_{,i}^{\prime i}+\frac{1}{2}a^{\prime i}a_{i}^{\prime }=\frac{\left(
d-2\right) ^{2}}{8}\delta ^{ij}\frac{\rho _{i}\rho _{j}}{\rho }-\frac{d^{2}-4%
}{4}\delta ^{ij}\frac{\rho _{i}\rho _{j}}{\rho }.  \tag{3.39}
\end{equation}%
In calculating this result we selected the sign "+" in Eq.(3.36). This is
caused by the following considerations. For spaces of constant curvature the
conformal factor $\rho $ was obtained in Appendix C. By adopting the
Euclidean version of the metric given in Eq.(C.7), small calculation using
Eq.s(3.36),(3.37) results in%
\begin{equation}
a_{,i}^{\prime i}+\frac{1}{2}a^{\prime i}a_{i}^{\prime }=\frac{d(d-2)}{2}K. 
\tag{3.40}
\end{equation}%
Going back to Eq.(3.32) and taking into account that $\dfrac{d-2}{2\left(
d-1\right) }=2\alpha (d)$ and \ that, according to [65$],$ for spaces of
constant scalar curvature $R(g)=-d(d-1)K,$ we obtain: $2\alpha (d)R(g)=-%
\dfrac{d(d-2)}{2}K$. By combining this result with Eq.(3.40) and using it in
Eq.(3.32) we obtain: 
\begin{equation}
H^{\prime }=-2c^{\prime }+a_{,i}^{\prime i}+\frac{1}{2}a^{\prime
i}a_{i}^{\prime }+2\alpha (d)R(g)=-2c^{\prime }.  \tag{3.41}
\end{equation}%
But since $H$ is invariant of $\left( \beta \right) $-transformations, by
construction, we must have $H^{\prime }=H.$ In view of Eq.(3.41), this is
only possible when $c^{\prime }$ is a constant. In such a case Eq.(3.35) is
\ the stationary Schr\"{o}dinger's equation for the hydrogen atom according
to Eq.s(3.19a) and (3.19b) where we have to select $\gamma =1.\bigskip $

Using Eq.s(3.12a) and (3.12b) we consider now the following equation 
\begin{equation}
F[u]=\square _{g}u=-\rho u_{,i}^{i}+\frac{d-2}{2}\rho ^{i}u_{i}+\alpha
(d)R(\rho )u=0.  \tag{3.42}
\end{equation}%
Clearly, it is of exactly the same type as Eq.(3.14). Therefore, in
Eq.(3,42) we must have $b^{i}=-\frac{d-2}{2}\rho ^{i},$ $c=-\alpha (d)R(\rho
).$ \ At first, we consider this equation for the space of constant scalar
curvature. We already know that in such a case $2\alpha (d)R(g)=-\dfrac{%
d(d-2)}{2}K.$ Using this result we can replace the term $c=-\alpha (d)R(\rho
)$ by $c=$ $\dfrac{d(d-2)}{4}K.$ Now we want to compare Eq.(3.42) with
Eq.(3.35) since they both have the same $\rho $ factor. Recall that the $%
\left( \beta \right) -$type transformations do not change the metric of the
underlying space. Since this is so, the equality of invariants $H=H^{\prime
} $ leads to the result: 
\begin{equation}
\text{ }a_{,i}^{i}+\frac{1}{2}a^{i}a_{i}=-2c^{\prime }+d(d-2)K=const. 
\tag{3.43}
\end{equation}%
To check this result we notice that the contravariant vector $a^{i}$ is
defined by Eq.(3.29a). The term $g^{kl}\Gamma _{kl}^{i}$ in this equation is
calculated in the Appendix D, Eq.(D.2). Therefore, by selecting the "+" \
sign we obtain,%
\begin{equation}
a^{i}=b^{i}+g^{kl}\Gamma _{kl}^{i}=-\frac{d-2}{2}\rho ^{i}+\frac{d-2}{2}\rho
^{i}=0,  \tag{3.44}
\end{equation}%
implying $c^{\prime }=\frac{d(d-2)}{2}K.$ Thus, $c^{\prime }=2c$ which makes
physical sense. Notice, in arriving at this result Eq.(3.44) \ was used in
which we have not made any specification of the conformal factor. The result 
$a^{i}=0$ was obtained \ by Ibragimov [60$],$ page120$,$ by different \ set
of arguments. Just obtained result gives us an opportunity to take also a
look at the Yamabe Laplacian for the metric \textsl{other} than that of
constant scalar curvature. To do so we put the factor $\psi $ in Eq.(D.7)
temporarily to zero in view of Eq.(3.44). Then, in view of Eq.s
(3.13),(3.27a), we write $e^{\theta }=\left( e^{2f}\right) ^{\frac{d-2}{4}}$%
. To utilize fully Eq.(3.13) we have to use Eq.(3.28a), that is we have to
use the $\left( \gamma \right) $ transformation which \ is connecting
equations with conformally related metrics. To facilitate matters, we notice
that the combination $e^{2f}e^{-\theta }F[e^{\theta }u]$ in which $e^{\theta
}=\left( e^{2f}\right) ^{\frac{d-2}{4}}$ produces exactly the same result $%
e^{-\left( \frac{d}{2}+1\right) f}F[e^{\left( \frac{d}{2}-1\right) f}u]$ as
was recorded already in Eq.(3.13) for the Yamabe Laplacian. This provides us
with an independent check of correctness of our computations since the
conformal factor $e^{2f}$ is present in the equation $\bar{g}%
_{ij}(x)=e^{2f(x)}g_{ij}(x)$ defining conformally related spaces.

We still have to comment on several topics.\bigskip\ First, \ we need to
comment on the role of $\left( \alpha \right) $-transformations. Such type
of transformations are described in detail in the monumental work, [66$],$
Chapter 3$.$ These are needed to bring the PDE's, Eq.(3.14), into the
canonical/standard form. Also these transformations had been used for
defining the contravariant vector $a^{i}.$ Since we are using only the
metric of the type $g^{ij}(x)=\rho (x)\delta ^{ij}$, this spares us from the
necessity to consider $\left( \alpha \right) $ transformations further.
Second, in using the $\left( \beta \right) $-type transformations we are
replacing the coefficients $b^{i}$ \ and $c$ by $b^{\prime i}$ and $%
c^{\prime }$ using the same factor $\theta .$ This means that by properly
selecting $\theta $ we can always replace the factor $c$ by the pre assigned
factor $c^{\prime }.$ However, in such a case the factor $b^{\prime i}$
cannot be forced to become zero. \ For this to happen, in addition to
parameter $\theta ,$ the parameter $\psi $ defined in Eq.(D.5) should be
used. If we would select the sign "-" initially in Eq.(3.29a), the result $%
a^{i}=0$ in Eq.(3.44) will change into 
\begin{equation}
a^{i}=-\left( d-2\right) \rho ^{i}.  \tag{3.45}
\end{equation}%
This is helpful if we want to use Eq.s(D.3)-(D.7) from Appendix D.

These equations should be used as follows.

a) Using the equation $c^{\prime }=c+b^{i}\theta _{i}+g^{ij}\theta
_{ij}+g^{ij}\theta _{i}\theta _{j}$ with pre assigned $c^{\prime }$ factor \
we can

determine \ the factor $\theta .$ For this purpose we need to solve the PDE
equation

$\ g^{ij}\theta _{ij}+g^{ij}\theta _{i}\theta _{j}+b^{i}\theta
_{i}=c^{\prime }-c$ . Its solution is greatly facilitated by our use

of spherical symmetry.

b) Use of this result in Eq.(D.7) determines the factor $\psi $.

c) After these two factors are determined, equations $b^{\prime
i}=b^{i}+2g^{ij}\theta _{j}$ and

$c^{\prime }=c+b^{i}\theta _{i}+g^{ij}\theta _{ij}+g^{ij}\theta _{i}\theta
_{j}$ yield the one-to-one relations between $c^{\prime }$

and $c$ and $b^{\prime i}$ and $b^{i}$ in such a way that we can put $%
b^{\prime i}=0.$

In treating the hydrogen atom case to keep $b^{\prime i}=0$ is essential. \
The hydrogen atom case was treated based on Eq.(3.19) where we had selected $%
\gamma =1.$ But for $\gamma \neq 1$ the same equation can also be brought
into canonical form 
\begin{equation}
\rho (\frac{\partial ^{2}}{\partial x^{2}}+\frac{\partial ^{2}}{\partial
y^{2}}+\frac{\partial ^{2}}{\partial z^{2}})u+\tilde{c}^{\prime }u=0 
\tag{3.46a}
\end{equation}%
in which 
\begin{equation}
\rho =\left[ \tilde{V}(x,y,z)\right] ^{-1},\text{ }\tilde{c}^{\prime }=const.
\tag{3.46b}
\end{equation}

In such a form it is used in section 5. Eq.(3.35) is our physical analog of
Eq.(3.27b). In Eq.(3.27b), we have: $\rho =\left[ V(x,y,z)\right] ^{-1},$ $%
b^{\prime i}=0$ and $\tilde{c}^{\prime }=const.$ Next, we use Eq.(3.42)
which is an analog of \ Eq.(3.14). In it, we use the $\rho -$ factor just
defined. Further, in this equation we have $b^{i}=-\frac{d-2}{2}\rho ^{i}$
and $c=-\alpha (d)R(\rho ),$ as required. Now, we need to relate Eq.(3.42)
with Eq.(3.46a) using $\left( \beta \right) $-type transformations. This is
achievable as we just described above. Next, by inverting these operations
and using transformation described in Eq.(3.13) \textbf{we conformally} 
\textbf{relate the D-O} Eq.(3.46a) \ \textbf{describing \ the multielectron
atoms \ with} Eq.(3.35) \textbf{describing the hydrogen atom}. To answer the
earlier posed question b), we follow analysis made by Ovsyannikov [59] in
Chr.8, section 10. In it he considers the Lie algebras of two versions of
Eq.(3.46a). One- with $\rho \neq 1$ and, another-with $\rho \equiv 1.$ In
the last case the Lie algebra is made of two types of generators: a)
translations: $\frac{\partial }{\partial x_{i}}$ and, b) rotations $%
x^{i}\partial _{j}-x^{j}\partial _{i}$. While in the first case it is 
\textbf{made of the same translations} as in the second, but the rotation
generators change now to ($2x_{i}x^{k}+(1-\rho ^{2})\delta _{i}^{k})\partial
_{k}-(d-2)x_{i}u\partial _{u},$ $i=1,...,d$. These two sets of Lie algebras
cannot be smoothly converted into each other. Only the Lie algebra for
Eq.(3.46a) with $\rho \neq 1$ makes physical sense. It conformally relates
the metric of constant scalar curvature (the hydrogen atom case) with the
modified metric \ induced by the presence of the rest of electrons.
\bigskip\ 

\textbf{4.} \textbf{From analysis to synthesis\bigskip }

Already in their first seminal work [$7$], Demkov and Ostrovsky discussed
method of solving Eq.s(3.19a),(3.19b) in full generality. Subsequently,
their results \ were further developed\ in [8, 17, 49, 67$],etc.$ These
papers differ from each other by the extent of generality of employing the
potential, Eq.(3.19b), in Eq.(3.19a)\footnote{%
E.g. some of the papers discussed only the physically relevant case: $\gamma
=1/2,$etc$.$}. In this section we shall entirely reconsider all these
results since none of them are without mistakes. One of the purposes of this
section is to correct these mistakes. As a by product new quantum mechanical
formalism of dealing with discrete spectrum of hydrogen and many-electron
atoms is developed. It is tested on known examples and, \ subsequently, it
is extended \ along the directions outlined in [7]. While doing so, we shall
connect results obtained in section 3 with those to be presented in this
section. We begin with \bigskip the following observation.

The 3-dimensional flat space Laplace equation%
\begin{equation}
\Delta \Psi (\mathbf{x})=0  \tag{4.1}
\end{equation}%
according to [68$],$[69$]$ admits 11 separable coordinate systems. At the
same time, there are 17 coordinate systems for the same Laplace equation
permitting R-separation of variables. \bigskip

\textbf{Definition 4.1. }An\textbf{\ }R-separable \ coordinate system \{%
\textit{u,v,w}\} for Eq.(4.1) is a coordinate system which permits a family
of solutions%
\begin{equation}
\Psi _{\lambda ,\mu }(\mathbf{x})=R(\mathit{u,v,w)}\text{\textit{U}}%
_{\lambda ,\mu }(u)V_{\lambda ,\mu }(v)W_{\lambda ,\mu }(w),  \tag{4.2}
\end{equation}%
where $\lambda $ and $\mu $ are separation constants \ and $R$ is a fixed
factor such that it is either $R\equiv 1$ (\textsl{pure separation}) or $%
R\neq 1$ and $R$ cannot then be written in the form $%
R=R_{1}(u)R_{2}(w)R_{3}(w).$\bigskip

This property of R-separation for the Laplace equation happens to be a
general trend. The PDE equations admit more \ non equivalent R-separable
coordinate systems than separable coordinate systems. Here we shall be
mainly concerned with the separation options for the conformal Laplacian,
Eq.s(3.12a),(3.15), and its classical Hamilton-Jacobi analog- the equation
for the\textsl{\ null geodesics} 
\begin{equation}
g^{ij}\partial _{i}W\partial _{j}W=0.  \tag{4.3}
\end{equation}%
Our interest in conformal (Yamabe) Laplacian and \ null geodesics equations
is coming from the results of Appendix B. In it we demonstrated that the
Reeb dynamics is always taking place on null geodesics. Quantum mechanically
this leads to study of solutions of the conformal (Yamabe) Laplacian
equation. The Eq.(4.3) is \textsl{additively separable. }This means that%
\textsl{\ } there is a system of coordinates $\{q_{i}\}$\ in which it can be
split into the system of ordinary differential equations \ implying the
solution in the form%
\begin{equation}
W=\dsum\limits_{i=1}^{d}W(q_{i};c_{\alpha }),c_{\alpha }\in \text{\textbf{R}%
, }\alpha =1,...,d.  \tag{4.4}
\end{equation}%
Here $c_{\alpha }$ are some known\ (pre assigned) constants. For the
conformal Laplacian and in Eq.(4.3) \ it is assumed that: a) $g^{ii}\neq 0$
and b) $g^{ij}=0$ for $i\neq j.$ This leads to separability in terms of
orthogonal coordinates. Both Eq.s(3.15) and (4.3) are \ manifestly
conformally invariant. Suppose now that $\Psi $ is R-separable solution of \
Eq.(3.15), i.e. of the equation $\square _{g}\Psi =0.$ Then, $e^{-\left( 
\frac{d}{2}-1\right) f}\Psi =\tilde{\Psi}$ is a also a solution of $\square
_{\tilde{g}}\tilde{\Psi}=0$.\bigskip

\textbf{Corollary 4.2.} If $\square _{g}\Psi =0$ \ possesses an R-separable
solution, then $\square _{\tilde{g}}\tilde{\Psi}=0$ also posses an
R-separable solution.\bigskip\ Thus, R-separability of the conformal
Laplacian is conformally invariant property. \bigskip

\textbf{Corollary 4.3. }The equation $\square _{g}\Psi =0$ is R-separable
only if the metric $g$ is conformally flat. A necessary condition for the
R-separability of Eq.(3.15) is for the Hamilton-Jacobi equation for the null
geodesic, Eq.(4.3), to be additively separable in the same separable
coordinates [70$]$.\bigskip \bigskip

\textbf{Remark 4.4}. Pure separation is a special case of the R-separation.
\bigskip This follows from Eq.(3.25).\bigskip

\textbf{Remark 4.5.} \bigskip R-separability is tightly linked with the
superintegrability. The superintegrability in turn is tightly linked with
the fact that for superintegrable systems all dynamical trajectories are
closed [71]\footnote{%
The latest applications of superintegrability in optics is given, for
example, in [72$].$Closed geodesics sometimes are called periodic geodesics
in mathematical literature [73$]$}. Superintegrable $2d$ dimensional
dynamical system has more than $d$ integrals of motion. \ Those systems
which have $d$ integrals of motion are just integrable. Since for
superintegrable systems all dynamical trajectories are closed brings us
instantly to the problematics of Betrand spacetimes [20$],$[74$]$. Dynamics
in such spaces is superintegrable by definition. Accordingly, such dynamics
admits quantization. Further details on superintegrability and Bertrand
spaces are presented in section 5.\bigskip

Based on the results of previous section these properties single out Schr%
\"{o}dinger's Eq.(3.19a) (in which we have to put $\gamma =1)$ for the
hydrogen atom. Other physically meaningful solutions are obtainable from
that for hydrogen atom with help of conformal transformations by using
Eq.(3.13).

In [7, 8$,17,49,67]$ the spherical symmetry of the potential, Eq.(3.19b)
(for any $\gamma )$ was taken into account at the outset leading to the
standard protocol of separation of variables just like that for the hydrogen
atom. Unfortunately, all these treatments are mathematically incorrect. This
is explained in detail below, in this section. Nevertheless, the account of
spherical symmetry was done correctly leading to separation of variables.
Use of Corollaries 4.2. and 4.3. allows then to obtain different integrable
cases from the one (seed) solution, e.g. that for the hydrogen atom.\bigskip

Relevant fundamentals of contact geometry are described in Appendix B. The
essence of the Eisenhart lift is nicely summarized in [75$].$ At the level
of classical mechanics the Eisenhart lift is a method of enlargement of the
phase space of dynamical system resulting in the associated system whose
dynamics is taking place on geodesics. Because of this, such a lift \ serves
as an alternative to the Jacobi- Maupertius \ formulation of classical
dynamics as described in Eq.(B.16) of Appendix B. A slight elaboration on
Jacobi- Maupertius \ formulation of mechanics leads to the equation \
describing the Reeb dynamics on null geodesics, that is to Eq.(4.3). From
here it follows that use of the Eisenhart lift is facilitating separability
and superintegrability [76]. In the formalism of Eisenhart lift Eq.(4.3) is
interpreted as the null constraint $\mathcal{H=}0$ \ imposed on new
Hamiltonian $\mathcal{H}$. This constraint is essential for recovery of the
original dynamics from the lifted one [77$].$ The equation $\mathcal{H=}0$
coincides with that for the Reeb vector field as explained in Appendix B
(after Eq.(B.17)). The additive solution for the null geodesics, Eq.(4.3),
depends on $c_{\alpha }$ constants defined in Eq.(4.4). As mentioned above,
the additivity of the solution of Eq.(4.3) is essential for the separability
\ (or R-separability) of the solution of$\ $equation $\square _{g}\Psi =0.$
Different sets of admissible constants for Eq.(4.3) cause different type of
R-separability in Eq.(3.15). It happens that existence of separable
coordinates has a coordinate-free characterization [78$].$ This property
leads to the following\bigskip

\textbf{Definition 4.6.} \ \textsl{St\"{a}ckel transform} \ maps the
constants of motion \ for one orthogonally separable system to the constants
of motion for another orthogonally separable system. Two systems related by
a sequence of St\"{a}ckel transforms \ are \textsl{St\"{a}ckel-equivalent}%
.\medskip \bigskip

\textbf{Remark 4.7.\bigskip }St\"{a}ckel equivalence and superintegrability
are inseparably linked to each other [79]. Recently Tsiganov demonstrated
[80] how to determine St\"{a}ckel matrices preserving superintegrability.
\bigskip Since the number of R-separable coordinates is always finite (for a
given dynamical system), the number of St\"{a}ckel transforms is also
finite. Evidently, none of them changes physics.\ But all of them are linked
with the topology of the underlying manifolds through the issue of
closed/periodic geodesics [81]. \ Although this topic is still open for
study, references [76],[82$]$ might serve as good starting point for
studying this issue. \bigskip

The superintegrability for Kepler-Coulomb mechanical problem was studied
recently in connection with Betrand spaces in [20$].$ This fact is helpful.
It will be further considered in section 5. It cannot be used immediately
though for reasons which will become evident upon reading. It appears, that
the topic of obtaining solutions for the Schr\"{o}dinger equation for
hydrogen atom is so well known that there is nothing which can be added to
this subject. This prevailing opinion happens to be incorrect. \ Indeed,
what is taught in any class on quantum mechanics is that the hydrogenic wave
functions $\Psi $ are given by the product [83$]$%
\begin{equation}
\Psi _{nlm}(r,\theta ,\varphi )=const\rho ^{l}e^{-\frac{\rho }{2}%
}L_{n+l}^{2l+1}(\rho )Y_{lm}(\theta ,\varphi ).  \tag{4.5}
\end{equation}%
Here $\rho =\frac{2r}{n},n=\sqrt{\frac{1}{-2E}},$ $r$ is the radial
coordinate, $E$ is the energy, $E<0$, $Y_{lm}(\theta ,\varphi )$ are the
spherical harmonics, $L_{n+l}^{2l+1}(\rho )$ are the generalized Laguerre
polynomials. The system of units was chosen in which Bohr's radius $a_{0}=1.$%
The constant in Eq$.(4.5)$ is determined by the normalization of wave
function. Since these results are standard, no more details are provided.
Much lesser known are four dimensional representations of the hydrogen wave
functions taking into account explicitly the hidden O(4) symmetry\footnote{%
Surely, the total dynamical symmetry of the hydrogen atom is SO(4,2). This
fact is discussed at length in [1$]$ and $[3].$ Nevertheless, the account of
O(4) symmetry is sufficient for now [8$4].$Once the results for wave
functions with O(4) symmetry are known, the SO(4,2) results are easily
obtainable with help of \ the wavefunctions with O(4) symmetry [1$%
],[85],[86].$}. According to [8$4]$,[87$]$ in 4 dimensions the wave function
is looking as follows%
\begin{equation}
Y_{nlm}(\alpha ,\theta ,\varphi )=2^{l+1}l![\frac{n(n-l-1)!}{2\pi (n+l)!}]^{%
\frac{1}{2}}\left[ \sin ^{l}\alpha \right] C_{n-l-1}^{l+1}(\cos \alpha
)Y_{lm}(\theta ,\varphi ).  \tag{4.6a}
\end{equation}%
Here the spherical harmonics $Y_{lm}(\theta ,\varphi )$ are the same as in
Eq.(4.5) while $C_{n-l-1}^{l+1}(\cos \alpha )$ are the Gegenbauer
polynomials to be discussed at length below. Unlike 3 dimensional spherical
system of coordinates which is well known, the 4 dimensional system of
coordinates 
\begin{eqnarray}
x_{1} &=&r\sin \alpha \sin \theta \cos \varphi ,  \notag \\
x_{2} &=&r\sin \alpha \sin \theta \sin \varphi ,  \notag \\
x_{3} &=&r\sin \alpha \cos \theta ,  \notag \\
x_{4} &=&r\cos \alpha .  \TCItag{4.7}
\end{eqnarray}%
is known in standard textbook literature much less. It should be noticed,
that already in 3 dimensions there is a number of separable coordinate
systems. In addition to spherical, there are cylindrical, parabolic, etc.
coordinate systems. These are St\"{a}ckel-equivalent and, hence, make the
hydrogen atom a superintegrable system [20$]$. The choice of coordinate \
system is dictated by the physics of the problem. In 4 dimensions the number
of separable coordinate systems increases remarkably. Many of them are
listed in [8$4$]. \ However, in [87$]$ yet another coordinate system, \
introduced by Bateman in 1905, is discussed\footnote{%
E.g. read [88$].$}. It will be described in section 5.

In the meantime, the normalization conditions for $Y_{nlm}(\alpha ,\theta
,\varphi )$ are 
\begin{equation}
\dint\limits_{0}^{\pi }\sin ^{2}\alpha d\alpha \dint\limits_{0}^{\pi }\sin
\theta d\theta \dint\limits_{0}^{2\pi }d\varphi \bar{Y}_{n^{\prime
}l^{\prime }m^{\prime }}Y_{nlm}=\delta _{n,n^{\prime }}\delta _{m,m^{\prime
}}\delta _{l,l^{\prime }}.  \tag{4.8}
\end{equation}%
Interestingly enough, Eq.s (4.5), (4.6) can be compared with each other
thanks to the result published as Appendix 5 of the book [89$].$ By
introducing the Fourier transform of $\Psi _{nlm}(r,\theta ,\varphi )$ via 
\begin{equation}
\Phi _{nlm}(\mathbf{p},\theta ,\varphi )=\left( 2\pi \right) ^{\frac{-3}{2}%
}\dint d^{3}re^{-i\mathbf{p}\cdot \mathbf{r}}\Psi _{nlm}(r,\theta ,\varphi ),
\tag{4.9}
\end{equation}%
where we had put $\hbar =1,$ after some very detailed calculation described
in [92], the final result is looking as follows: $\Phi _{nlm}(\mathbf{p}%
,\theta ,\varphi )=F_{nl}(\mathbf{p})Y_{lm}(\theta ,\varphi ),$%
\begin{equation}
F_{nl}(\mathbf{p})=\left[ \frac{2}{\pi }\frac{(n-l-1)!}{(n+l)!}\right] ^{%
\frac{1}{2}}\text{ }n^{2}2^{2l+2}l!\frac{\left( np\right) ^{l}}{\left[
\left( np\right) ^{2}+1\right] ^{l+2}}C_{n-l-1}^{l+1}(\frac{1-\left(
np\right) ^{2}}{1+\left( np\right) ^{2}}).  \tag{4.10a}
\end{equation}%
To compare this result with Eq.(4.6a) we need to consult [8$4],$ page 649.
According to this reference we should get an equality\footnote{%
It should be noted that in [8$4]$ the combination $%
p_{0}^{-1}(p^{2}+p_{0}^{2})^{2}$ was written as $p_{0}^{-\frac{5}{2}%
}(p^{2}+p_{0}^{2})^{2}.$ This is caused by the fact that, instead of \ the
constant factor $[\frac{n(n-l-1)!}{2\pi (n+l)!}]^{\frac{1}{2}}$ \ taken from
[87$]$ in Eq.(4.6a) the factor $[\frac{(l+1)(n-l-1)!}{2\pi (n+l)!}]^{\frac{1%
}{2}}$ was used erroneously.}if we multiply the 3-dimensional result,
Eq.(4.9), by the $p-$factor defined by 
\begin{equation}
Y_{nlm}(\alpha ,\theta ,\varphi )=\frac{p_{0}^{-1}(p^{2}+p_{0}^{2})^{2}}{4}%
\Phi _{nlm}(\mathbf{p},\theta ,\varphi ).  \tag{4.10b}
\end{equation}

\bigskip\ To check the correctness of this proposition, we adopt the
spherical coordinates defined by Eq.(4.7) to the case of 3-sphere of unit
radius and use the Fock parametrization of this 3-sphere [37],[93$]\footnote{%
See also Eq.s(2.2),(2.3) in addition.}.$ Specifically, in this
parametrization 
\begin{equation}
x_{4}=\frac{p_{0}^{2}-p^{2}}{p_{0}^{2}-p^{2}}=\frac{1-\left( np\right) ^{2}}{%
1+\left( np\right) ^{2}}=\cos \alpha .  \tag{4.11}
\end{equation}%
We also took into \ account that $p_{0}=\sqrt{-E},E<0,-E=1/n,m=1/2,$ $m$ is
the electron mass in selected system of units. Based on this information, we
obtain:%
\begin{equation}
\left( np\right) ^{2}=\frac{1-\cos \alpha }{1+\cos \alpha }.  \tag{4.12}
\end{equation}%
Next, using this result, we obtain as well: 
\begin{equation*}
\left( np\right) ^{2}+1=\frac{2}{1+\cos \alpha },\text{ and }\left( \frac{%
1-\cos \alpha }{1+\cos \alpha }\right) ^{l}\left( \frac{1+\cos \alpha }{2}%
\right) ^{l+2}
\end{equation*}%
and, therefore, 
\begin{equation}
p_{0}^{-3}\left[ \left( np\right) ^{2}+1\right] ^{2}\frac{\left( np\right)
^{l}}{\left[ \left( np\right) ^{2}+1\right] ^{l+2}}=p_{0}^{-3}\left( \frac{%
1-\cos \alpha }{1+\cos \alpha }\right) ^{\frac{l}{2}}\left( \frac{1+\cos
\alpha }{2}\right) ^{l}=p_{0}^{-3}\sin ^{l}\alpha .  \tag{4.13}
\end{equation}%
Using this result in Eq.(4.10a), accounting for Eq.(4.10b) and \ the
comparing the result with Eq.(4.6) we obtain: 
\begin{equation}
\text{ }Y_{nlm}(\alpha ,\theta ,\varphi )=\text{ }2^{l+1}l![\frac{n(n-l-1)!}{%
2\pi (n+l)!}]^{\frac{1}{2}}\left[ \sin ^{l}\alpha \right] C_{n-l-1}^{l+1}(%
\cos \alpha )Y_{lm}(\theta ,\varphi ).  \tag{4.6b}
\end{equation}%
The obtained result is in perfect agreement with Eq.(4.6a), as required.
These results allow us now to move further. In particular,we would like to
consider a solution of the Schr\"{o}dinger Eq.(3.19a) with the potential
Eq.(3.19b). D-O formally obtained its solution \ already in their first
work, [7$],$ for any $\gamma .$We found, however, more convenient to discuss
the same equation and its solution following [49$].$ In Eq.s(3.19a),(3.19b)
and in [7$],$[49$]$ this equation was treated as 3-dimensional while in [17$%
] $ Ostrovsky did consider the 4 dimensional version of the same equation,
but \ the obtained solution is not consistent with mathematically rigorous
results of [87$].$Therefore his results are to be reconsidered below.
\bigskip

In section 1 we noticed that initially Demkov and Ostrovsky [7$]$ believed
that "Maxwell's fish-eye problem is \textbf{closely related to} the Coulomb
problem". In section 3 \ we discussed the geometrical meaning \ of \ the
stationary Schr\"{o}dinger's equation for the hydrogen atom\bigskip . Now it
is time to reproduce Eq.(4.6a) with help of Eq.s(3.19a),(3.19b) ($\gamma
=1). $ By doing so we are going to replace the statement of D-O about "close
relationship" by the statement "isomorphic". Following [49$],$ we reproduce
Eq.s(3a),(3b) and (6) from this reference. Thus, we have%
\begin{equation}
\Psi (\rho ,\theta ,\varphi )=const\frac{\rho ^{l+1}}{(1+\rho ^{2})^{\frac{%
2l+1}{2}}}C_{n-l-1}^{l+1}(\xi (\rho ))Y_{lm}(\theta ,\varphi ),  \tag{4.14}
\end{equation}%
where $\xi (\rho )=\dfrac{1-\rho ^{2}}{1+\rho ^{2}}.$ A comparison between
Eq.s(4.6a) and (4.14) suggests that we have to rewrite the factors
containing the variable $\rho $ accordingly. Fortunately, this can be done
almost immediately by analogy with results displayed in Eq.s(4.11)-(4.14).
Now we have to write $\left( np\right) ^{2}=\rho ^{2},$ so that we obtain: 
\begin{equation}
\cos \alpha =\frac{1-\rho ^{2}}{1+\rho ^{2}};\rho ^{2}=\frac{1-\cos \alpha }{%
1+\cos \alpha };\rho ^{2}+1=\frac{2}{1+\cos \alpha }.  \tag{4.15a}
\end{equation}%
With help of these results we establish that $\xi (\rho )=\cos \alpha ,$ 
\begin{equation}
\left( \frac{1+\cos \alpha }{2}\right) ^{\frac{2l+1}{2}}\left( \frac{1-\cos
\alpha }{1+\cos \alpha }\right) ^{\frac{l+1}{2}}=2^{\frac{2l+1}{2}}\sin
^{l}\alpha  \tag{4.15b}
\end{equation}%
Using these results in Eq.(4.14) brings us back to Eq.(4.6a) so that needed
isomorphism is established, provided that we resolve the apparent paradox.
Eq.s(3.19a),(3.19b) as well as Eq.s(3a),(3b) and (6) of [49$]$ are
manifestly 3 dimensional while the result, Eq.(4.6a), is 4 dimensional. It
is also rather weird to have the 3-dimensional spherical harmonics $%
Y_{lm}(\theta ,\varphi )$ in the 4-dimensional result, Eq.(4.6). These
circumstances prompt us to dig deeper into the issues just mentioned.\bigskip

In the entire section 3 we dealt exclusively with the 3 dimensional versions
of Eq.s (3.19a),(3.19b). In the already cited references, e.g.[7$],$[49$],$
the $3$ dimensional versions of Eq.s (3.19a),(3.19b) were treated and
solved. Just above we noticed that these solutions carry information about
the 4th dimension. It was left unnoticed in [7$],$[49] while in [17$]$
Ostrovsky made an attempt to solve the 3-dimensional Eq.s(3.19a),(3.19b) by
using the 3-sphere living in 4 dimensional Euclidean space (without
justification for this step whatsoever). His results are not mathematically
rigorous for reasons just described. Below, we provide a detailed
explanation. Notice that our demonstration has almost nothing in common with
the celebrated 4-dimensional solution of the Sch\"{o}dinger equation for the
hydrogen atom developed by Fock [37$]$. Nevertheless, \ the end result of
our calculation will reproduce the 4-dimensional Eq.(4.6a) taken from [87$].$
Thus, we are about to uncover the "invisible" \ influence of the 4th
dimension on solutions \ of 3-dimensional equations (3.19a), (3.19b). Our
arguments are complementary to those by Fock [37$]$.

We begin with description of pentaspherical coordinates introduced in
section 7 of our work [3$].$This time, however, we shall provide additional
details following [69$]$ and [90$].$ The pentaspherical coordinates had been
discussed in accessible form in the monograph by Felix Klein cited in [3$]$.
\ Mathematical usefulness of these coordinates \ was \ recognized already by
B\"{o}cher at the end of 19th century and was \ rediscovered at the end of
the 20th century and further developed by Miller and collaborators [69$]$.
Physical usefulness of pentaspherical coordinates \ was explained in [3$]$.
For uninterrupted reading, we \ provide here some excerpts from [3$]$ on
this topic$.$We begin with consideration of a sphere $S^{2}$ in \textbf{R}$%
^{3}.$ Analytically, the $S^{2}$ is described as%
\begin{equation}
x^{2}+y^{2}+z^{2}-2ax-2by-2cz+D=0.  \tag{4.16a}
\end{equation}%
The radius $r$ of the sphere is defined via%
\begin{equation}
r^{2}=a^{2}+b^{2}+c^{2}-D.  \tag{4.16b}
\end{equation}%
The introduced parameters are considered as coordinates in the space of
spheres. In addition, it is convenient to introduce the homogenous
(projective) coordinates%
\begin{equation}
a=\frac{\xi }{\nu },b=\frac{\eta }{\nu },c=\frac{\zeta }{\nu },r=\frac{%
\lambda }{\nu },D=\frac{\mu }{\nu }.  \tag{4.17}
\end{equation}%
These projective coordinates allow \ us to embed $S^{2}$ into projective
space \textbf{RP}$^{5}$ resulting in the equation 
\begin{equation}
\xi ^{2}+\pi ^{2}+\zeta ^{2}-\lambda ^{2}-\nu \mu =0  \tag{4.18a}
\end{equation}%
or, with $\nu \mu =\alpha ^{2}-\beta ^{2},$ its analog 
\begin{equation}
\xi ^{2}+\pi ^{2}+\zeta ^{2}+\beta ^{2}-\lambda ^{2}-\alpha ^{2}=0. 
\tag{4.18b}
\end{equation}%
Although in pentaspherical coordinates $(x_{1},...,x_{5})$ one uses a
collection of 5 orthogonally intersecting spheres, just described
mathematics remains very much the same. Specifically, the quadric, Eq.
(4.18), now is written as the cone equation 
\begin{equation}
\dsum\limits_{i=1}^{5}x_{i}^{2}=0.  \tag{4.19}
\end{equation}%
Let $(x,y,z)\in \mathbf{R}^{3}$ . Via stereographic projection (e.g. see
Eq.s (2.2),(2.3)) these Euclidean coordinates are related to $\ $the
spherical coordinates $(s_{1},...,s_{4})\in S^{3}$, $\dsum%
\limits_{i=1}^{4}s_{i}^{2}=1.$ \ As in Eq.s (2.2) and (2.3), we obtain
analogously: 
\begin{equation}
x=\frac{s_{1}}{1+s_{4}},y=\frac{s_{2}}{1+s_{4}},z=\frac{s_{3}}{1+s_{4}} 
\tag{4.20a}
\end{equation}%
so that 
\begin{equation}
s_{1}=\frac{2x}{r^{2}+1},\text{ }s_{2}=\frac{2y}{r^{2}+1},s_{3}=\frac{2z}{%
r^{2}+1},s_{4}=\frac{1-r^{2}}{1+r^{2}},\text{ }r^{2}=x^{2}+y^{2}+z^{2}. 
\tag{4.20b}
\end{equation}%
Next, it is convenient to introduce the analog of Eq.s(4.17) by taking into
account Eq.(4.19). Thus, we write: 
\begin{equation}
x_{1}=2XT,x_{2}=2YT,x_{3}=2ZT,x_{4}=X^{2}+Y^{2}+Z^{2}-T^{2},x_{5}=i(X^{2}+Y^{2}+Z^{2}-T^{2}).
\tag{4.21}
\end{equation}%
Accordingly,%
\begin{eqnarray}
x &=&\frac{X}{T}=\frac{-x_{1}}{x_{4}+ix_{5}},\text{ }y=\frac{Y}{T}=\frac{%
-x_{2}}{x_{4}+ix_{5}},\text{ }z=\frac{Z}{T}=\frac{-x_{3}}{x_{4}+ix_{5}}, 
\notag \\
x^{2}+y^{2}+z^{2}-1 &=&\frac{-2x_{4}}{x_{4}+ix_{5}},\text{ }%
x^{2}+y^{2}+z^{2}+1=\frac{2ix_{5}}{x_{4}+ix_{5}}.  \TCItag{4.22}
\end{eqnarray}%
Eq.s(4.21) imply (via duality) the following chain of equalities (taken from
[90$],$page 403):%
\begin{equation}
\partial _{X}=2T\partial _{x_{1}}+2X\partial _{x_{4}}+2iX\partial
_{x_{5}},\partial _{Y}=2T\partial _{x_{1}}+2Y\partial _{x_{4}}+2iY\partial
_{x_{5}}\text{, }\partial _{Z}=2T\partial _{x_{1}}+2Z\partial
_{x_{4}}+2iZ\partial _{x_{5}}.  \tag{4.23}
\end{equation}%
Also, 
\begin{equation}
\partial _{x}=T\partial _{X},\partial _{y}=T\partial _{Y},\partial
_{z}=T\partial _{Z}  \tag{4.24}
\end{equation}%
and, following [69$],[90],$ we obtain as well:%
\begin{eqnarray}
\partial _{x} &=&-\left( x_{4}+ix_{5}\right) \partial
_{x_{1}}+x_{1}(\partial _{x_{4}}+i\partial _{x_{5}}),\partial _{y}=-\left(
x_{4}+ix_{5}\right) \partial _{x_{2}}+x_{2}(\partial _{x_{4}}+i\partial
_{x_{5}}),  \notag \\
\partial _{z} &=&-\left( x_{4}+ix_{5}\right) \partial
_{x_{3}}+x_{3}(\partial _{x_{4}}+i\partial _{x_{5}}).  \TCItag{4.25a}
\end{eqnarray}%
By replacing the quantum $-i\partial _{x}$ by the classical $p_{x},etc.$we
obtain the classical analog of Eq.(4.25a)%
\begin{eqnarray}
p_{x} &=&-\left( x_{4}+ix_{5}\right)
p_{x_{1}}+x_{1}(p_{x_{4}}+ip_{x_{5}}),p_{y}=-\left( x_{4}+ix_{5}\right)
p_{x_{2}}+x_{2}(p_{x_{4}}+ip_{x_{5}}),  \notag \\
p_{z} &=&-\left( x_{4}+ix_{5}\right) p_{x_{3}}+x_{3}(p_{x_{4}}+ip_{x_{5}}) 
\TCItag{4.25b}
\end{eqnarray}%
The conformally flat space classical Hamiltonian $\mathcal{H}$ now reads 
\begin{equation}
\mathcal{H}=p_{x}^{2}+p_{y}^{2}+p_{z}^{2}+V(\mathbf{x})=0=\left(
x_{4}+ix_{5}\right)
^{2}\{p_{x_{1}}^{2}+p_{x_{2}}^{2}+p_{x_{3}}^{2}+p_{x_{4}}^{2}+p_{x_{5}}^{2}+%
\tilde{V}(x)\}=0  \tag{4.26a}
\end{equation}%
so that $\left( x_{4}+ix_{5}\right) ^{2}\tilde{V}(x)=V(\mathbf{x}).$ Now we
adopt the quadric, Eq.(4.19), to $S^{3}.$ This is done by requiring $%
x_{5}=-i.$ Accordingly, using Eq.(4.19), now we obtain:%
\begin{equation*}
\dsum\limits_{i=1}^{4}\frac{x_{i}^{2}}{-x_{5}^{2}}=1,\text{ }
\end{equation*}%
implying $s_{1}=x_{1},etc.$ In view of this, Eq.(4.26a) can be rewritten as 
\begin{equation}
\left( x_{4}+1\right)
^{2}\{p_{x_{1}}^{2}+p_{x_{2}}^{2}+p_{x_{3}}^{2}+p_{x_{4}}^{2}+\tilde{V}%
(x)\}=0.  \tag{4.26b}
\end{equation}

By comparing Eq.s(4.26a),(4.26b) with Eq.s(3.19a),(3.19b) adapted to the
case $\gamma =1,$ we have to rewrite the potential, Eq.(3.19b) taking also
into account Eq.(2.14b) and discussion which follow after this equation.
Thus, we obtain: 
\begin{equation}
V(\mathbf{x})=\frac{\left( Ze^{2}\right) ^{2}}{\left\vert E_{n}\right\vert }%
\left( \frac{1}{1+r^{2}}\right) ^{2}\mid _{S^{3}}=\left( x_{4}+1\right) ^{2}%
\frac{\left( Ze^{2}\right) ^{2}}{4\left\vert E_{n}\right\vert }  \tag{4.27}
\end{equation}%
In Eq.(4.26b) the combination $%
p_{x_{1}}^{2}+p_{x_{2}}^{2}+p_{x_{3}}^{2}+p_{x_{4}}^{2}$ must be multiplied
by the factor 1/2 in accord with the discussion following Eq.(2.14b).
Furthermore, this combination represents not the flat Laplacian. This fact
is fundamentally nontrivial. Its origin is based on the following chain of
arguments.

Using Eq.(4.26a) we have to require that $p_{x_{i}}$ and $x_{i}$ to be the
canonically conjugate \ pair of variables. If, indeed, this is so then, say,
at the classical level of description we have to calculate the Poisson
bracket (P.B). Thus, we obtain: 
\begin{equation}
\{\dsum\limits_{i=1}^{5}x_{i}^{2},\mathcal{H}\}_{P.B.}=2\left(
x_{4}+ix_{5}\right) ^{2}\dsum\limits_{i=1}^{5}x_{i}p_{i}=0.  \tag{4.28}
\end{equation}%
Adopted to $S^{3}$, Eq.(4.28) reads:%
\begin{equation}
\dsum\limits_{i=1}^{4}x_{i}p_{i}=0.  \tag{4.29}
\end{equation}%
Thus, we have to solve Eq.(4.26b) subject to the constraint, Eq.(4.29). The
solution of this problem is nontrivial. It is presented in the Appendix E.
With help of obtained results, below we shall recover the 4-dimensional
hydrogen wave function, Eq.(4.6a).\bigskip

The 4-dimensional wave function, Eq.(4.6a), is presented in [8$4]$,[87$].$
In both references it was given without derivation. Therefore, we would like
to restore needed details being guided by the results of our Appendix E,
Ref.[91$],$ and known quantum mechanical results for the 3-dimensional rigid
rotator. At this point we need to extend these 3 dimensional results to the
4th dimension. In 3 dimensions the well known Casimir operator \textbf{\^{L}}%
$^{2}$=\^{J}$_{x}^{2}$+\^{J}$_{y}^{2}$+\^{J}$_{z}^{2}$ is used together with 
\^{J}$_{z}$ to label the spherical wavefunctions $Y_{lm}(\theta ,\varphi )$.
The first task now lies in finding the 4-d Casimir operator (or
operators)-the analog(s) of \textbf{\^{L}}$^{2}$. Fortunately, the results
can be found in [87$],$ pages 112-113. These are taken from [91$],$page 248.
In Appendix E we argue that the 4-dimensional analogs of \^{J}$_{x}$,\^{J}$%
_{y}$ and \^{J}$_{y}$ are the quantized Pl\"{u}cker coordinates, e.g. $\hat{l%
}_{12}$, \textit{\^{l}}$_{13},$\textit{\^{l}}$_{14}$ $,$\textit{\^{l}}$%
_{23}, $\textit{\^{l}}$_{24}$ and \textit{\^{l}}$_{34}$ so that \textbf{\^{L}%
}$^{2}$ is replaced now by $\mathcal{\hat{L}}^{2}=\hat{l}_{12}^{2}$+ \textit{%
\^{l}}$_{13}^{2}+$\textit{\^{l}}$_{14}^{2}+$\textit{\^{l}}$_{23}^{2}+$%
\textit{\^{l}}$_{24}^{2}+$\textit{\^{l}}$_{34}^{2}$. Explicitly, in terms of
spherical coordinates defined in Eq.(4.7) we obtain: 
\begin{equation}
\mathcal{\hat{L}}^{2}=-\frac{\partial ^{2}}{\partial \alpha ^{2}}-2\cot
\alpha \frac{\partial }{\partial \alpha }-\frac{1}{\sin ^{2}\alpha }[\frac{%
\partial ^{2}}{\partial \theta ^{2}}+\cot \theta \frac{\partial }{\partial
\theta }+\frac{1}{\sin ^{2}\theta }\frac{\partial ^{2}}{\partial \varphi ^{2}%
}].  \tag{4.30a}
\end{equation}%
Notice, however, that \textbf{\^{L}}$^{2}Y_{lm}(\theta ,\varphi )=\hslash
^{2}l(l+1)Y_{lm}(\theta ,\varphi ).$ Because of the known relation 
\begin{equation*}
\frac{\mathbf{\hat{L}}^{2}}{\hslash ^{2}}=-[\frac{1}{\sin \theta }\frac{%
\partial }{\partial \theta }(\sin \theta \frac{\partial }{\partial \theta })+%
\frac{1}{\sin ^{2}\theta }\frac{\partial ^{2}}{\partial \varphi ^{2}}]=-[%
\frac{\partial ^{2}}{\partial \theta ^{2}}+\cot \theta \frac{\partial }{%
\partial \theta }+\frac{1}{\sin ^{2}\theta }\frac{\partial ^{2}}{\partial
\varphi ^{2}}],
\end{equation*}%
Eq.(4.30a) should be rewritten accordingly as 
\begin{equation}
\mathcal{\hat{L}}^{2}=-\frac{\partial ^{2}}{\partial \alpha ^{2}}-2\cot
\alpha \frac{\partial }{\partial \alpha }+\frac{1}{\sin ^{2}\alpha }\frac{%
\mathbf{\hat{L}}^{2}}{\hslash ^{2}}.  \tag{4.30b}
\end{equation}%
By analogy with 3 dimensions, in 4 dimensions we write instead: 
\begin{equation}
\mathcal{\hat{L}}^{2}Y_{nlm}(\alpha ,\theta ,\varphi )=I_{nl}Y_{nlm}(\alpha
,\theta ,\varphi ).  \tag{4.31a}
\end{equation}%
The eigenvalue $I_{nl}$ is to be determined from the equation 
\begin{equation}
\lbrack \frac{l(l+1)}{\sin ^{2}\alpha }-\frac{\partial ^{2}}{\partial \alpha
^{2}}-2\cot \alpha \frac{\partial }{\partial \alpha }]\Psi _{nl}(\alpha
)=I_{nl}\Psi _{nl}(\alpha ).  \tag{4.32}
\end{equation}%
in which, in view of Eq.s(4.30b),(4.31a), we have to present $Y_{nlm}(\alpha
,\theta ,\varphi )$ as $\Psi _{nl}(\alpha )Y_{lm}(\theta ,\varphi ).$ Since
the obtained equation is equivalent to Eq.(4.26b), we should make an
identification: $I_{nl}=\frac{-\left( Ze^{2}\right) ^{2}}{2\left\vert
E_{n}\right\vert }.$ \ In arriving at this result we took into account that
the combination $p_{x_{1}}^{2}+p_{x_{2}}^{2}+p_{x_{3}}^{2}+p_{x_{4}}^{2}$ is
coming along with the factor of 1/2. This was explained after Eq.(2.14b).
The combined use of Eq.s (4.26b(,(4.30b), (4.31),(4.32) then leads to $%
I_{nl}=\frac{-\left( Ze^{2}\right) ^{2}}{2\left\vert E_{n}\right\vert }.$

Next, we let $x=\cos \alpha $ in Eq.(4.32). Then, we write $\Psi
_{nl}(\alpha )=(1-x^{2})^{\frac{l}{2}}F_{nl}(\alpha ).$ With help of these
substitutions Eq.(4.32) acquires the following look%
\begin{equation}
(1-x^{2})\frac{d^{2}}{dx^{2}}F_{nl}-(2l+3)x\frac{d}{dx}%
F_{nl}+[I_{nl}-l(l+2)]F_{nl}=0.  \tag{4.33}
\end{equation}%
This equation should be compared against the canonical equation for the
Gegenbauer polynomials [92$]:$%
\begin{equation}
(1-x^{2})\frac{d^{2}}{dx^{2}}C_{n}^{\lambda }(x)-(2\lambda +1)x\frac{d}{dx}%
C_{n}^{\lambda }(x)+n(n+2\lambda )C_{n}^{\lambda }(x)=0.  \tag{4.34}
\end{equation}%
Upon comparison between these two equations, we obtain: $(2l+3)%
\rightleftarrows (2\lambda +1)$ implying $\lambda =l+1.$ Also, $n(n+2\lambda
)\rightleftarrows I_{nl}-l(l+2)$ implying $I_{nl}=(n+l+1)^{2}-1.$ Let now $%
n+l+1=2F+1$ then, $(n+l+1)^{2}-1=4F(F+1)$. This result should be compared
with those presented in [8$4],[87].$ For this, it is sufficient to relabel $%
n+l+1$ as $\hat{n}$. By comparing with standard physics literature on
hydrogen atom, we have to relabel $n$ as $n_{r},$ where $n_{r}$ is the
radial quantum number. Such an identification between $n$ as $n_{r}$ is
plausible but misleading even though it is present in all works by D-O and
those who used the D-O works. Explanation is given at the end of section 5.
With such a background, we obtain: $\hat{n}^{2}-1=4F(F+1).$On page 5 of [87$%
] $ we find that the 4 dimensional wave \ functions $Y_{\hat{n}lm}$ (e.g.
see Eq.(4.31)) are belonging to the eigenvalue $\hat{n}^{2}-1$as required.
Next, we obtain : $n_{r}=2F-l$ implying $\ F_{\hat{n}l}(\alpha
)=C_{2F-l}^{l+1}(\cos \alpha ).$ Since $\Psi _{\hat{n}l}(\alpha )=(1-x^{2})^{%
\frac{l}{2}}F_{\hat{n}l}(\alpha ),$ we obtain as well $\Psi _{\hat{n}%
l}(\alpha )=const(\sin \alpha )^{l}C_{2F-l}^{l+1}(\cos \alpha ).$ Finally,
since $Y_{\hat{n}lm}(\alpha ,\theta ,\varphi )=\Psi _{\hat{n}l}(\alpha
)Y_{lm}(\theta ,\varphi ),$ we end up (accounting for the normalization,
Eq.(4.8)) with \ the wave function, Eq.(4.6a), as required.\bigskip

Calculation of the eigenvalue spectrum for the hydrogen atom can be found in
any textbook on quantum mechanics. Here, we approach this problem
nontraditionally. In doing so, we borrow some ideas from the obscure paper
by Schr\"{o}dinger [93$]$ (circa 1940) superimposed with results of our
Appendix F. Schr\"{o}dinger was interested in finding the extent to which
the effects of curvature of Universe \ may affect the spectrum of hydrogen
atom. Schr\"{o}dinger choose to consider \ the Keplerian motion on the
hypersphere $S^{3}$ of radius $R.$ Fundamentally interesting (for us) is to
read about the rationale of his calculation. In his paper we find the
following passages:

"It may appear foolish to pay attention \ to the extremely feeble curvature
of the Universe in dealing with the hydrogen atom, because even the
influence of those much stronger fields of gravitation in which all our
observations are actually made (if the frame is properly chosen) entirely
negligible. But this problem, by obliterating the sharp cut between "
elliptic and hyperbolic orbits"(the classical orbits are all closed) and by
resolving the continuum spectrum into intensely crowded line spectrum, has
extremely interesting features, well worth investigating..."

By looking for spaces \ \textsl{in which all classical orbits are closed},
Schr\"{o}dinger actually was looking at

a) Bl\"{a}shke manifolds discussed in Appendices A and B,

b) Superintegrable dynamical systems discussed in this and next sections,

c) Bertrand spaces \ to be discussed in the next section.\bigskip

He begins his calculations with the equation on $S^{3}$ (not on \textbf{R}$%
^{3}!)$%
\begin{equation}
\frac{\hslash ^{2}}{2mR^{2}}[-\frac{1}{\sin ^{2}\alpha }\frac{d}{d\alpha }%
(\sin ^{2}\alpha \frac{dS}{d\alpha })+\frac{l(l+1)}{\sin ^{2}\alpha }S]-%
\frac{e^{2}}{R}S\cot \alpha -ES=0.  \tag{4.35a}
\end{equation}%
A quick look at our Eq.(4.30b) allows us to rewrite Eq.(4.35a) into more
familiar form as follows%
\begin{equation}
\frac{\hslash ^{2}}{2mR^{2}}\mathcal{\hat{L}}^{2}S=ES+[\frac{e^{2}}{R}\cot 
\frac{\alpha }{2}]S.  \tag{4.35b}
\end{equation}%
In this equation we corrected Scrodinger's result: in going from \textbf{R}$%
^{3}$ to $S^{3}$ of radius $R$ he replaced the Coulombic 3-dimensional \
term $\dfrac{e^{2}}{R}$ by $\dfrac{e^{2}}{R}\cot \alpha $. The correct
result is: $\dfrac{e^{2}}{R}\cot \dfrac{\alpha }{2}.$ It is worthy of
demonstrating how this result is actually obtained. Using Eq.(4.20b) we
write: 
\begin{equation}
r^{2}=\frac{1-s_{4}}{1+s_{4}}.  \tag{4.36}
\end{equation}%
Hence,%
\begin{equation}
\frac{1}{r}=\left( \sqrt{\frac{1-s_{4}}{1+s_{4}}}\right) ^{-1}=\sqrt{\frac{%
\left( 1+s_{4}\right) ^{2}}{1-s_{4}^{2}}}  \tag{4.37a}
\end{equation}%
Taking into account that $s_{4}=x_{4}=R\cos \alpha ,$ let $R=1$, then we
obtain:%
\begin{equation}
\frac{1}{r}=\frac{1+\cos \alpha }{\sin \alpha }=\frac{2\cos ^{2}\frac{\alpha 
}{2}}{2\sin \frac{\alpha }{2}\cos \frac{\alpha }{2}}=\cot \frac{\alpha }{2}.
\tag{4.37b}
\end{equation}%
In our calculations we want to take advantage of the Coulomb-fish-eye
isomorphism. Therefore, being guided by this isomorphism, \ instead of the
factor $\dfrac{e^{2}}{R}\cot \dfrac{\alpha }{2}$\ we systematically used \
the factor $-\frac{\left( Ze^{2}\right) ^{2}}{2\left\vert E_{n}\right\vert }%
. $ This caused us to obtain the eigenvalue spectrum $I_{nl}$ correctly in
agreement with [84$],[87].$ However, this is not yet the spectrum for
hydrogen atom. To obtain this spectrum correctly,we use the results of
Appendix F. These are also allowing us \ to make a connection between our
current results and those by Schr\"{o}dinger. This is achieved with the
amended Eq.(4.31a), that is, in view of Eq.(4.35b), with 
\begin{equation}
\mathcal{\hat{L}}^{2}Y=(I_{nl}-E)Y  \tag{4.31b}
\end{equation}%
The choice of $E$ is determined by replacing $I_{nl}=\hat{n}^{2}-1=4F(F+1)$
by $I_{nl}-E=4F(F+1).$ From Appendix F we know that the parameter $E$ can be
chosen once and for all based on physical considerations. In our case, it is
sufficient to chose$-E=-1$. With this choice we obtain:%
\begin{equation}
-\frac{\left( Ze^{2}\right) ^{2}}{2\left\vert E_{n}\right\vert }=(2F+1)^{2}%
\text{ or }E_{n}=-\frac{\left( Ze^{2}\right) ^{2}}{2(2F+1)^{2}},F=0,\frac{1}{%
2},1,...  \tag{4.38}
\end{equation}%
This result coincides with that given on p.645 of [84$]$ as required.
\bigskip

\textbf{Remark 4.8. }In the limit $R\rightarrow \infty $ the spectrum
obtained by Schr\"{o}dinger correctly reproduces the spectrum of hydrogen
atom. However, the calculated wavefunctions differ from the standard,
Eq.(4.6a). Since in this section we obtained both the eigenfunctions and
eigenvalues for hydrogen atom correctly using methods and \ arguments
different from those used by Fock [37], we leave for our readers to
investigate further two topics. a) What logic made Schr\"{o}dinger to use
4-dimensional calculations instead of 3-dimensional\footnote{%
Bertrand \ curved spaces of section 5 are 3 dimensional.}? Unlike Fock, who
clearly stated the reason for using 4 dimensional calculations for hydrogen
atom, Schr\"{o}dinger, being apparently not aware of Fock's results,
bypassed this issue altogether. b) It is of interest to recover the
wavefunctions for hydrogen atom, Eq.(4.6a), using Eq.(4.35a). \bigskip

The entire section 3 and a large part of Appendix F is devoted to the issues
related to conformal invariance. Previously we mentioned the work by
Makowski [53$].$ He took advantage of the fact that in 2 dimensions the
Laplacian is conformally invariant. This fact enabled him to solve the
2-dimensional Sturmian Eq.(3.1b) for the fish-eye-type potentials exactly.
In 3 dimensions the Laplacian no longer possess the property of conformal
invariance. To bypass this difficulty the conformally invariant (Yamabe)
Laplacian should be used instead. This fact created a lot of technical
difficulties but allowed us to demonstrate in detail the full power of
Hadamard-style calculations. These were only outlined in [3$].$ Such
calculations are of significance not only for our work. To our knowledge,
they have not been exposed in sufficient detail in physics literature,
perhaps with exception of gravitation. But even in this field the
alternative \ route of by- passing the issue of conformal invariance in 3
dimensions was not discussed also. \ It is associated with the systematic
uses of \ the Hopf mapping. This mapping is not a novel item in physics
literature [94]. Nevertheless, to our knowledge, the aspects we are willing
to discuss are new. Because of this, they will be studied further in detail
in Part II.

The idea of Hopf mapping is simple. Take the equation for the 3-sphere $%
S^{3} $ living in \textbf{C}$^{2},e.g.$ $\left\vert z_{1}\right\vert
^{2}+\left\vert z_{2}\right\vert ^{2}=1,z_{i}=x_{i}+iy_{i},$ $i=1,2$, and
look for the ratio $Z=\dfrac{z_{1}}{z_{2}}$ (or $Z=\dfrac{z_{2}}{z_{1}}).$
While $S^{3}\in $ \textbf{C}$^{2},Z\in \mathbf{C}$ . But \textbf{C} $\cup $ $%
\infty =S^{2}!$ From this observation the Hopf map follows: $S^{3}$ $%
\rightarrow S^{2}$. Subsequently the map $S^{2}$ $\rightarrow S^{3}$ was
also designed and used, e.g. read [9$5],$etc. Above, using the pentasperical
coordinates we mapped the 3 dimensional Sturmian problem involving Eq.(3.19)
into $S^{3}$. We \ want now to reanalyze the 2 dimensional results by
Makowski [53$]$ using the notion of Hopf map$.$ It should be noticed,
though, that Makowski used the 2 dimensional plane \textbf{C }in his work.
By adding a point at infinity his results can be lifted to $S^{2}.$ E.g$.$%
look again at Eq.s(2.15)-(2.22) and (3.1a,b). The \ first thing we have to
check: will his results reproduce known 2 dimensional results for the
hydrogen atom on \textbf{C}?\ The answer depends upon our ability to solve
the analogous problem on $S^{2}$. On $S^{2}$ the potential, Eq.(3.19b), is
constant. In fact, it is the same constant as we had calculated already: $%
\tilde{V}(x)=-\frac{\left( Ze^{2}\right) ^{2}}{2\left\vert E_{n}\right\vert }%
.$ The Laplacian restricted to $S^{2}$ is describing the rigid rotator.
Therefore, we again run into the familiar equation: \textbf{\^{L}}$%
^{2}Y_{lm}(\theta ,\varphi )=\hslash ^{2}l(l+1)Y_{lm}(\theta ,\varphi ).$
The question remains: will this equation reproduce the 2-d hydrogen
spectrum? Very fortunately, positive answer was found in [95$].$ These
authors had adapted Pauli's \ dynamical group-theoretic method for solving
hydrogen atom in 2 dimensions. By repeating the same steps, they indeed
arrived at the rigid rotator equation. \ The Pauli-like group algebra
enabled them to arrive at the equation for the spectrum 
\begin{equation}
l(l+1)=-(\frac{1}{4}+\frac{1}{E})  \tag{4.39a}
\end{equation}%
leading to the known result:%
\begin{equation}
E_{l}=-\frac{1}{(l+\frac{1}{2})^{2}},\text{ }l=0,1,2,...  \tag{4.39b}
\end{equation}%
For the future reference, the 3-dimensional result taken from [84$]$ and
used in Eq.(4.38) is given as follows:%
\begin{equation}
F(F+1)=-(\frac{1}{4}+\frac{1}{8E}),\text{ }F=0,1/2,1,...  \tag{4.40}
\end{equation}%
leading to $E_{F}=-\dfrac{1}{2n^{2}},$ $n=2F+1=1,2,3,...$

Both results can be made identical if one notices that Eq.(4.39a) uses only
the angular momentum interpretation of the standard commutator algebra $[%
\hat{J}_{j},\hat{J}_{k}]$ $=i\varepsilon _{jkl}\hat{J}_{l}$. In the case if
the same commutator algebra is interpreted spin-theoretically, the results
Eq.(4.39b) and (4.40) will coincide. This is becoming permissible in the
case if the 2-dimensional results for $S^{2}$ are lifted to $S^{3}$ using
the inverse Hopf map\footnote{%
Some technical details of this mapping are discussed at the end of section 5.%
}. We expect that the rest of 2-dimensional results to be discussed shortly
below will follow the same trend. \bigskip

To prepare our readers for this discussion we recall that in mathematics the
3-dimensional rigid rotator eigenvalue problem is known under the name of
eigenvalue equation problem for the associated Legendre polynomials $%
P_{l}^{m}$($\cos \theta )$. In terms of the variable $x=\cos \theta $ \ the
equation for these polynomials reads:%
\begin{equation}
(1-x^{2})\frac{d^{2}}{dx^{2}}P_{l}^{m}(x)-2x\frac{d}{dx}P_{l}^{m}(x)+[l(l+1)-%
\frac{m^{2}}{1-x^{2}}]P_{l}^{m}(x)=0.  \tag{4.41}
\end{equation}%
It should be compared with Eq.s(4.32)-(4.34). By analogy, we shall look for
a solution of Eq.(4.41) in the form $P_{l}^{m}(x)=(1-x^{2})^{\frac{m}{2}}%
\bar{P}_{l}^{m}(x).$ Substitution of this ansatz into Eq.(4.41) leads to [97$%
]$%
\begin{equation}
(1-x^{2})\frac{d^{2}}{dx^{2}}\bar{P}_{l}^{m}(x)-2(m+1)x\frac{d}{dx}\bar{P}%
_{l}^{m}(x)+[l(l+1)-m(m+1)]\bar{P}_{l}^{m}(x)=0.  \tag{4.42}
\end{equation}%
But this is an equation for the Gegenbauer polynomials, Eq.(4.34)! In the
simplest case, $m=0$, we obtain: $\bar{P}_{l}^{0}(x)=C_{n}^{\frac{1}{2}%
}(x),n=l.$ For $m\neq 0,$ we have to put $\lambda =m+1/2$ in Eq.(4.34) and $%
l-m=n$ \ in \ Eq.(4.42) implying $\bar{P}_{l}^{m}(x)=C_{n}^{m+\frac{1}{2}%
}(x)=C_{l-m}^{m+\frac{1}{2}}(x)$. These results we would like to compare now
against Eq.(9) of Makowski [53]. To do so, it is sufficient to make an
identifications : Makowski's $\lambda \rightleftarrows m,$ also his $%
n_{r}+\lambda \rightleftarrows n+m.$ We are warning our readers not to
confuse the factor $\lambda $ in Gegenbauer's Eq.(4.34) and the factor $%
\lambda $ in Makowski's paper [53]. Evidently, our $n$ is the same as his $%
n_{r}$. Up to a complex factor exp( $\pm im\varphi )$ the solution obtained
by Makowski is given by 
\begin{equation}
\psi _{n_{r}}(x,y)=const(1-\xi ^{2})^{\frac{\lambda }{2}}C_{n_{r}}^{\lambda +%
\frac{1}{2}}(\xi ),\text{ }\xi =\frac{1-r^{2}}{1+r^{2}}=\frac{1-z^{2}}{%
1+z^{2}}.  \tag{4.43a}
\end{equation}%
Here we used our Eq.s (2.15)-(2.22) to write his Eq.(9) in our notations.
This allows us to take into consideration that $\xi =\dfrac{1-r^{2}}{1+r^{2}}%
=Z=\cos \theta $ resulting in%
\begin{equation}
\psi _{n_{r}}(\theta ,\varphi )=N(\sin \theta )^{m}C_{n_{r}}^{m+\frac{1}{2}%
}(\cos \theta ).  \tag{4.43b}
\end{equation}%
Using Eq.(2.15), it is clear, that $x=\cos \theta \rightleftarrows Z.$ In
addition, $N$ is the normalization constant. We would like to compare
Makowski's 2 dimensional result, Eq.(4.43a) (or Eq.(4.43b)), against that
coming from the equation for the rigid rotator, Eq.(4.42). This equation is
leading us to the solution 
\begin{equation}
\psi _{n}(\theta ,\varphi )=N^{\prime }(\sin \theta )^{m}C_{l-m}^{m+\frac{1}{%
2}}(\cos \theta ).  \tag{4.44}
\end{equation}

\ Since we just defined $l-m=n$ and $n=n_{r},$ this identification causes us
to write $N^{\prime }=N.$ Makowski's result, Eq.(4.43b) was obtained \ using
planar geometry while the result, Eq. (4.44), was obtained \ using the
geometry/topology of $S^{2}$. Use of stereographic projection connects these
two geometries. When it comes to the connection \ between the respective
differential equations -on \textbf{C} and on $S^{2}$\ -use of conformal
invariance in Eq.(3.1b) is essential since the stereographic projection is a
conformal mapping. Should the potential in the planar Schr\"{o}dinger
equation be zero, the mapping \textbf{C} $\rightleftarrows S^{2}$ \ would
transform the flat Laplacian to the Laplacian on $S^{2}.$ The Sturmian
problem on \textbf{C} is the eigenvalue problem for the Laplacian on $S^{2}.$
It is the familiar eigenvalue problem for the rigid rotator in quantum
mechanics.

It is always possible to adapt the 3-dimensional Eq.(3.19) to $S^{2}$ as
long as the conformal transformations are not involved. When they are
involved, we only can lift Eq.(3.1b) to $S^{2}$ but we cannot make a
reduction from 3 dimensions to 2 since in 3 and higher dimensions we are
dealing with the conformal (Yamabe) Laplacian while in dimension 2- \ with
the ordinary Laplacian which is conformally invariant. But we can use the
stereographic projection from: from $S^{3}$ to \textbf{R}$^{3}.$ Because of
this, it is of interest to compare conformally modified planar results by
Makowski and Peplowski [48] against the analogous 3-dimensional (actually
the 4-dimensional, as demonstrated in previous subsection) results of Ref.[49%
$].\bigskip $ Such a comparison will give us an idea about the extent to
which results on $S^{2}$ by Makowski [53], actually done on \textbf{C},
change when they are lifted (via the Hopf map) to $S^{3}.\bigskip $ It
should be noted though that we have not yet constructed rigorously the Hopf
map. This is done in mathematical literature already and will be adapted for
our needs in Part II. \bigskip

To complete our calculation we need to reproduce Eq.s(4.39b). For this
purpose we have to replace now the equation 
\begin{equation}
\frac{\mathbf{L}^{2}}{\hslash ^{2}}Y_{lm}(\theta ,\varphi
)=I_{lm}Y_{lm}(\theta ,\varphi )  \tag{4.45a}
\end{equation}%
by 
\begin{equation}
\frac{\mathbf{L}^{2}}{\hslash ^{2}}Y_{lm}(\theta ,\varphi )=\left(
I_{lm}-E\right) Y_{lm}(\theta ,\varphi )  \tag{4.45b}
\end{equation}%
where we have to choose, as before, $I_{lm}=-\frac{\left( Ze^{2}\right) ^{2}%
}{2\left\vert E_{n}\right\vert }$ and select$-E=-\frac{1}{4}.$ This choice
follows from the identification 
\begin{equation}
l(l+1)=-\frac{\left( Ze^{2}\right) ^{2}}{2\left\vert E_{n}\right\vert }-%
\frac{1}{4}  \tag{4.46a}
\end{equation}%
done with help of Eq.(4.42) and the property $l(l+1)+\frac{1}{4}=(l+\frac{1}{%
2})^{2}.$ Using these results, we finally arrive at the equation for
spectrum 
\begin{equation}
E_{l}=-\frac{\left( Ze^{2}\right) ^{2}}{\left( l+\frac{1}{2}\right) ^{2}}%
,l=0,1,2,...  \tag{4.46b}
\end{equation}%
Obtained result is in accord with Eq.(4.39b) as required. Furthermore, in
the system of units $\hslash =1,m=1,$ the obtained result exactly coincides
with that reported in [98$].\bigskip \bigskip $

Now we want to discuss conformally deformed 2-dimensional results presented
in [48$]$. They are immediately linked with results by \ the same authors
just \ discussed. Instead of duly reproducing the results of [48$]$, we
approach the whole problem differently. That is, first, in the absence of
potential \ we lift the flat Laplacian from \textbf{C} to $S^{2}$. The
conformal invariance is\ compatible \ with \ such a lift as we already know,
e.g. see Eq.(3.1). Before performing any conformal transformation we notice
that the potential $\tilde{V}(x)$ must be replaced by $l(l+1)$ in Eq.(4.45)
in accord with theory of 2 dimensional rigid rotator. To study the effects
of conformal transformations we analyze Eq.(5) of [48] by the same methods
as we\ used in Eq.s(4.11)-(4.15). \ In the present case we start with 
\begin{equation}
F_{nl}^{\left( \gamma \right) }(\xi )=N_{nl}^{\left( \gamma \right) }(1-\xi
^{2})^{\frac{l}{2\gamma }}C_{n}^{l/\gamma +1/2}(\xi ),\xi =\frac{%
1-r^{2\gamma }}{1+r^{2\gamma }}.  \tag{4.47a}
\end{equation}%
To be compatible with our notations we replaced the $k$- factor in [48] by
the $\gamma $ factor defined in our Eq.(3.19). Before performing any
conformal transformation we put $\gamma =1$ in the above result to check it
against Makowski's Eq.(4.43b) (see above)$.$ By comparing Eq.s (4.43) and
(4.47a) we notice that $l$ in Eq.(4.47a) must be replaced by $m.$ If we
treat $r^{\gamma }=\rho $ as independent variable, we \ still can use $\xi
=\cos \theta =x.$ Accordingly, we obtain: 
\begin{equation}
F_{nm}^{\left( \gamma \right) }(\theta )=N_{nm}^{\left( \gamma \right)
}\left( \sin \theta \right) ^{\frac{m}{\gamma }}C_{n}^{m/\gamma +1/2}(\cos
\theta ),n=n_{r}.  \tag{4.47b}
\end{equation}%
For $\gamma =1$ Eq.s(4.43b) and (4.47b) coincide as required. Use of
conformal transformations causes us to make a replacement $m\rightarrow
m/\gamma $ in Eq.(4.43b) to reach an agreement with Eq.(4.47b). Such a
replacement makes perfect physical sense because in this work the conformal
transformation $f(z)=z^{\gamma }$ (e.g. see Eq.(2.22)) is used consistently.
Since $m$ is the azimutal quantum number, the wave function boundary
condition $\Psi (0)=\Psi (2\pi )$ \ prior to conformal mapping must be
preserved \ after the conformal mapping is performed. This requirement is
the cause of the transformation: $m\rightarrow m/\gamma .$

Now we are ready to analyze the 4-dimensional results of Ref.[49],page 34%
\footnote{%
We would like to remind to our readers that the authors of Ref.[49] were not
aware of 4 dimensional nature of their calculations.}. For $\gamma =1$ this
was done already in Eq.s (4.14),(4.15). Therefore, only the conformally
deformed case is treated in this subsection. It \ is treated by analogy with
that we just discussed when we considered the case of 2 dimensions. The
additional help is\ coming from Eq.s(4.14),(4.15). Thus, now we obtain: 
\begin{equation}
\cos \alpha =\frac{1-\rho ^{2\gamma }}{1+\rho ^{2\gamma }};\rho ^{2\gamma }=%
\frac{1-\cos \alpha }{1+\cos \alpha };\rho ^{2\gamma }+1=\frac{2}{1+\cos
\alpha };  \tag{4.48a}
\end{equation}%
\begin{equation}
\left( \frac{1+\cos \alpha }{2}\right) ^{\frac{2l+1}{2\gamma }}\left( \frac{%
1-\cos \alpha }{1+\cos \alpha }\right) ^{\frac{l+1}{2\gamma }}=2^{\frac{2l+1%
}{2\gamma }}\sin ^{\frac{l}{\gamma }}\alpha  \tag{4.48b}
\end{equation}%
Eq.(4.14) is replaced now by%
\begin{equation}
\Psi (\alpha ,\theta ,\varphi )=const^{\prime }(\sin ^{\frac{l}{\gamma }%
}\alpha )C_{M-\frac{l}{\gamma }-1}^{\frac{2l+1}{2\gamma }+\frac{1}{2}}(\cos
\alpha ))Y_{lm}(\theta ,\varphi ).  \tag{4.49a}
\end{equation}%
Here, in accord with authors of [17] and [49], $M=n+$($\gamma ^{-1}-1)l.$
For $\gamma =1$ Eq.(4.49a) coincides with Eq.s (4.14),(4.15) as required.
Next, we compare this (4-dimensional) result against\ the 2-dimensional
result, Eq.(4.47b), that is with 
\begin{equation}
F_{nm}^{\left( \frac{1}{2}\right) }(\theta )=N_{nm}^{\left( \frac{1}{2}%
\right) }\left( \sin ^{2m}\theta \right) C_{n_{r}}^{2m+1/2}(\cos \theta ). 
\tag{4.47c}
\end{equation}%
In such a form we compare it now against Eq.(4.49a) in which we should put $%
\gamma =1/2:$%
\begin{equation}
\Psi (\alpha ,\theta ,\varphi )=const^{\prime }(\sin ^{2l}\alpha
)C_{n_{r}}^{2l+\frac{3}{2}}(\cos \alpha ))Y_{lm}(\theta ,\varphi ). 
\tag{4.49b}
\end{equation}%
This result is written in terms of notations used in [17] and [49]. In these
references $M=n+$($\gamma ^{-1}-1)l$ \ and, we \ have\ to write now $\gamma
=1/2,$ and to use the standard result: $n=n_{r}+l+1,n_{r}=0,1,2,...$ The
comparison implies that for $\gamma =1/2$ the 4 dimensional wave function $%
\Psi (\alpha ,\theta ,\varphi )$ can be obtained with help of 2 dimensional $%
F_{n_{r}m}^{\left( \frac{1}{2}\right) }(\theta )$ \ if we take into account
that 3-dimensional $z=\cos \theta $ is equal to $t=\cos \alpha $ in 4
dimensions, e.g. see Eq.(4.7). Therefore we should make a replacement $%
\theta $ by $\theta \rightleftarrows \alpha $ wherever this is appropriate$.$%
Next, we have to make a replacement $m\rightleftarrows l$ \ and, finally, we
have to use the addition formula [97$],$page$121,$%
\begin{equation}
C_{n}^{\alpha +\beta }(x)=\dsum\limits_{m=0}^{n}C_{m}^{\alpha
}(x)C_{n-m}^{\beta }(x).  \tag{4.50}
\end{equation}%
After these replacements and use of the addition formula the final result
should be multiplied by $Y_{lm}(\theta ,\varphi ).$ In the end, the obtained
result must be properly normalized with help of the normalization condition
for the Gegenbauer polynomials%
\begin{equation}
\dint\limits_{0}^{\pi }d\alpha \left( \sin ^{2p}\alpha \right)
[C_{l}^{p}(\cos \alpha )]^{2}=\frac{\pi \Gamma (2p+l)}{2^{2p-1}l!(l+p)\Gamma
^{2}(p)}.  \tag{4.51}
\end{equation}%
In writing the above results, following works by D-O and those who extended
their works, we silently assumed that everything is fine with the result,
Eq.(4.49a). But this is not at all the case. Mathematically correct results
are presented below. \bigskip

Calculation of the deformed spectrum is technically easy but, for the
results to make physical sense, some prior discussion is absolutely
essential. For readers convenience, needed background material is given in
the Appendix G. In it we explain what is wrong with the existing "proofs" of
the Madelung rule. We also describe the unexpected Hartree-Fock origins of
the conformally deformed fish-eye potential, Eq.(3.19), with $\gamma =1/2.$
The description of this connection with the Hartree-Fock contribution to the
effective potential $V_{eff}(\mathbf{r}),$ Eq.(1.3), cannot be found among
D-O papers cited thus far. Nevertheless, D-O did mention this connection in
[99] and [100$]$ but only in the context of \ the Thomas-Fermi (T-F) theory.
The Appendix G does contain an information about the connection with the
Hartree-Fock results. Yes, it is true that the contributions to the T-F
theory made by Tietz resulted in the potential, Eq.s(G.8),(G.9), bearing his
name. It is also true that, up to a constant factor, analytically this
potential is the same as the deformed fish-eye potential, Eq.(3.19) with $%
\gamma =1/2.$ \ But D-O have not cited the results by other authors in which
it is demonstrated that \textsl{the combined use of the Coulombic and
Hartree-Fock contributions to the effective potential} $V_{eff}(\mathbf{r}),$
Eq.(1.2), \textsl{results in the deformed fish-eye potential with} $\gamma
=1/2$.

The problem of solving Eq.(1.2) \ had began with the work by Slater. In his
1932 paper [101$]$ he writes " For any detailed calculations dealing with
atomic or molecular structure, good approximations to the atomic wave
functions are essential. The most satisfactory method, in general, for
building up such functions to be by the use of one-electron functions which
are solutions of the problem \ of an electron moving in a central field..."
A noticeable advancement in this direction was made by Latter in 1955 [26$].$
He \ numerically solved Eq.(1.2) using the Thomas-Fermi-Dirac type potential
for all elements \ with $Z=1\div 92.$ Latter noticed that" The important
error in this application of a fixed central field potential arises just
from the use of the same potential for all configurations of the atom. This
error clearly grows with increasing excitation of the atom above its ground
state....The statistical potentials give acceptable values \ for ground
state energies and, presumably, \ will therefore give \ semiquantitative
predictions for states of low excitation." \ Results by Latter had been
further analyzed by March. In his book [27$]$ on p.76 March writes " In a
Coulomb field we would , of course, fill the K, L, M shells successively.
The closed shell of principal quantum number $n$ holds $2n^{2}$ electrons
and hence shells close, for a Coulomb field, for 2,10, 28, etc. electrons
.... Now, in real atoms, the closed shell configurations occur of He (Z=2)
and Ne (Z=10) which fit the Coulomb field scheme. But the next rare gas is
Ar, with 18 electrons.... a screened Coulomb field such as given by Hartree
or a Thomas-Fermi theory leads for an atomic number Z=18 to the filing of 3p
sub shell which is rather far from the 3d level. Also, for\textsl{\ larger
atomic number , level crossing occurs. The principal quantum number is no
longer precise enough to group the energy levels: for \ a screened Coulomb
field these depend upon n and the orbital angular momentum quantum number l".%
} \ Thus, without using the word "Madelung rule", March recognized that the
accurate calculations by Latter [26$]$ \ provide the strongest possible
justification of the validity of Madelung rule! It should be kept in mind
that Latter compared the Thomas -Fermi and the Thomas-Fermi-Dirac solutions
for energies of individual electrons \ against the available Hartree and
Hartree-Fock results for respective atoms. All results by Latter \ are
obtained numerically and \ are based on numerical solution of Eq.(1.2) with
the Thomas-Fermi-Dirac type effective potential $V_{eff}(\mathbf{r}).$ They
demonstrate a very remarkable agreement with the Hartree and Hartree-Fock
results. The solutions were obtained numerically because Latter was not
using the Tietz potential. As discussed in the Appendix G, the Tietz
potential is of Thomas-Fermi type but is of convenient analytical form and
fits very nicely the Hartree-Fock data for the effective potential [25],
p.664, Fig.10. Contrary to D-O descriptions of the Tietz work, Tietz
recognized the qualities of his potential immediately after he discovered
it. This caused him to apply the methodology developed by Kerner [102] in
1951 to the potential Tietz discovered later on. At the same time, Kerner,
after consulting with\ Bethe and Feynman, obtained the exact analytical
solution for the stationary Schr\"{o}dinger equation having the potential $%
V_{eff}(\mathbf{r})=-\dfrac{Ze}{r}\sqrt{\varphi (\dfrac{r}{a})}$ . \ Here $%
\varphi (x)$ is the same as in Eq.(G.8), except that Kerner was fiddling
with the constant $\alpha $ (to get the best possible agreement with
experiment) while in Eq.(G.8) this constant is fixed by the logic of the T-F
calculations. Kerner tested his effective potential against\ the numerically
obtained Hartree potential and found astonishingly good agreement as
demonstrated in [102], Fig.1, page 71. Subsequently, Titz adapted
word-for-word Kerner's methodology \ for solving the Schr\"{o}dinger's
equation with the potential\ $V_{eff}(\mathbf{r})=-\dfrac{Ze}{r}\varphi (%
\dfrac{r}{a}).$ In [29$]$ the latest attempt by Tietz at solving this type
of \ Schr\"{o}dinger \ equation \ is presented along with the list of all
his previous attempts at exact solution. Unlike Latter, Kerner had not
calculated the spectrum for all $Z$'s and for many $l^{\prime }s$. He fixed
his attention at the Mercury ($Z=80$) and made \ a comparison \ of a single
electron eigenvalues (for various $n$'s and $l$'s) calculated by his exact
method against the relevant Hartree-type calculations. He obtained a good
agreement with the Hartree results summarized in the Table II, page 74, of
[102]. Kerner has no mention of the Madelung rule in his paper. Following
into footsteps of Kerner, Tietz \ also selected Mercury for his study and
compared his exact (Kerner-style) calculations against those published in
Kerner's paper. Naturally, a good agreement with experimental data was also
obtained.

\ D-O entered into research on this topic only in 1971-1972. \ Unlike other
authors, they recognized the importance of \ restrictions \ caused by the
Bertrand theorem and they reinterpreted the Tietz potential, without
mentioning Tietz's name and his works, as the conformally deformed Maxwell's
fish-eye potential. This is explained above, in section 2. This caused them
to restrict their treatment [7],[17],[25] aimed at the proof of Madelung
rule to studies of the Schr\"{o}dinger equation (with Tietz potential ) to
the $E=0$ case only. \ The rationale for doing so is discussed above, in
sections 1$-$3. Already in section 2 we explained that the Schr\"{o}dinger
equation with the fish-eye potential \textbf{should }be used with $E=0$ by
design. This requirement has \textbf{nothing to do} with the energy spectrum
for hydrogen atom though. It was calculatedabove, in this section. \textbf{%
But it has everything to do with the conformal invariance of the underlying }%
Eq.(3.19) \textbf{allowing us to conformally deform the Maxwell fish-eye} 
\textbf{potential} \textbf{thus allowing us to study the multielectron atoms
with help of the same formalism.} Being unaware of just described facts,
especially those, in the Appendix F, D-O studied only the $E=0$ solution of
the Schr\"{o}dinger equation with the Tietz potential. \ At the same time,
following the methodology developed by Kerner, Tietz obtained the exact
solution of the same equation as studied by D-O in his 1956 paper [28] 
\textbf{without any restrictions} \textbf{on} $E$. \ The D-O method of
solution of Eq.(3.19) differs from that by Tietz and, therefore, by Kerner.
The analytical form of \ the wave function, Eq.(4.49b), is absent in\ the
paper by Tietz.\bigskip

Now, we are in the position\textbf{\ }to calculate the deformed spectrum
using the above background information as well as results of Appendix C. We
proceed by analogy with results just discussed. This brings us back to the
deformed wave function Eq.(4.49b). The question emerges: Will this function
be an eigenfunction of Eq.(4.31a)? A quick look at the analogous
egenfunction, Eq.(4.6a), and taking into account that $Y_{nlm}(\alpha
,\theta ,\varphi )$= $\Psi _{nl}(\alpha )Y_{lm}(\theta ,\varphi ),$ with $%
\Psi _{nl}(\alpha )=(1-x^{2})^{\frac{l}{2}}F_{nl}(\alpha ),x=\cos \alpha $
leads us to the following conclusions.

Using the factor $(1-x^{2})^{\frac{l}{2}}=\sin ^{l}\alpha $ in $\Psi
_{nl}(\alpha )$ and substitution of the ansatz $\Psi _{nl}(\alpha
)Y_{lm}(\theta ,\varphi )$ into Eq.(4.31a) allows us to get rid of the
denominator 1/$\sin ^{2}\alpha $ in Eq.(4.30b) (see also the analogous steps
in Eq.s(4.41) and (4.42)). In view of Eq.(4.49b), in the deformed case the
factor $(1-x^{2})^{\frac{l}{2}}$ is replaced by the factor $%
(1-x^{2})^{l}=\sin ^{2l}\alpha $ in $\Psi _{n2l}(\alpha )$. When substituted
back into Eq.(4.31a), calculation shows that only if we replace $l(l+1)$ by $%
2l(2l+1)$ in this equation the denominator 1/$\sin ^{2}\alpha $ will
disappear. Next, we need to adapt the fundamental Eq.(4.34) for the
Gegenbauer polynomials to $C_{n_{r}}^{2l+\frac{3}{2}}(x)$ to our current
needs. Such an adaptation produces: 
\begin{equation}
(1-x^{2})\frac{d^{2}}{dx^{2}}C_{n_{r}}^{\lambda }(x)-(4l+4)x\frac{d}{dx}%
C_{n_{r}}^{\lambda }(x)+n_{r}(n_{r}+4l+3)C_{n_{r}}^{\lambda }(x)=0,\lambda
=2l+3/2.  \tag{4.52}
\end{equation}%
At the same time we should compare Eq.(4.52) with Eq.(4.33) in which we have
to replace $l$ by $2l.$ Comparison indicates that the match is achieved only
if we replace $C_{n_{r}}^{2l+\frac{3}{2}}(x)$ by $C_{n_{r}}^{\frac{l}{\gamma 
}+1}(x)$ and take $\gamma =1/2.$ The combination $\left( \sin ^{\frac{i}{%
\gamma }}\alpha \right) C_{n_{r}}^{\frac{l}{\gamma }+1}(x)$ is fully
consistent with two dimensional result, Eq.(4.47b). This is very plausible
if we would like to apply the Hopf mapping to the two dimensional result,
Eq.(4.47b), to get \ the result on $S^{3}.$ Furthermore, under such
circumstances we can repeat word-for-word the discussion following
Eq.(4.34), that is if $\lambda =2l+1,$ then $2\lambda +1\rightleftarrows
4l+3.$ Also, $n(n+2\lambda )\rightleftarrows I_{nl}-2l(2l+2)$ implying the
relationship $n+2l+1=2F+1.$ If, as before, we let $\hat{n}=n+l+1,$then $2F+1=%
\hat{n}+l.$ By analogy with the derivation of the hydrogen spectrum,
Eq.(4.38), we now obtain:%
\begin{equation}
E_{n}=-\frac{Z^{\frac{7}{3}}e^{4}}{w(2F+1)^{2}},\text{ }F=0,\frac{1}{2},1,...%
\text{ or, }E_{n}=-\frac{Z^{\frac{7}{3}}e^{4}}{w(\hat{n}+l)^{2}},w=const. 
\tag{4.53}
\end{equation}%
In arriving at Eq.(4.53) we took into consideration the scaling analysis
presented in Appendix G and in section 2. We also used the Remark 5.3. of
the following section.\ The obtained result is fully consistent with the
Madelung rule and group-theoretical analysis of periodic system [1$],[103]$
done in close analogy with the Gell-Mann-Zweig theory of quark-hadron
matter.\bigskip \bigskip\ It should be noted though that the discussion of
likely sources of anomalies of the Madelung rule given in [30] indicates
that the group-theoretic methods are not capable of explaining these
anomalies while the results presented thus far are fully compatible with the
formalism describing the anomalies.\bigskip \medskip\ 

\bigskip \textbf{5.} \textbf{Bertrand spacetimes in the atomic
physics\bigskip\ and the Madelung rule}

The title of our paper involves the phrase "modified Bertrand theorem". This
terminology was introduced by Voler Perlick [23].Perlick was interested
infinding the relativisticanalogue of The Bertrand theorem.That is he wanted
to find all spherically symmetric and static spacetimes (in the sense of
general relativity) all of whose bounded trajectories are periodic. Although
Perlick succeeded in finding such spacetimes (which he calls "Bertrand
spacetimes"), he was puzzled by the fact that the metric of such spacetimes
is \ not reduced to the Schwartzshield metric used for description of
spacetimes around massive objects, e.g. \ of our Sun. Such a metric leads to
the precession of planetary trajectories rotating around the Sun. \ This
result was explained by Einstein using his theory of general relativity.
Classical Bertrand theorem forbids such a precession \ and, by design, its
relativistic \ extension also forbids precession. 2 types of Brertrand
metric found by Perlick caused Perlick to wonder about kind of a body
(isolated from all other gravitational sources) \ that produces a metric in
which all bounded trajectories are periodic. He concluded that "such a body
must be a rather exotic object". \ The authors of [104] suggested that such
a body should be identified with the dark matter. A quick introduction into
this topic is given in our work [105], page 27. Further studies of dynamics
in Bertrand spacetimes revealed [20] that dynamics in such spacetimes is
superintegrable\footnote{%
E.g.read \ the Remark 4.5.} since the orbits are closed by design. Closure
of orbits had prompted study of the quantum Bertrand systems [106$],[107].$
In these references no attempt was made to connect the obtained results with
atomic physics. The purpose of this section is to demonstrate that the
central object of our study, the D-O Eq.(3.19), justly belongs to the set of
quantum Bertrand systems.

\ \ As before, this demonstration is essential to present in the proper
context. Specifically, we begin with the Groenewold-van Hove (G-vH) theorem
nicely explained \ in [108$],[109].$ Basically, it is telling us that only
the harmonic oscillator-type dynamical systems can be unambiguously
quantized. The St\"{a}ckel transform\footnote{%
Definition 4.6 and Remark 4.7 .} connects the harmonic oscillator with the
Kepler-Coulomb systems [106$].$ Only these two systems are permissible by
the classical Bertrand theorem and, formally, can be quantized without
violating the G-vH theorem. \ Extension of the Bertrand theorem to the
curved spacetimes results in producing a class of superintegrable systems,
including the Kepler-Coulomb -type [20], also quantizable [106$],[107].$
There are two independent ways to develop quantum mechanics without
violating G-vH. The first one is using the Morse theory for development of
the sophisticated asymptotic (semiclassical) methods of analysis of quantum
mechanical problems [110$].$ The latest monograph [111$]$ \ aimed at
physically educated readers presents a detailed panoramic view of successful
applications of semiclassical methods to all branches of quantum mechanics.

The idea of the second method belongs to Eisenhart. Although his results are
summarized in section 4, because of their use in Perlick's work [23], we
need to add additional details at this point. These are coming from the
recognition of the fact that the addition of time transformations into
symplectomorphisms of classical mechanics makes it effectively
general-relativistic [105]. Indeed, suppose that the Lagrangian $\mathcal{L}$
of a dynamical system is given by 
\begin{equation}
\mathcal{L}=T(p,q)-V(q),T(p,q)=\frac{1}{2}g_{ij}(q)v^{i}v^{j}.  \tag{5.1}
\end{equation}%
The configurational space $Q$ of this system is a Riemannian manifold whose
metric $ds^{2}=g_{ij}(q)dq^{i}dq^{j}$ is determined by the kinetic energy
term in Eq.(5.1). The Hamiltonian $\mathcal{H}(q,p)$ is obtainable in a
usual way as 
\begin{equation}
\mathcal{H}(q,p)=\tilde{T}(p,q)+V(q),\tilde{T}(p,q)=\frac{1}{2}%
g^{ij}(q)p_{i}p_{j},\text{ }p_{i}=g_{ij}(q)v^{j},g^{ij}g_{jk}=\delta
_{k}^{i}.  \tag{5.2}
\end{equation}%
By extending the configurational space $Q$ to $\tilde{Q}=R\times Q$ and
adding a new momentum, say, $p_{z}$ leads to the replacement of $\mathcal{H}%
(q,p)$ by 
\begin{equation}
\mathcal{\tilde{H}}(q,p)=\frac{1}{2}g^{ij}(q)p_{i}p_{j}+\frac{1}{2}%
V(q)p_{z}^{2}.  \tag{5.3}
\end{equation}%
Accordingly, the metric $ds^{2}$ changes to%
\begin{equation}
d\sigma ^{2}=g_{ij}(q)dq^{i}dq^{j}+\frac{dz^{2}}{V(q)}.  \tag{5.4a}
\end{equation}%
If such an extension is caused by changes of time variable making it
dynamical \footnote{%
Section 3 and Appendix B.}, we obtain the metric $d\sigma ^{2}$ used in
Perlick's paper [23], e.g. 
\begin{equation}
d\sigma ^{2}=g_{ij}(q)dq^{i}dq^{j}-\frac{dt^{2}}{V(q)}.  \tag{5.4b}
\end{equation}%
In section 4.3.we noticed that the Eisenhart lift of a\ dynamical system
involves the condition $\mathcal{\tilde{H}}(q,p)=0$ essential for recovery
of the original dynamics. Such a condition leads to Eq.(4.3) associated with
the equation for null geodesics. Eq.(4.3) is conformally invariant\footnote{%
Theorem 3.1.} Hamilton-Jacobi (H-J) equation. In spite of its simple look,
the additive separation of variables\footnote{%
Eq.(4.4).} is highly nontrivial. The nontriviality of separation problem in
classical H-J equation leads to the nontriviality of quantization problem
associated with the conformal Laplacian, Eq.(3.12). References [69$],[90]$
illustrate the essence of this problem. \ Results of these references should
be considered as alternative to the G-vH theorem\footnote{%
Perhaps, the Bertrand-Darboux theorem should be linked with the G-vH theorem
in deciding which dynamical system is admitting quantization.}. In addition
to the references just mentioned, we recommend to our readers excellent
lectures by Benenti [112$]$ as well as the thesis [11$3$] by Bruce. Both are
containing many practical examples. From Benenti lectures as well as from $%
[90]$ and Appendix F it follows that analysis of the H-J equation can be
done the same way even if $\mathcal{\tilde{H}}(q,p)=E\neq 0,$e.g. for
Eq.(B.16). With such a background we are ready to demonstrate that, in
accord with expectations of D-O [7$]$, Eq.(3.19) \ indeed belongs to the set
of equations describing quantum Bertrand systems in the sense of [106$%
],[107] $.\bigskip

We begin by rewriting Eq.(3,19) in the equivalent form%
\begin{equation}
\left( \frac{r}{a}\right) ^{2}\left[ \left( r/a\right) ^{-\gamma }+\left(
r/a\right) ^{\gamma }\right] ^{2}(\frac{\partial ^{2}}{\partial x^{2}}+\frac{%
\partial ^{2}}{\partial y^{2}}+\frac{\partial ^{2}}{\partial z^{2}})\psi
+n_{0}^{2}\psi =0  \tag{5.5a}
\end{equation}%
For $\gamma =1$ , that is for hydrogen atom, this equation coincides with
Eq.(3.35) as required. Accounting for scaling arguments presented in
Eq.(2.14), classical mechanics version of this equation coincides with
Eq.(2.29). For arbitrary $\gamma $ classical version of Eq.(5.5) reads ($%
\hslash =1,m=1)$ 
\begin{equation}
\left( \frac{r}{a}\right) ^{2}\left[ \left( r/a\right) ^{-\gamma }+\left(
r/a\right) ^{\gamma }\right] ^{2}\left\vert \mathbf{p}\right\vert
^{2}=n_{0}^{2}\text{ }  \tag{5.5b}
\end{equation}%
or, in the case if $\gamma =1/2,$ we get%
\begin{equation}
\left( \frac{r}{a}\right) [1+\left( r/a\right) ]^{2}\left\vert \mathbf{p}%
\right\vert ^{2}=n_{0}^{2}\text{ .}  \tag{5.5c}
\end{equation}%
In such a form (without applications to atomic physics)\ this result reads
as Eq.(2.119), page 49, of the PhD thesis [108$]$ on quantum Bertrand
spaces.\ In Eq.(2.119) we have to put $K=-1$, $\beta =\gamma ,$and $\dfrac{%
\mathbf{P}^{2}}{4}=\left\vert \mathbf{p}\right\vert ^{2}.$ It remains to
demonstrate how this result emerges from Perlick's results [23].

We begin with some comments on two types of metric obtained \ in [23]. We
describe only type I metric, that is the Kepler-Coulomb (KC) type. The type
II metric, the oscillator-like, is obtainable from type I via St\"{a}ckel
transform as demonstrated in [106$].$ The type I metric is described as 
\begin{equation}
d\sigma _{\text{I}}^{2}=\frac{dr^{2}}{\beta ^{2}(1+Kr^{2})}+r^{2}(d\theta
^{2}+\sin ^{2}\theta \text{ }d\varphi ^{2})-\frac{dt^{2}}{G+\sqrt{r^{-2}+K}}
\tag{5.6}
\end{equation}%
Here $G$ and $K$ are some constants while $\beta $ is \textsl{rational}
number. By comparing Eq.s (5.3),(5.4) and letting $G=K=0$ and $\beta =1$ we
recover \ the KC Hamiltonian%
\begin{equation}
\mathcal{\tilde{H}}_{CK}=p_{r}^{2}+\frac{L^{2}}{r^{2}}-\frac{1}{r}. 
\tag{5.7}
\end{equation}%
In general, the associated Hamiltonian, Eq.(5.2), reads\footnote{%
Since $G$ is arbitrary constant thus far, we are free to change its sign.}%
\begin{equation}
\mathcal{H}_{\text{I}}(q,p)=\frac{\beta ^{2}}{2}(1+Kr^{2})p_{r}^{2}+\frac{%
L^{2}}{r^{2}}-\sqrt{r^{-2}+K}+G.  \tag{5.8}
\end{equation}%
The task now is to relate Eq.(5.5b) to Eq.(5.8). This requires several
steps. The Perlick metric, Eq.(5.4b), defines the kinetic energy realized on
null geodesics of 3+1 dimensional space-time of Lorentzian signature.
Alternatively, it can be used for description of the motion in 3 dimensional
spherically symmetric space with the metric 
\begin{equation}
d\Sigma _{\text{I}}^{2}=\frac{dr^{2}}{\beta ^{2}(1+Kr^{2})}+r^{2}(d\theta
^{2}+\sin ^{2}\theta \text{ }d\varphi ^{2})  \tag{5.9}
\end{equation}%
describing the classical motion on 3 manifold with metric $d\Sigma _{\text{I}%
}^{2}$ in the presence of spherically symmetric potential $V(r)$. To move
forward, following [114] we need to look at several alternative ways of
writing $d\Sigma _{\text{I}}^{2}.$ Specifically,%
\begin{equation}
d\Sigma _{\text{I}}^{2}=g(r)dr^{2}+r^{2}d\Omega ^{2}=F^{2}(\rho )(d\rho
^{2}+d\Omega ^{2})=f(\left\vert \mathbf{q}\right\vert )d\mathbf{q}^{2}=f^{2}(%
\tilde{r})(d\tilde{r}^{2}+\tilde{r}^{2}d\Omega ^{2})  \tag{5.10}
\end{equation}%
Here $d\Omega ^{2}=d\theta ^{2}+\sin ^{2}\theta $ $d\varphi ^{2},d\mathbf{q}%
^{2}=dq_{1}^{2}+dq_{2}^{2}+dq_{3}^{2},$ $q_{1}=\tilde{r}\cos \theta ,$ $%
q_{2}=\tilde{r}\sin \theta \cos \varphi ,q_{3}=\tilde{r}\sin \theta \sin
\varphi .\left\vert \mathbf{q}\right\vert =\tilde{r}.\rho =\ln \tilde{r}.$
Use of the metric \ written in the form $f(\left\vert \mathbf{r}\right\vert
)d\mathbf{r}^{2}$ immediately brings us to Eq.(3.35). But we still can do
more since out task\ is to relate Eq.(5.5b) to Eq.(5.8). Therefore, we have
to take into account that $f(\left\vert \mathbf{r}\right\vert )=f(\tilde{r}%
). $ To determine $\tilde{r}$ and $f(\tilde{r})$ we have to take into
account that : a) $F(\rho )=\tilde{r}f(\tilde{r}),$ b) $\dfrac{d\rho }{dr}%
=r^{-1}g(r),c)$ $r=F(\rho ).$ From Eq.(5.10) and just defined results we
obtain: 
\begin{equation}
a)\frac{r}{\tilde{r}}=f(\tilde{r});\text{ b) }g(r)=\frac{1}{\beta }\frac{1}{%
\sqrt{1+Kr^{2}}};c)\dfrac{d\rho }{dr}=\frac{1}{\beta r\sqrt{1+Kr^{2}}}\text{ 
}  \tag{5.11}
\end{equation}%
Integrating Eq.(5.11c)) yields%
\begin{equation}
\rho =\frac{1}{\beta }\ln \left( \frac{r}{1+\sqrt{1+Kr^{2}}}\right) . 
\tag{5.12}
\end{equation}%
Since $\rho =\ln \tilde{r}$ we obtain as well 
\begin{equation}
\tilde{r}^{\beta }=\frac{r}{1+\sqrt{1+Kr^{2}}}.  \tag{5.13}
\end{equation}%
Since according to Eq.(5.11a)) $\dfrac{r}{\tilde{r}}=f(\tilde{r}),$ by
resolving Eq. (5.13) with respect to $r$ we obtain: 
\begin{equation}
\left( \dfrac{r}{\tilde{r}}\right) ^{2}=f^{2}(\tilde{r})=\frac{4}{\tilde{r}%
^{2}}\left( \frac{1}{\tilde{r}^{-\beta }-K\tilde{r}^{\beta }}\right) ^{2}. 
\tag{5.14}
\end{equation}%
Taking into account Eq.s (5.1)-(5,4), (5.6)-(5.10) (and omitting tildas) we
obtain the (null) Hamiltonian 
\begin{equation}
r^{2}(r^{-\beta }-Kr^{\beta })^{2}\left\vert \mathbf{p}\right\vert
^{2}+\alpha =0  \tag{5.15}
\end{equation}%
Here K and $\alpha $ are some constants. They are fixed by comparing Eq.s
(5.15)and (5.5c). This is achieved by selecting respectively $\beta
=1/2,K=-a^{-1}$ and $\alpha =-n_{0}^{2}.$ Thus, we just established the
desired correspondence between the D-O Eq.(3.19) and the generalized \
Bertrand problem.

By looking at Eq.s (5.6),(5.8) and (5.10) it remains to demonstrate that
presence of the potential $-\sqrt{r^{-2}+K}+G$ in Perlick's Hamiltonian,
Eq.(5.8), is harmless for the correspondence we just established. For this
purpose we need to take into account that $\sqrt{r^{-2}+K}=\frac{1}{2}(%
\tilde{r}^{-\beta }+K\tilde{r}^{\beta })$. \ This result is obtained on page
8 of [114]. In terms of these variables the Hamiltonian, Eq.(5.8), acquires
the following form%
\begin{equation}
\mathcal{H}_{\text{I}}(r,p)=\frac{1}{4}r^{2}(r^{-\beta }-Kr^{\beta
})^{2}\left\vert \mathbf{p}\right\vert ^{2}-\frac{1}{2}(r^{-\beta
}+Kr^{\beta })+G.  \tag{5.16}
\end{equation}%
Using the conformal St\"{a}ckel transform described in the Appendix F the
Hamiltonian $\mathcal{H}_{\text{II}}(r,p)$ is obtainable from $\mathcal{H}_{%
\text{I}}(r,p)$ as follows. Relabel $\mathcal{H}_{\text{I}}(r,p)$ as \ $%
\mathcal{H}_{U\text{I}}(r,p),$ where $U=-\frac{1}{2}(r^{-\beta }+Kr^{\beta
})+G.$ Let%
\begin{equation}
\mathcal{H}(r,p)=r^{2}(r^{-\beta }-Kr^{\beta })^{2}\left\vert \mathbf{p}%
\right\vert ^{2}+\alpha .  \tag{5.17a}
\end{equation}%
The Hamiltonian $\mathcal{H}_{\text{II}}(r,p)$ is defined now as%
\begin{equation}
\mathcal{H}_{\text{II}}(r,p)=\frac{\mathcal{H}(r,p)}{U}=\frac{1}{2}\frac{%
r^{2}(r^{-\beta }-Kr^{\beta })^{2}\left\vert \mathbf{p}\right\vert ^{2}}{%
r^{-\beta }+Kr^{\beta }-2G}+\frac{2\alpha }{r^{-\beta }+Kr^{\beta }-2G}. 
\tag{5.17b}
\end{equation}%
By introducing new variables $r^{\prime }$ and $p^{\prime }$ so that $r=%
\mathcal{F}(r^{\prime }),p=F(p^{\prime })$ with known\footnote{%
E.g. read page 49 of [106$].$} functions $\mathcal{F}$ and $F$ \ converts $%
\mathcal{H}_{\text{II}}(r,p)$ into the standard form. It is obtainable with
help of Perlick's oscillator-type metric, Eq.(13) of [23], using
Eq.s(5.3),(5.4). Thus, by using results of section 4, we just established
the correspondence between the D-O Hamiltonian $\mathcal{H}(r,p)$ and that
obtained with help of type II Bertrand metric. \ Since in the conformally
flat spaces of constant curvature there is a Coulomb-oscillator duality
[115], this duality does exist between the Bertrand spaces of types I and
II. This result is supported by the following two theorems\bigskip

\textbf{Theorem 5.1}. \footnote{%
[116$],$Theorem $17.$}\textit{Every nondegenerate second-order quantum
superintegrable system on }

\textit{a 3d conformally flat space is St\"{a}ckel equivalent to a
superintegrable system on constant }

\textit{curvature space\bigskip }

\textbf{Theorem 5.2.\footnote{%
[116$],$Theorem 4.}} \textit{Every superintegrable system with nondegenerate
potential on 3d }

\textit{conformally flat space is St\"{a}ckel equivalent to a
superintegrable system on either 3d flat }

\textit{space or on }$S^{3}\bigskip $

\textbf{Remark 5.3.} Theorem 5.2. provides justification of the result,
Eq.(4.53),

obtained in previous section.\bigskip

For reader's convenience we notice that Perlick's spacetimes are, in fact,
warped products. This means the following. Consider a static (spherically
symmetric in Perlick's case) spacetime ($\mathcal{M},d\sigma ^{2})$ made of
the product of spherically symmetric 3-manifold $\mathcal{M}$ (with metric $%
dg^{2})$ and the line (time) thus making the Lorentzian warped product. The
varping function $V(q)$ must be strictly positive. The projection of each
timelike geodesic on a constant time leaf $M_{0}=\mathcal{M}\times \{t_{0}\}$
is a trajectory of the dynamical system with Hamiltonian, Eq.(5.3), with $%
p_{z}^{2}=1.$ Thus the trajectory in Perlick's spacetime ($\mathcal{M}%
,d\sigma ^{2})$ is just a projection of the timelike geodesic onto $M_{0}.$
For more details on warping products one should consult [118$%
],[119].\bigskip $

In section 1 we mentioned work by Little [19$].$In it he obtained a
remarkable result. On $S^{2}$ there are only 3 distinct regular homotopy
classes for oriented closed curves: a) those without self -intersections, b)
those with just one self-intersection and, c) those with 2
self-intersections. Because the Bertrand spaces $M_{0}$ are spherically
symmetric and, in view Theorems 5.1. and 5.2. and, of Hopf mapping, $\ $they
are topologically either $S^{2}$ $\ $or $S^{3}.$ Based on these observations
we anticipate Little's results to be fully enforced. This is indeed the
case. In [25$]$ Kirzhnitz (with associates) studied Newtonian dynamic of a
particle of unit mass moving in the presence of the Tietz potential\footnote{%
In earlier publication [120$]$ \ by the same authors more focused exposition
is given.}, Eq.(G.9), under standard initial conditions. Results were
parametrized in terms of the angular momentum $L$ of the particle, charge Z,
and dimensionless distance $x=r/a$ defined with help of Eq.(G.7) of Appendix
G. The obtained trajectory is described in polar coordinates with help of
the equation 
\begin{equation}
x+1/x=\Delta +1+(\Delta -1)\cos \varphi  \tag{5.18a}
\end{equation}%
describing a closed curve with just one self-interesction. The observed
period $T$ is described as 
\begin{equation}
T=\frac{\pi \left( \Delta +1\right) (3\Delta -1)a^{2}}{L}.  \tag{5.18b}
\end{equation}%
Here $\Delta =\frac{Za}{L^{2}}-1.$ Exactly the same type of once
self-intersecting \ closed (periodic) motion was obtained by Wheeler
[14],[16] and his assistant Powers [15$].$ At the time of these studies the
concept of Bertrand spacetime was nonexistent. Only in 2017 a systematic
study of dynamical trajectories of type I $\ ($Kepler-Coulomb$)$ Perlick
systems $\ $with Hamiltonian $\mathcal{H}_{\text{I}}(q,p),$ Eq.(5.8), was
performed by Kuru and associates [18$]$. \ In complete agreement with
previously described results, for $\beta =1/2$ the closed trajectories were
obtained with either none or one self-intersection. These are displayed
respectively in Fig.5 and Fig.7 on page 3362.\bigskip

We conclude this section with some geometrical and topological remarks which
would not be required should results of D-O papers be free of inaccuracies
to be described momentarily. In atomic physics it is common to write $%
n=n_{r}+l+1.$ By referring to Appendix G and referring to Eq.(G.5) we
recognize that it holds only if $n_{r}=0.$From standard textbooks on quantum
mechanics this result is looking suspicious. However, the results of section
4 indicate that, in fact, to use $n_{r}\neq 0$ is suspicious if calculations
are made on $S^{2}$ and, via Hopf mapping on $S^{3}.$

Below we explain why this is so using topological arguments.We begin by
recalling the 2-dimensional results of section 4. Use of the conformal
mapping permits us to solve the Sch\"{o}dinger equation on \textbf{R}$^{2},$
to lift it to $S^{2}$ and, then to lift the obtained result to $S^{3}$ using
Hopf mapping$.$In 2 dimensions the solution of the Kepler-Coulomb quantum
problem is reduced to the solution of the standard rigid rotator equation as
we have just demonstrated. By doing so known results were reproduced.Thus,
we demonstrated that there is no room for $n_{r}$ on $S^{2}$ in the 2d
Kepler-Coulomb problem. Surely, it is possible to still use $n_{r}$ on 
\textbf{R}$^{2}$ as it was done by Makowski [53$].$Since use of \textbf{R}$%
^{2}$ (or $S^{2})$ allows us to perform the conformal transformations
legitimately, it comes as no surprise that the result Eq.(4.49) for the
deformed wave function in which $n_{r}$ was used is not correct. The
corrected result, Eq.(4.53), is analytical expession of the Madelung rule%
\footnote{%
In Appendix G we demonstrate that this mathematically correct result is
consisent with much earlier results by Enrico Fermi obtained in 1928 [121] !}%
. In Appendix G \ this fact allows us to use in Eq.(G.5) $n=l+1$ (instead of 
$n=n_{r}+l+1)$as required.

The wave function $Y_{nlm}(\alpha ,\theta ,\varphi )=\Psi _{nl}(\alpha
)Y_{lm}(\theta ,\varphi )$ is not reflecting the underlying symmetry
(geometry) and topology. This could be the likely cause for the D-O to use
in the \ remnant of 3 dimensional calculations, the $n_{r}.$ By using the
above representation for $Y_{nlm}(\alpha ,\theta ,\varphi ),$the symmetry
becomes broken. Its \ breakage is obscuring the relationship between \ the
results on $S^{2}$and $S^{3}$.Group-theoretically, this issue is best
understood following [87$]$ and $[123].$ The Lie algebra $so(4)$ (working%
\footnote{%
Surely, the $su(2)$ is working of $S^{3}.$ However, the $so(4)$ of is
capable of working on $S^{3}$ very much like the rotation group $so(3)$ on $%
S^{2}$.} on $S^{3})$ is a direct product of two $so(3)$ Lie algebras%
\footnote{%
E.g.read the Appendix E}, each describing the rigid rotators with standard
spectrum $j_{i}(j_{i}+1),i=1\div 2.$ The Pl\"{u}cker embedding condition
described in Appendix E leads to the requirement: $j_{1}=j_{2}$ $[122].$
Thus, the Lie algebra so(4)$\simeq $so(3)$\oplus $so(3). This result is
achieved with help of the requirement $j_{1}=j_{2}.$ Topologically, $S^{3}$
is obtainable in two ways: a) either by gluing 2 solid tori to each other
(Heegard splitting) or, b) by gluing two 3- balls pointwise to each other
[123$].$ The last construction can be easily explained. Begin with $S^{2}.$
It can be made out of 2 discs glued to each other along the disc boundaries.
Consider a stereographic projection- from the 2 sphere whose equator lies in
the plane. Then, the projection can be made from the Northern pole or from
the Southern pole. Thus, we obtain the correspondence between the half
sphere -the disc- and the plane. It is possible then to map a plane onto the
half plane and gluing of 2 half spheres -to make an $S^{2}-$ can be
identified with gluing of 2 half planes along their common boundary. By
extending these ideas to $S^{3}$ we have to glue together 2 half $\left( 
\mathbf{R}^{3}\right) ^{\prime }s$ along their common plane. This is exact
topological equivalent of the group-theoretic condition: $j_{1}=j_{2}!$

It is possible to still extend just obtained results. Following [88$]$ the
wavefunction $Y_{nlm}(\alpha ,\theta ,\varphi )$ \ can be replaced by its
manifestly symmetric analog. \ For this , use of the manifestly symmetric 4
dimensional spherical coordinates is essential. These are given by 
\begin{equation}
x_{1}=R\cos \xi \cos \varphi ,x_{2}=R\cos \xi \sin \varphi ,x_{3}=R\sin \xi
\cos \eta ,x_{4}=R\sin \xi \sin \eta  \tag{5.19}
\end{equation}%
to be contrasted with those, given by Eq.(4.7). In terms of such coordinate
parametrization the eigenvalue Eq.(4.31a) acquires the following look%
\begin{equation}
\mathcal{\hat{L}}^{2}Z_{nfg}(\xi ,\varphi ,\eta )=I_{nfg}Z_{nfg}(\xi
,\varphi ,\eta )  \tag{5.20}
\end{equation}%
Here [87$]$%
\begin{equation}
Z_{nfg}(\xi ,\varphi ,\eta )=\left( 2\pi \right) ^{-1}\left( 2n\right) ^{%
\frac{1}{2}}(-1)^{F+g}d_{gf}^{F}(2\xi )\exp \{i\varphi (f+g)+i\eta (f-g)\} 
\tag{5.21}
\end{equation}%
and, as before, $2F+1=n.$ The detailed description of Wigner's coefficients $%
d_{gf}^{F}(2\xi )$ is given in [124].\bigskip

\textbf{Note added in proof: Existence of Bertrand spaces favors
quantization of dynamics at all scales.}

\textbf{\ It also helps to understand and to describe the exceptions to
Madelung's rule } \bigskip

Initially, the Bertrand spaces developed by Perlick [23] were \textbf{not}
meant for description of microscopic systems. Nevertheless the results of
this paper, e.g. sections 2 and 3, indicate that, by definition,
superintegrable systems of physical interest do posses closed orbits
irrespective to the actual size of these systems. If this is so, the
immediate question emerges: Since quantum mechanically closure of orbits is
linked with quantization, can one expect to encounter the macroscopic
Bertrand dynamical systems which are quantum? Although we had initiated
study of this topic previously [105], the results of \ this paper prompt us
to make additional remarks.

First, the assumption \ that all planets of our solar systems lie in the
same (Sun's equatorial) plane and move in the same direction coinciding with
the direction of rotation of the Sun around its axis was used essentially by
Henri Poincar$e^{\prime }$ in his "Les Methodes Nouvelles de la Mechanique
Celeste" written between 1892 and 1899. \ Poincar$e^{\prime }$s assumptions
\ are consistent with popular hypothesis\footnote{%
With lots of numerical simulations made in its support through years} that
our solar system was made out of pancake-like rotating cloud.\ 

Second, this plausible hypothesis was confronted with the discovery (in
1898-1899)\footnote{%
https://en.wikipedia.org/wiki/Phoebe\_(moon)} of Phoebe-the ninth moon of
Saturn rotating \ in the direction opposite to all other satellites of
Saturn. Recall that the Jupiter, Saturn, Uranus-all\ have their own
satellite systems analogous to the planetary system of our solar system.
Therefore, apparently, the pancake models should be applicable to the
satellite systems of heavy planets as well. \ But they are not! Since
Phoebe's discovery many other Phoebe-like satellites were found. The
satellites rotating in the "normal " direction \ had received the name "%
\textit{prograde}" while those in the opposite direction had been called "%
\textit{retrograde}". By 2018 in our solar system 125 retrograde satellites
were found\footnote{%
E.g.read https://en.wikipedia.org/wiki/Irregular\_moon}. Interestingly, the
retrograde motion takes place in the same planes as prograde motion.
Newton's mechanics does not imply whatsoever such a motion for retrogrades.
Not only this fact breaks down the pancake model but it and other facts of
satellite motion in solar system discussed in [105] point in favor of
quantum mechanical Bertrand-type dynamical model of solar system dynamics.

Third, the description of multielectron quantum Bertrand models of
multielectron atoms developed in this paper involves uses of conformal
invariance. This fact makes the obtained results scale insensitive. However,
general relativity together with classical mechanical Le Verrier's
calculations\footnote{%
https://en.wikipedia.org/wiki/Urbain\_Le\_Verrier} makes the assumption
about the quantum nature of solar system dynamics very problematic. The
observed planetary orbits are not closed, they precess. In atomic systems
this fact was discovered by Arnold Sommerfeld in 1916 [30] immediately after
seminal 1915 \ work by Einstein on general relativity. Unlike the case of
celestial mechanics, Sommerfeld came up with ingenous way to by pass the
issue of orbit nonclosure resulting in his result for the fine structure of
hydrogen atom spectrum rediscovered by Dirac in 1928.

Forth, these facts are not detrimental, however, to the assumption that the
Bertrand model spaces are capable of describing reality at all scales. This
is so because of the following observations. In 1966 Greenberg\footnote{%
Accidental degeneracy \textit{Am.J.Phys}\textbf{.}34 (1966) 1101-1109}
noticed that: a) the accidental degeneracy is equivalent to
superintegrability, b) by applying perturbations the superintegrability can
be removed. As an example Greenberg looked at the relativistic effects
(e.g.velocity-dependent mass ) in the Kepler-Coulomb problem.The same
problem as studied by Sommerfeld! This effect is making\ Runge-Lentz vector
nonconserved [30]. By considering the relativistic effects as small
perturbations, the celebrated Einstein's general relativistic result for the
precession of a planetary orbit was obtained. Much later, Riglioni (in the
Appendix A of his PhD thesis [106]) had reobtained the same result as
perturbation of the Bertrand-type motion caused by (incomplete) mapping of
the Bertrand-like metric onto the Schwarzschild-like. These results surely
require further study since, for instance, the exceptions to the Madelung
rule\footnote{%
E.g. read https://en.wikipedia.org/wiki/Aufbau\_principle} are consequences
of relativistic -type perturbations of superintegrable systems.

Apparently, these physics-inspired results are unknown to mathematicians
thus far. \ At the same time, physics practitioners are not aware of the
very important recent developments in mathematics related to the theory of
perturbations of superintegrable systems. \ Extremely readable introduction
into this topic can be found in papers by Francesco Fasso\footnote{%
Notes on Finite dimensional integrable Hamiltonian systems (Padova:
University of Padova, Department of Mathematics) 1999.}and Anthony Blaom%
\footnote{%
A geometric setting for Hamiltonian perturbational theory\textit{,} AMS
Memoir 153, 2001.}. \ From both sources it follows that for the
superintegrable systems theory of perturbations should be developed using
Nekhoroshev's results rather than the KAM results\footnote{%
The book by Gutzwiller [12] and, with it all other books on classical and
quantum chaos written in physics literature, does not contain any mention of
Nekhoroshev's results.}. General principles of Nekhoroshev's theory are
recently illustrated on the benchmark example of hydrogen atom in crossed
electric and magnetic fields\footnote{%
Fasso F., Fontanari D. and Sadovskii D. , An application of Nekhoroshev
theory to the sudy of the perturbed hydrogen atom Math.Phys Anal Geom \ 18%
\textbf{\ (}1\textbf{) (}2015\textbf{) }Art.30}. This example was also used
by Gutzwiller [12] as well as by many other authors without references to
Nekhoroshev's results. Nevertheless, among these publications we notice the
paper by Fl\"{o}thman and Welge\footnote{%
Crossed-field hydrogen atom and the three-body Sun-Earth-Moon problem,
Phys.Rev.A 54 (1996) 1884-1888} in which the connection (isomorphism)
between this quantum problem and the classical 3-body problem is developed
in detail. These authors notice that such quantum problem is the benchmark
problem for study of classical and quantum chaos. Rigorous study of
manifestations of adiabatic classical and quantum chaos in perturbed
superintegrable systems had just began\footnote{%
Fontanari D. Quantum manifestations of the adiabatic chaos of perturbed
superintegrable Hamiltonian systems (Padova: University of Padova,
Department of Mathematics), 2016.}. \ 

\bigskip

\textbf{Acknowledgements\bigskip }

Both authors are indebted to the members of the Department of Mathematics of
the Noviosibirsk State University for supplying us with unpublished lecture
notes by L.Ovsyannikov. Work by Louis Kauffman's was supported in part by
the Laboratory of Topology and Dynamics, Novosibirsk State University
(contract no. 14.Y26.31.0025 with the Ministry of Education and Science of
the Russian Federation).

\bigskip

\bigskip \textbf{\ Appendix A. \ Contributions of Constantin Caratheodory to
the design of \ absolute optical instruments}

\bigskip

In this appendix we shall comment on the appropriate places in remarkable
book by Caratheodory "Geometrical Optics" [22]. In it, he begins with Fermat
principle, Eq.(2.1), which he rewrites as 
\begin{equation}
\dint\limits_{\gamma ^{\prime }}L^{\prime }(t^{\prime },x_{i}^{\prime }\frac{%
dx_{i}^{\prime }}{dt})dt^{\prime }=\dint\limits_{\gamma }L(t,x_{i}\frac{%
dx_{i}}{dt})dt+\dint\limits_{\gamma }d\Psi ,i=1\div 3.  \tag{A.1}
\end{equation}%
This equation is interpreted as follows. The difference between the optical
lengths of two curve segments $\gamma $ and $\gamma ^{\prime }$ that
correspond to each other \ by means of the stigmatic map is equal to the
difference between the values of $\Psi (t,x_{i})$ at the endpoints of the
curve $\gamma .$ This difference is then independent of the shape of the
curves $\gamma $ and $\gamma ^{\prime }$ and depends only upon the position
of their endpoints. Under such circumstances Eq.(A.1) implies that the
relation 
\begin{equation}
L^{\prime }(t^{\prime },x_{i}^{\prime }\frac{dx_{i}^{\prime }}{dt}%
)dt^{\prime }=L(t,x_{i}\frac{dx_{i}}{dt})dt+\left( \frac{\partial }{\partial
t}\Psi \right) dt+\left( \frac{\partial }{\partial x_{i}}\Psi \right) dx_{i}
\tag{A.2}
\end{equation}%
must be fulfilled identically when \ one replaces the old variables $t^{{}}$%
, $x_{i}^{{}}$ by the new ones%
\begin{equation}
t^{\prime }=t^{\prime }(t,x_{i}),x_{i}^{\prime }=x_{i}^{\prime }(t,x_{i}) 
\tag{A.3}
\end{equation}%
so that, 
\begin{equation}
dt^{\prime }=\frac{\partial t^{\prime }}{\partial t}dt+\frac{\partial
t^{\prime }}{\partial x_{i}}dx_{i}\text{ and }dx_{i}^{\prime }=\frac{%
\partial x_{i}^{\prime }}{\partial t}dt+\frac{\partial x_{i}^{\prime }}{%
\partial x_{j}}dx_{j}.  \tag{A.4}
\end{equation}%
Caratheodory proves then that $\frac{\partial }{\partial t}\Psi \equiv \frac{%
\partial }{\partial x_{i}}\Psi \equiv 0$ causing Eq.(A.1) \ to be replaced
by 
\begin{equation}
\dint\limits_{\gamma ^{\prime }}L^{\prime }(t^{\prime },x_{i}^{\prime }\frac{%
dx_{i}^{\prime }}{dt})dt^{\prime }=\dint\limits_{\gamma }L(t,x_{i}\frac{%
dx_{i}}{dt})dt.  \tag{A.5}
\end{equation}%
We are not reproducing the proof since Eq.(A.5)\ was treated by Dirac whose
arguments we are presenting in section 2 of the main text. Here we only
notice that the map-from the object space to the image space, given by
Eq.s(A.3), must be conformal. This fact causes us to bring some facts from
our earlier work [3] superimposed with results of section 2. Caratheodory
notices that "From a celebrated theorem of Liouville\footnote{%
Discuseed in our work [3], page 50, Theorem B.8.} in contrast to the planar
conformal maps which depend upon infinitely many constants there is only
restricted class of conformal maps of three-dimensional space. It can always
be represented as \ a sequence of transformations through reciprocal radii
of at most five spheres\footnote{%
Caratheodory is having in mind the M\"{o}bius and Lie sphere-type
transformations discussed in our work [3] on pages 38-40 and \ the
pentaspherical transformations discussed in section 4.}. \textsl{It follows
then that the circles and lines in object space are transformed into curves
in image space that will always be either the circles or lines. Maxwell \
treated the simplest case of such a ray map (when one ignores a reflection)
on occasion"\footnote{%
Here the emphasis is ours}. }Furthermore\textsl{, }" Maxwell imagined the
entire space to be occupied \ by the medium whose index of refraction is
described by Eq.(2.5) of section 2. The light rays themselves are then
either circular or rectilinear. One confirms this result the most simply
when one recalls the equation\footnote{%
Eq.s (2.4a) and (2.4b) of the main text}

\begin{equation}
d\sigma =\frac{2ab}{b^{2}+r^{2}}\sqrt{dx^{2}+dy^{2}+dz^{2}},\text{ }%
r^{2}=x^{2}+y^{2}+z^{2}.  \tag{A.6}
\end{equation}%
The differential $d\sigma $, which defines the optical length of a line
element inside the interior of the Maxwell fish-eye, can be also interpreted
\ as a line element in the three-dimensional boundary of a four-dimensional
sphere of radius $a$ that is projected stereographically onto a space of $%
x,y,z$ that should be found at a distance $b$ from the center of the
projection." Explicitly, 
\begin{equation}
\xi =\frac{2abx}{b^{2}+r^{2}},\eta =\frac{2aby}{b^{2}+r^{2}},\zeta =\frac{%
2abz}{b^{2}+r^{2}},\tau =a\frac{b^{2}-r^{2}}{b^{2}+r^{2}}  \tag{A.7}
\end{equation}%
so that $\xi ^{2}+\eta ^{2}+\zeta ^{2}+\tau ^{2}=a^{2}$ and, reciprocally,%
\begin{equation}
x=\frac{b\xi }{a+\tau },y=\frac{b\eta }{a+\tau },z=\frac{b\zeta }{a+\tau }%
,r^{2}=\frac{b^{2}(a-\tau )}{a+\tau }.  \tag{A.8}
\end{equation}%
By differentiating either Eq.(A.7) or (A.8) we reobtain Eq.s(2.4a),(2.4b) of
section 2. Let now $x,y,z$ and, respectively, $x^{\prime },y^{\prime
},z^{\prime }$ denote the stereographic \ projections of two opposite points 
$\xi ,\eta ,\zeta $ (respectively, $-\xi ,-\eta ,-\zeta )$ on the
four-sphere, then we obtain%
\begin{equation}
x^{\prime }=-\frac{b^{2}x}{r^{2}},y^{\prime }=-\frac{b^{2}y}{r^{2}}%
,z^{\prime }=-\frac{b^{2}z}{r^{2}}  \tag{A.9}
\end{equation}%
implying 
\begin{equation}
r^{\prime }r=b^{2}  \tag{A.10}
\end{equation}%
easily recognizable \ as the main result defining the M\"{o}bius geometry.
This is achieved when $b^{2}=r^{2}=x^{2}+y^{2}+z^{2}$ so that $x^{\prime
}=-x,$ $y^{\prime }=-y$ and $z^{\prime }=-z$ and, finally, $\tau =0$. With
these results in our hands we are in the position to interpret them using
either the language of Lie sphere geometry or, which is equivalent, of
twistors\footnote{%
E.g. read page 41 of [3].} By analogy with Eq.s (6.30a,b) of [3], page 30,
and, in accord with Caratheodory, we obtain the great circles on the three
sphere as intersection of two hyperplanes 
\begin{equation}
A_{k}\xi +B_{k}\eta +C_{k}\zeta +bD_{k}\tau =0,k=1,2.  \tag{A.11}
\end{equation}%
Their intersection, when projected into $x,y,z$ space is described by 
\begin{equation}
D_{k}(x^{2}+y^{2}+z^{2}-b^{2})-2A_{k}x-2B_{k}y-2G_{k}z=0,k=1,2.  \tag{A.12}
\end{equation}%
These equations coincide exactly with Eq.s (7.16) of [3], page 41, as
required. Recall, that these equations had been used in [3] to establish
twistor- Lie sphere geometry correspondence. In the present case, following
Caratheodory, the interpretation is as follows. Eq.s (A.12) are images of
the great circles on the $S^{3}$. Due to conformal property given by
Eq.(A.10) the great circles on $S^{3}$ are in one-to one correspondence with
great circles on $S^{2}$ ($S^{2}:b^{2}=x^{2}+y^{2}+z^{2})$ so that if \ for $%
S^{3}$ they intersect in diametrically opposite points, the same is
happening in $S^{2}.$ Therefore, Eq.s.(A.1) and (A.5) can be interpreted as
follows. The requirement $d\Psi =0$ is coming from the equality of optical
lengths of the corresponding curves. That is the spherical length of two
diametrically opposite curve segments is the same for both curves. Having in
mind additional applications mentioned in the main text, it is useful to
think about just described results as follows. Suppose that \ there is a
manifold $M$ (Blashke manifold) with the property that for each point $P\in
M $ all geodesics passing through $P$ will meet again \ at the conjugated
point $P$ then, such a medium can be used as perfect optical instrument.
Arthur Besse published a book entitled " Manifolds all of whose geodesics
are closed" [125]. Blashke conjectured that the manifolds with such a
property are $S^{n}$. The current status of this conjecture is described in
[126]. For the latest optical applications of geometric and topological
ideas of Besse's book, e.g. the geodesic lenses and similar devices, please,
consult [127] and [128]. Results of Besse's book are further discussed in
section 5 from the point of view of Bertrand spaces. Interesting enough,
another construction of Blashke (e.g. Blashke products) had been used in
studies of gravitational lensing [129$].$ Gravitational lensing can be
connected with results of this Appendix as an elaboration on \ the
discussion in section 2 of the main text. \ The topic of photon sphere known
from general relativity falls also in the same category as that of Maxwell's
fish -eye [130$],[131].\medskip $

\textbf{\ Appendix B.} \ \textbf{Many uses of time changes in classical
mechanics\bigskip }

Newtonian mechanics is invariant under Galilei transformations. Such
transformations assume that time is nonchangeable parameter. However, this
assumption breaks down in Newtonian mechanics as soon as mathematically
rigorous treatment of collisions becomes of interest. Occurring mechanical
singularities typically are being treated by the appropriate time changes.
Short but excellent introduction to this topic \ is given in Ref.[132$].$
The need for variable time also emerges in theory of contact
transformations. For a quick introduction, please, consult [94$].$ In
mechanics the contact 1-form $dS$ is given by 
\begin{equation}
dS=\dsum\limits_{i=1}^{n}p_{i}dq_{i}-\text{H}(\{q_{j}\},\{p_{j}\})dt 
\tag{B.1}
\end{equation}%
where \ $\{q_{j}\}=[q_{1},...,q_{n}],$ etc$.$ For the conservative system $%
E= $H$(\{q_{j}\},\{p_{j}\})$. At least in that case it is possible to treat
time $t$ and energy $E$ as canonically conjugate variables. So, we can make
the following identifications: $t=q_{n+1},-E=-$H$=p_{n+1}.$ With this
identification the action $S$ can be written in the Maupertuis form%
\begin{equation}
S=\dint \dsum\limits_{i=1}^{n+1}p_{i}dq_{i}.  \tag{B.2}
\end{equation}%
We would like to connect this result with that discussed in the Appendix A.
For this purpose consider an equality\footnote{%
Summation over repeated indices is always being assumed}%
\begin{equation}
dS^{\prime }-p_{i}^{\prime }dx_{i}^{\prime }=\rho
(S,\{x_{j}\},\{p_{j}\})(dS-p_{i}dx_{i})  \tag{B.3a}
\end{equation}%
where $\rho (S,\{x_{j}\},\{p_{j}\})\neq 0$ and $S^{^{\prime }}=\hat{S}%
(S,\{x_{j}\},\{p_{j}\}),x_{i}^{\prime
}=x_{i}(S,\{x_{j}\},\{p_{j}\}),p_{i}^{\prime }=p_{i}(S,\{x_{j}\},\{p_{j}\}).$

The equality, Eq.(B.3a), is defining a contactomorphism [$94]$,[$133$],[$134$%
]. In contact geometry \ one always begins with the 1-form $\alpha
=dS-p_{i}dx_{i}$ and studies groups of contactomorphisms. Clearly, the
symplectic 2-form is readily obtainable via prescription: $\omega =d\alpha
=dx_{i}$ $\wedge dp_{i}\ $ These results create an impression that the
contact and symplectic geometries are different in general since it is not a
priory obvious that the contactomorphic transformations are isomorphic to
symplectic. To shed some light on this issue, we multiply both sides of
Eq.(B.3a) by new variable $\lambda \neq 0.$ Then, Eq.(B.3a) acquires the
following form%
\begin{equation}
\frac{\lambda }{\rho }\left( dS^{\prime }-p_{i}^{\prime }dx_{i}^{\prime
}\right) =\lambda (dS-p_{i}dx_{i}).  \tag{B.3b}
\end{equation}%
By introducing new notations:%
\begin{eqnarray*}
S &=&x_{0};\lambda =y_{0},y_{i}=-\lambda p_{i}; \\
S^{\prime } &=&x_{0}^{\prime };\frac{\lambda }{\rho }=y_{0}^{\prime
};y_{i}^{\prime }=-\frac{\lambda p_{i}}{\rho }
\end{eqnarray*}%
it is possible to rewrite Eq.(B.3.b) equivalently as 
\begin{equation}
y_{0}^{\prime }dx_{0}^{\prime }+y_{i}^{\prime }dx_{i}^{\prime
}=y_{0}dx_{0}+y_{i}dx_{i}.  \tag{B.4}
\end{equation}%
Using this result along with Eq.s(B.1),(B.2) it should be obvious that
Eq.(B.4) is exactly the same as Eq.(A.2) (with $\Psi =0)$ implying Eq.(A.5)
and, hence, the Dirac theory of constraints, and time changes \ discussed in
section 2. We would like to do more in this appendix.

First, we want to emphasize the differences between the contact and
symplectic geometries. \ The major difference between these two geometries
takes place when they are confronted with the description of dynamical
systems subject to the constraint H$=E$. At the quantum level this topic is
discussed in the Appendix F.

As soon as we are having such a constraint, the choice between two
geometries is broken in favor of contact geometry. The equation H$%
(\{q_{j}\},\{p_{j}\})=E$ defines the hypersurface $\Sigma $ of dimension $%
2n-1$ in $2n$ dimensional symplectic phase space $\mathcal{M}$. \ To relate
the vector fields on $\Sigma $ and $\mathcal{M}$ \ the concept of Liouville
vector field $Y$ is essential. Given that the Lie derivative $\mathcal{L}%
_{Y} $ is defined via 
\begin{equation}
\mathcal{L}_{Y}=d\circ i_{Y}+i_{Y}\circ d,  \tag{B.5}
\end{equation}%
where $d$ is the exterior differential and $i_{Y}$ is the contraction (both
applied to differential forms $\omega $), the Liouville vector field $Y$ is
defined with help of the equation 
\begin{equation}
\mathcal{L}_{Y}\omega =\omega .  \tag{B.6}
\end{equation}%
Based on this definition, the vector field $Y$ can be defined, for example,
as $Y=p_{i}\frac{\partial }{\partial p_{i}}.$

The Liouville vector field $Y$ connects the symplectic and contact
geometries with help of the equation 
\begin{equation}
\alpha =i_{Y}\omega .  \tag{B.7}
\end{equation}%
At the same time, in symplectic geometry there is a Hamiltonian vector field 
$X_{\text{H}}$ such that 
\begin{equation}
i_{X_{\text{H}}}\omega =-d\text{H or }\omega (X_{\text{H}},-)=-dH(-). 
\tag{B.8a}
\end{equation}%
Here $d$H$=\dfrac{\partial \text{H}}{\partial x_{i}}dx_{i}+\dfrac{\partial 
\text{H}}{\partial p_{i}}dp_{i}$ implying%
\begin{equation}
X_{\text{H}}=\dfrac{\partial \text{H}}{\partial p_{i}}\partial _{x_{i}}-%
\dfrac{\partial \text{H}}{\partial x_{i}}\partial _{p_{i}}  \tag{B.8b}
\end{equation}%
\ since $\omega =dx_{i}$ $\wedge dp_{i}$. With such definition of $X_{\text{H%
}}$ we obtain: 
\begin{equation}
\omega (X_{\text{H}},X_{\text{H}})=-dH(X_{\text{H}})=0  \tag{B.8c}
\end{equation}%
as required. Consider now what will happen to $\mathcal{L}_{X_{H}}\omega .$
Using Eq.(B.5) we obtain 
\begin{equation}
\mathcal{L}_{X_{\text{H}}}\omega =d\circ i_{X_{\text{H}}}\omega +i_{X_{\text{%
H}}}\circ d\omega =0  \tag{B.9}
\end{equation}%
since $d\omega =0$ and, because $i_{X_{\text{H}}}\omega =-d$H$,$we also have 
$d\circ i_{X_{\text{H}}}\omega =0.$ Thus, the Lie derivative of the
Hamiltonian vector field $X_{\text{H}}$ preserves the symplectic two-form $%
\omega .$ The question emerges: what vector field preserves the contact form 
$\alpha ?$ Let us denote this field as $R_{\alpha }$, so that%
\begin{equation}
\mathcal{L}_{R_{\alpha }}\alpha =d\circ i_{R_{\alpha }}\alpha +i_{R_{\alpha
}}\circ d\alpha =0.  \tag{B.10}
\end{equation}%
The Reeb vector field $R_{\alpha }$ is defined as the field that obeys two
conditions:%
\begin{equation}
i_{R_{\alpha }}\alpha =const,\text{ }i_{R_{\alpha }}\circ d\alpha =0. 
\tag{B.11}
\end{equation}%
Typically in the literature one encounters the condition, $const=1,$ but it
is clear that one can choose as well the condition $const>0.$ The question
of central importance is the following\bigskip

\textsl{Is there any connection between }$X_{\text{H}}$\textsl{\ and }$%
R_{\alpha }?$\textsl{\ and, if there is, how to describe such a connection?
\bigskip }

We begin with the observation [94$]$ that the very same manifold $\mathcal{M}
$ has both the symplectic and the Riemannian structure. The Riemannian
structure is described in terms of the metric tensor 2-form: $%
g=g_{ij}dx^{i}\otimes dx^{j}.$ Introduce now the (musical) $\flat -$
operator converting the vector field $X=X^{i}\frac{\partial }{\partial x^{i}}
$ into 1-form $X^{\flat }=i_{X}g=g(X,\cdot )=g_{ij}X^{i}dx^{j}.$ Introduce
also the inverse operator $\sharp $ via $\left( X^{\flat }\right) ^{\sharp
}=g^{ij}X_{i}\frac{\partial }{\partial x^{j}}=X.$ Suppose now that instead
of Eq.(B.7) we can write 
\begin{equation}
\alpha =i_{X}g.  \tag{B.12}
\end{equation}%
Suppose also that some vector field $X=\tilde{X}$ is Reeb-like, that is,
looking at Eq.(B.11) we have to require 
\begin{equation}
i_{\tilde{X}}\alpha =i_{\tilde{X}}i_{\tilde{X}}g=1.  \tag{B.13a}
\end{equation}%
Explicitly, 
\begin{equation}
i_{\tilde{X}}[g_{ij}\tilde{X}^{i}dx^{j}]=g_{ij}\tilde{X}^{i}\tilde{X}^{j}=1.
\tag{B.13b}
\end{equation}%
But, since $\alpha =i_{\tilde{X}}g$ we also must have 
\begin{equation}
\alpha =i_{\tilde{X}}g=i_{Y}\omega .  \tag{B.14}
\end{equation}%
In view of Eq.s(B5-B7) it should be clear that $d\circ i_{Y}\omega =\omega $%
. Accordingly, in view of Eq.(B.14) we also should have $d\circ $ $i_{\tilde{%
X}}g=d\alpha .$ If $\tilde{X}$ is Reeb-like, then we have to have $i_{\tilde{%
X}}\circ d\alpha =0$. But 
\begin{equation}
i_{\tilde{X}}\circ d\alpha =\omega (\tilde{X},-)=-d\text{H}(-)  \tag{B.15}
\end{equation}%
And this result becomes the same as Reeb's 2nd condition, Eq.(B.11), only
when $d$H$\equiv 0.$ Fortunately, this is the case. This follows from the
observation that Eq.(B.13b) can be reinterpreted as an equation for the
Jacobi-Maupertius Hamiltonian \~{H} for the fixed energy surface H$%
-E=0,[133].$ That is 
\begin{equation}
\text{\~{H}}=\frac{1}{2}\tilde{g}^{ij}(x)p_{i}p_{j}=1\text{,}  \tag{B.16}
\end{equation}%
where 
\begin{equation*}
\tilde{g}^{ij}(x)=\frac{g^{ij}(x)}{m[E-U(x)]}\text{ or }\tilde{g}%
_{ij}(x)=m[E-U(x)]g_{ij}(x).
\end{equation*}%
Alternatively, the results can be rewritten in the Lagrangian form as follows%
\begin{equation}
\mathcal{\tilde{L}=}\frac{1}{2}\tilde{g}_{ij}(x)\xi ^{i}\xi ^{j}\equiv \frac{%
1}{2}<\xi ^{i},\xi ^{j}>=2  \tag{B.17}
\end{equation}%
where $\xi ^{i}=\dfrac{dx^{i}}{d\sigma }$ and $p_{i}=\tilde{g}_{ij}(x)\xi
^{j}.$ \ The Euler- Lagrange equations obtainable with Lagrangian $\mathcal{%
\tilde{L}}$ are equations for geodesics $[134],$[$36]$ relative to the \
metric $\tilde{g}_{ij}(x).$ Equations for geodesics do not change if we
replace $\mathcal{\tilde{L}}$ by $\mathcal{\bar{L}}$ = $\sqrt{\mathcal{%
\tilde{L}}}$ . But this Lagrangian leads to the reparametrization-invariant
action and is homogenous of \ the first degree in $\xi ^{i}.$ That is $%
\mathcal{\bar{L}}(x,\lambda \xi )=\lambda \mathcal{\bar{L}}(x,\xi ).$ Use of
Euler's theorem produces then: $\xi ^{i}\dfrac{\partial \mathcal{\bar{L}}}{%
\partial \xi ^{i}}$ =$\mathcal{\bar{L}}$. \ From here we obtain the
Hamiltonian H=$\xi ^{i}\dfrac{\partial \mathcal{\bar{L}}}{\partial \xi ^{i}}-%
\mathcal{\bar{L}\equiv }0.$

Thus, the Reeb vector field $\tilde{X}$ $\ \QTR{sl}{always}$ describes the
(null) geodesics of energy $E=0$ in Riemannian space with metric $\tilde{g}%
_{ij}(x).$

Obtained result is in agreement with that obtained by a very different set
of arguments in [134$],$ page 27$.$Now we are in the position to formulate
the following\bigskip\ (e.g. read [134$],$page 25):

\textbf{Theorem B.1}.\textit{\ If a codimension 1 submanifold }$\mathcal{M}$%
\textit{\ is both a hypersurface of contact type (that is with }$\alpha
=i_{Y}\omega )$\textit{\ and the level set of a Hamiltonian H, then the Reeb
flow}

\textit{is reparametrization of the Hamiltonian flow.\bigskip }

\textbf{Proof}: \ on one hand we have $i_{R_{\alpha }}\circ d\alpha \mid
_{\Sigma }=0,$on another hand, we have $i_{\tilde{X}}\circ d\alpha \mid
_{\Sigma }=\omega (\tilde{X},-)\mid _{\Sigma }=-dH(-)\mid _{\Sigma
}=0.\bigskip $

Obtained results still requires some improvement. Let us return to the
1-form $\alpha =dS-p_{i}dx_{i}$ defined after \ Eq.(B.3a). A \textsl{contact
structure} on a manifold $M$ of odd dimension (say, $2n+1$) is a hyperplane
field $\zeta =\ker \alpha $ \ such that $\alpha \wedge \left( d\alpha
\right) ^{n}\neq 0.$ The pair $(M,\zeta )$ is called \textsl{contact manifold%
} [94$],$[134$].$ Now, notice that $\ker \alpha $ is known in mechanics as a
statement: $p_{i}=\partial S/\partial x_{i}.$ Thus, the action $%
S=S(\{x_{i}\})$ is a surface for which $\omega =dx_{i}\wedge dp_{i}=0.$ Such
surfaces are called \textsl{Lagrangian} surfaces. On Lagrangian surfaces the
Bohr-Sommerfel quantization condition reads: $\doint pdx=0.$The
infinitesimal automorphisms of $\zeta =\ker \alpha $ are also described in
terms of the vector fields. \ \ Suppose that $\tilde{X}$ introduced after
Eq.(B.12) is one of these fields. Then, Eq.(B13b) can be equivalently
rewritten as $\alpha (\tilde{X})=1$. But we already know from Eq.(B.16) that 
\~{H}$=1.$ Therefore we can write $\alpha (\tilde{X})=$\~{H}$(\tilde{X})$ so
that \~{H}$(\tilde{X})$ defines now the \textsl{contact Hamiltonian}. Unlike
the Liouville vector field for which Eq.(B.6) holds, it makes sense to
require 
\begin{equation}
\mathcal{L}_{\tilde{X}}\alpha =\mu \alpha  \tag{B.18}
\end{equation}%
The physical meaning of $\mu $ multiplier is to be determined momentarily.
For this purpose \ we follow [$35$], page 57. According to this reference
for the Hamiltonian H we should discuss the dynamics $\ $under constraint: H$%
=c.$ Suppose that initially such a Hamiltonian is used for description of
dynamical system \ described by%
\begin{equation}
\frac{dx}{dt}=\text{H}_{y}(x,y),\frac{dy}{dt}=-\text{H}_{x}(x,y). 
\tag{B.19a}
\end{equation}%
Changing time 
\begin{equation}
s=\dint\limits_{0}^{t}d\tau \lambda (\phi ^{\tau }(x,y))\text{ or }\frac{ds}{%
d\tau }=\lambda  \tag{B.19b}
\end{equation}%
converts the equations of motion into the following form:%
\begin{equation}
\frac{dx}{dt}=\lambda ^{-1}(x,y)\text{H}_{y}(x,y),\frac{dy}{dt}=-\lambda
^{-1}(x,y)\text{H}_{y}(x,y).  \tag{B.19c}
\end{equation}%
This \ causes the Hamiltonian Eq.s(B.19a) to loose their Hamiltonian form.
Nevertheless the function \~{H}$=\lambda ^{-1}($H$-c)$ can be taken as new
(actually contact) Hamiltonian on hypersurface $H=c$ where 
\begin{equation}
d\text{\~{H}}=\lambda ^{-1}d\text{H}  \tag{B.20}
\end{equation}%
Accordingly, for the vector fields on such hypersurface, $X_{\text{\~{H}}%
}=\lambda ^{-1}X_{\text{H}}.$ Earlier we obtained: $\alpha (\tilde{X})=$\~{H}%
$(\tilde{X}).$ Therefore, in view of Eq.(B.5), Eq.(B.18) can be rewritten
now as 
\begin{equation}
d\text{\~{H}}+i_{\tilde{X}}d\alpha =\mu \alpha .  \tag{B.21}
\end{equation}%
This result was obtained with help Eq.s(B.13) and (B.16). Taken the Reeb
vector field $R_{\alpha }$,this equation becomes: $d$\~{H}$(R_{\alpha })=\mu
.$ By rewriting Eq.(B.21) as%
\begin{equation}
d\text{\~{H}}(R_{\alpha })\alpha -d\text{\~{H}}(\tilde{X})=i_{\tilde{X}%
}d\alpha  \tag{B.22a}
\end{equation}%
and, since $\tilde{X}$ is Reeb-like, we get $i_{\tilde{X}}d\alpha =0$ \
causing us to arrive at the final (equivalent) form of Eq.(B.20), that is $d$%
\~{H}$=\mu \alpha $

\bigskip \bigskip

\textbf{Appendix} \textbf{C.} \textbf{\ Some auxiliary facts about conformal
motions}

\bigskip

Although there is a number of good sources on this topic, e.g. $[62],[63],$%
for the sake of uninterrupted reading, we present in this appendix some
essentials.

Specifically, let $x$ and $x^{\prime }$ be two different points on the
Riemannian manifold $\mathcal{M}_{d}$ of dimension $d$. Typically, the
manifold $\mathcal{M}_{d}$ is covered by a system of charts. To avoid
complications, we shall only consider $x$ and $x^{\prime }$ placed in just
one chart. Then, the equation 
\begin{equation}
g_{ij}(x)=\frac{\partial x^{\prime k}}{\partial x^{i}}\frac{\partial
x^{\prime l}}{\partial x^{j}}g_{kl}(x^{\prime })  \tag{C.1}
\end{equation}%
relates the components $g_{ij}(x)$ and $g_{kl}(x^{\prime })$ of the metric
tensor at points $x$ and $x^{\prime }$ of $\mathcal{M}_{d}.$ Suppose that
for $\varepsilon \ll 1$ we have 
\begin{equation}
x^{\prime i}=x^{i}+\varepsilon \xi ^{i}(x),  \tag{C.2}
\end{equation}%
then the Killing vector field \textbf{X} is defined by 
\begin{equation}
\mathbf{X=}\xi ^{i}(x)\frac{\partial }{\partial x^{i}}.  \tag{C.3}
\end{equation}%
Substituting of Eq.(C.2) into Eq.(C.1), Taylor expanding and keeping only
the 1st order in $\varepsilon $ terms results in the Killing Eq.(3.21) of
the main text. Explicitly, 
\begin{equation}
\nabla _{i}\xi _{j}+\nabla _{j}\xi _{i}=0.  \tag{C.4}
\end{equation}%
Here $\xi _{i}=\xi ^{j}g_{ji}$ and the covariant derivative $\nabla _{i}\xi
_{j}$ is defined as usual by%
\begin{equation}
\nabla _{i}\xi _{j}=\partial _{i}\xi _{j}-\Gamma _{ij}^{k}\xi _{k}. 
\tag{C.5}
\end{equation}%
Familiar isometries are translations and rotations. \ For $d-$dimensional
space these can be represented by the $\ d\times d$ matrices decomposable
into diagonal $d$-dimensional matrix and skew symmetric $d\times d$ matrix
having $\dfrac{d^{2}-d}{2}$ components. Thus, in the case of isometries
there are $\dfrac{d^{2}-d}{2}+d=\dfrac{d(d+1)}{2}$ isometry generators.
There is a very important theorem by Eisenhart [65$],$ Chr.6, which reads as
follows\bigskip

\textbf{Theorem C.1. }\textit{Isometric group of motion for }$\mathcal{M}%
_{d} $\textit{\ can have no more than }$\dfrac{d(d+1)}{2}$\textit{\
elements. The maximum number is achieved \ only for spaces of constant \
scalar curvature}.\bigskip

\bigskip For description of these spaces we follow the classical book by
Petrov [52$].$ In\ chapter 1 of this reference we find the following \bigskip

\textbf{Theorem C.2. }\textit{In order for }$\mathcal{M}_{d}$\textit{\ to be
a space of constant curvature it is necessary and sufficient that the
Riemannian curvature tensor }$R_{ijkl}$\textit{\ can be represented in the
form}%
\begin{equation}
R_{ijkl}=K(g_{ik}g_{jl}-g_{jk}g_{il})  \tag{C.6}
\end{equation}%
\textit{where K is constant.\bigskip }

Eq.(C.6) is satisfied if the metric can be presented in the form 
\begin{equation}
ds^{2}=\frac{1}{\sigma ^{2}}\dsum\limits_{i=1}^{d}\varepsilon _{i}\left(
dx^{i}\right) ^{2}  \tag{C.7}
\end{equation}%
where $\varepsilon _{i}=\pm 1$ are in accord with the signature of $\mathcal{%
M}_{d}$ and, the conformal factor $\sigma $ is determined based on Eq.(C.6)
and known definition of $R_{ijkl}$ via Christoffel's symbols. This leads to
the following set of equations 
\begin{equation}
\partial _{ij}\sigma =0,\sigma \lbrack \varepsilon _{i}\partial _{jj}\sigma
+\varepsilon _{j}\partial _{ii}\sigma ]=\varepsilon _{i}\varepsilon
_{j}[K+\dsum\limits_{i=1}^{d}\varepsilon _{i}\left( dx^{i}\right)
^{2}],i\neq j.  \tag{C.8}
\end{equation}%
The solution of these equations is 
\begin{equation}
\sigma =1+\frac{K}{4}\dsum\limits_{i=1}^{d}\varepsilon _{i}\left(
dx^{i}\right) ^{2}.  \tag{C.9}
\end{equation}%
\bigskip

\textbf{Corollary C.3. }In view of the Definition 3.6. and of Eq.s(C.7),
(C.9), all spaces of constant curvature are conformally flat$.$

\bigskip

\textbf{\ Appendix D. } \textbf{Calculation of Christoffel's symbols \ and
related things\bigskip }

For the metric tensor in the form $g^{ij}(x)=\rho (x)\delta ^{ij}$( or $%
g_{ij}(x)=\dfrac{1}{\rho (x)}\delta _{ij})$ calculation of Christoffel's
symbols proceeds as follows.%
\begin{eqnarray}
\Gamma _{kl}^{i} &=&\frac{1}{2}\rho \delta ^{ij}[\delta _{kj}\left( \frac{%
\partial }{\partial x^{l}}\frac{1}{\rho }\right) +\delta _{lj}\left( \frac{%
\partial }{\partial x^{k}}\frac{1}{\rho }\right) -\delta _{kl}\left( \frac{%
\partial }{\partial x^{j}}\frac{1}{\rho }\right) ]  \notag \\
&=&\frac{1}{2\rho }[\delta _{k}^{i}\left( \frac{\partial }{\partial x^{l}}%
\rho \right) +\delta _{l}^{i}\left( \frac{\partial }{\partial x^{k}}\rho
\right) -\left( \frac{\partial }{\partial x^{j}}\rho \right) \delta
^{ij}\delta _{kl}].  \TCItag{D.1}
\end{eqnarray}%
From here, in view of Eq.(3.29a), we obtain:%
\begin{equation}
g^{kl}\Gamma _{kl}^{i}=-\frac{1}{2}[\delta ^{il}\left( \frac{\partial }{%
\partial x^{l}}\rho \right) +\delta ^{ik}\left( \frac{\partial }{\partial
x^{k}}\rho \right) -d\left( \frac{\partial }{\partial x^{j}}\rho \right)
\delta ^{ij}]=\frac{d-2}{2}\delta ^{ij}\left( \frac{\partial }{\partial x^{j}%
}\rho \right)  \tag{D.2}
\end{equation}%
Using Eq.(3.29a) and selecting sign "+" we obtain for \textsl{covariant}
components of the vector $b^{i}$ 
\begin{equation}
b_{i}=a_{i}-\frac{d-2}{2}\frac{\partial }{\partial x^{i}}\ln \rho . 
\tag{D.3a}
\end{equation}%
In view of Eq.(3.44) we need to consider as well the sign"-" resulting in 
\begin{equation}
b_{i}=a_{i}+\frac{d-2}{2}\frac{\partial }{\partial x^{i}}\ln \rho . 
\tag{D.3b}
\end{equation}%
But after Eq.(3.27b) we obtained for the contravariant components of $b^{i}$
the following transformation law: $b^{\prime i}=b^{i}+2g^{ij}\theta _{j}.$
Written in the covariant form this relation looks as follows 
\begin{equation}
b_{i}^{\prime }=b_{i}+2\theta _{i}.  \tag{D.4}
\end{equation}%
Since the identically nonzero vector $a_{i}$ \ always can be represented as 
\begin{equation}
a_{i}=\frac{\partial }{\partial x^{i}}\psi  \tag{D.5}
\end{equation}%
it is possible by combining Eq.s(D.3b) and (D.4) to represent vector $%
b_{i}^{\prime }$ as 
\begin{equation}
b_{i}^{\prime }=\frac{\partial }{\partial x^{i}}[\psi +\frac{d-2}{2}\ln \rho
+2\theta ].  \tag{D.6}
\end{equation}%
From here, it follows that if in Eq.(3.27a) the transformation is performed
with 
\begin{equation}
\theta =-\frac{d-2}{4}\ln \rho -\frac{1}{2}\psi ,  \tag{D.7}
\end{equation}%
the transformed Eq.(3.27b) will have $b^{\prime i}=0,i=1,...,d.$ In
calculating of Eq.(3.37) we needed to evaluate $\Gamma _{ki}^{i}.$ \ In view
of Eq.(D.1), this can be done as follows.%
\begin{equation}
\Gamma _{ki}^{i}=-\frac{1}{2\rho }(\rho _{i}\delta _{k}^{i}+\rho _{k}\delta
_{i}^{i}-\rho _{j}\delta ^{ij}\delta _{ki})=-\frac{d}{2}\frac{\rho _{k}}{%
\rho }.  \tag{D.8}
\end{equation}%
\bigskip

\textbf{\ Appendix E. \ \ Solution of the constraint Eq.(4.29). SO(4)
rotational group and Pl\"{u}cker matrices} \bigskip

In this appendix we are discussing the role of the constraint, Eq.(4.29), in
finding solutions of Eq.(4.26b). This equation is obtained by mapping of the
3 dimensional Schr\"{o}dinger's Eq.(3.19a) with potential Eq.(3.19b) ($%
\gamma =1)$ onto $S^{3}.$ We need to demonstrate that the constraint,
Eq.(4.29), is \ enforcing \ the spherical topology of $S^{3}$.

We begin with using some results from our work [3]. Specifically, in section
6 of [3$]$ we defined the moment \textbf{L} and the Laplace-Runge-Lentz
vector \textbf{A }for the hydrogen\textbf{\ }atom model Hamiltonian\textbf{.}
These vectors are subject to the 3-dimensional constraint \textbf{L}$\cdot 
\mathbf{A}$\textbf{=}0\textbf{. }To connect this result with Fock's results
on hydrogen atom [37], we found in [3$]$ that it is convenient to look at
this constraint from the point of view of the 4-dimensional space. Both in 3
and 4 dimensions this equation represents the Pl\"{u}cker embedding
condition. Because of this, the constraint, Eq.(4.29), is of the same
category as previously studied constraint \textbf{L}$\cdot \mathbf{A}$%
\textbf{=}0. Quantization of \textbf{L} and \textbf{A} leads to 2 decoupled
so(3) Lie algebras as demonstrated in [3$]$ and $[87]$. From group theory
[87] is is known that the Lie algebra so(4)$\simeq $so(3)$\oplus $so(3).
This observation is the starting point of our analysis.

Let the vector \textbf{X} $\in $ \textbf{R}$^{4}$ be \ described as \textbf{X%
}= [U,V,W,T]$^{\text{T}}.$ Using this notation we consider all points 
\textbf{X}$\in $\textbf{R}$^{4}\diagdown \mathbf{0.}$ They can be used for
design of projective space \textbf{P}$^{3}.$ It can be constructed out of
equivalence classes made out of vectors \textbf{X }and \textbf{Y }$\in $%
\textbf{R}$^{4}\diagdown \mathbf{0}$ such that \textbf{X=}$\lambda \mathbf{Y,%
}$ $\lambda \neq 0.$ To distinguish between the \textbf{R}$^{4}$ and \textbf{%
P}$^{3}$ the \textsl{homogenous coordinates} in \textbf{P}$^{3}$ should be
introduced, e.g. via notation: \textbf{X}= [U:V:W:T]. \ This helps us to
identify the points of \textbf{P}$^{3}$ with the equivalence classes.
Clearly, the 3 dimensional vectors, e.g. \textbf{v, } are obtained now as 
\textbf{v}=[$\frac{U}{T}$,$\frac{V}{T}$,$\frac{W}{T}$]$^{\text{T}}.$
Reciprocally, a homogenous presentation of the 3-vector \textbf{v=[}U,V,W%
\textbf{]}$^{T}$ is given by v=[U:V:W:1]. Thus, \textbf{P}$^{3}$ is made out
of all points \textbf{[}U,V,W\textbf{]}$^{T}$ of \textbf{R}$^{3}$ plus the
points [U:V:W:0] at infinity comprising a projective plane \textbf{P}$^{2}.$
Planes in \textbf{P}$^{3}$ can be defined with help of the concept of
duality defined \ with help of the scalar product \TEXTsymbol{<}$,>$ in
Euclidean space. Specifically, consider the equation 
\begin{equation}
<\mathbf{\pi },X>=\pi _{1}X_{1}+\pi _{2}X_{2}+\pi _{3}X_{3}+\pi _{4}X_{4}=0.
\tag{E.1a}
\end{equation}%
It is looking exactly the same as Eq.(4.29). Its interpretation is the
following. The point X=[X$_{1}$:X$_{2}$:X$_{3}$:X$_{4}$] lies on the
hyperplane $\mathbf{\pi }$=[$\pi _{1}$:$\pi _{2}$:$\pi _{3}$:$\pi _{4}$] if
and only if Eq.(E.1a) is satisfied. Thus, in projective spaces the planes
and the points are equivalent in view of the duality just defined. Clearly,
in a more familiar 3 dimensional setting \ the same result is presented as%
\begin{equation}
\pi _{1}X+\pi _{2}Y+\pi _{3}Z+\pi _{4}=0,  \tag{E.1b}
\end{equation}%
where X=X$_{1}$/X$_{4}$,Y=X$_{2}$/X$_{4}$, Z=X$_{3}$/X$_{4}.$ Eq.(E,1a) can
be conveniently rewritten as 
\begin{equation}
\mathbf{\pi }^{\text{T}}\mathbf{X}=0.  \tag{E.1c}
\end{equation}%
Because of the duality, there is a compelling reason to \ look as well at
the equation 
\begin{equation}
\left[ 
\begin{array}{c}
\mathbf{X}_{1}^{\text{T}} \\ 
\mathbf{X}_{2}^{\text{T}} \\ 
\mathbf{X}_{3}^{\text{T}}%
\end{array}%
\right] \mathbf{\pi }=0.  \tag{E.2a}
\end{equation}%
This should be understood as follows. Three points $\mathbf{X}_{1}^{{}},%
\mathbf{X}_{2}^{{}},\mathbf{X}_{3}^{{}}$ \ in general position define a
plane. Each of these points is satisfying Eq.(E.1c), that is, $\mathbf{\pi }%
^{\text{T}}\mathbf{X}_{i}=0,i=1\div 3.$ Therefore, Eq.(E.2a) is defining a
plane determined by 3 prescribed points in general position. By applying
duality, we also get 
\begin{equation}
\left[ 
\begin{array}{c}
\mathbf{\pi }_{1}^{\text{T}} \\ 
\mathbf{\pi }_{2}^{\text{T}} \\ 
\mathbf{\pi }_{3}^{\text{T}}%
\end{array}%
\right] \mathbf{X}=0.  \tag{E.2b}
\end{equation}%
To proceed, we need to define the notion of a line in \textbf{P}$^{3}$ and
to connect it with the results pertinent to the Lie algebra so(4). In doing
so, we follow the original Pl\"{u}cker ideas described in [3$]$ and [94$]$.
Some results from lecture notes by William Goldman [139$]$ and those of
[140] are also helpful$.$ Specifically, we need to recall that the Lie
algebra so(4) is made out of the set of 4$\times $4 skew symmetric matrices.
These can be built as follows 
\begin{eqnarray}
\mathbf{R}^{4}\times \mathbf{R}^{4} &\rightarrow &so(4)\text{ or,
explicitly, }  \notag \\
(\mathbf{v},\mathbf{w}) &\rightarrow &\mathbf{v}\wedge \mathbf{w:=wv}^{T}-%
\mathbf{vw}^{T}:=\mathbf{L.}  \TCItag{E.3}
\end{eqnarray}%
Here \textbf{L} is defining \ the Pl\"{u}cker matrix describing the line $%
\mathit{l}$ in \textbf{P}$^{3}$. The notion of Pl\"{u}cker matrix is absent
in [3$],[94]$ and [136$].$But it can be found in [137]. It is essential,
however, and requires some explanation.

\ To do so, we follow Pl\"{u}cker's ingenuous idea that 3 dimensional lines
can be described in terms of 3 dimensional vectors \textbf{y} and \textbf{x, 
}taken from\textbf{\ \ }the\textbf{\ }same origin\textbf{\ o, }whose ends%
\textbf{\ }are located on the line \textit{l}. In 3-space the direction for
such a line is \textbf{r}: \textbf{r}=\textbf{y}-\textbf{x. }The shortest
distance \textbf{z} between $\mathit{l}$ and \textbf{o,} the cross product 
\textbf{n}=\textbf{y}$\times $\textbf{x} and \textbf{r} are forming the
right handed tripod. Any 2 of these 3 vectors are sufficient to determine
the 3rd one. Typically the vectors \textbf{r} and \textbf{n} are chosen.
With these results in our hands, following Pl\"{u}cker, we introduce 
\begin{equation}
p_{01}=y_{1}-x_{1}\text{ ; }p_{23}=x_{2}y_{3}-x_{3}y_{2},  \tag{E.4a}
\end{equation}%
\begin{equation}
p_{02}=y_{2}-x_{2}\text{ ; }p_{31}=x_{3}y_{1}-x_{1}y_{3},  \tag{E.4b}
\end{equation}%
\begin{equation}
p_{03}=y_{3}-x_{3}\text{ ; }p_{12}=x_{1}y_{2}-x_{2}y_{1}.  \tag{E.4c}
\end{equation}%
From these results it follows that $\mathbf{x}=(x_{1},x_{2},x_{3})^{\text{T}%
},\mathbf{y}=(y_{1},y_{2},y_{3})^{\text{T}},$ $p_{ij}=x_{i}y_{j}-x_{j}y_{i},$
$x_{0}=y_{0}=1.$The Pl\"{u}cker relation, Eq.(6.18c), of [3$]$ in terms of
just defined notations acquires the form \textbf{n}$\cdot \mathbf{r}=0,$
where 
\begin{equation*}
\mathbf{r}=\left[ 
\begin{array}{c}
p_{01} \\ 
p_{02} \\ 
p_{03}%
\end{array}%
\right] \text{ and }\mathbf{n}=\left[ 
\begin{array}{c}
p_{23} \\ 
p_{31} \\ 
p_{12}%
\end{array}%
\right] ^{\text{T}}.
\end{equation*}%
Explicitly, 
\begin{equation}
p_{01}p_{23}+p_{02}p_{31}+p_{03}p_{12}=0.  \tag{E.5}
\end{equation}%
Suppose now that \textbf{q} is the point lying on the line \textit{l}
determined by \textbf{n} and \textbf{r. }Then\textbf{, }the condition\textbf{%
\ }that the point \textbf{q} is indeed lying on the line \textit{l \ }is
given by%
\begin{equation}
\mathbf{r\times q=n.}  \tag{E.6}
\end{equation}%
To prove this, \ we assume that \textbf{q}=\textbf{y}+$\lambda \mathbf{r}$
where $\lambda $ is some nonzero arbitrary multiplier$\mathbf{.}$ By
substituting this result into Eq.(E.6) we indeed obtain the identity. \
Next, we need to convert this result into manifestly 4 dimensional form.
This can be achieved as follows. We begin with representation for the Lie
algebra so(3). A suitable basis is made out of matrices $\left( \mathcal{A}%
_{i}\right) _{jk}=-\epsilon _{ijk}.$ Here $i,j$ and $k$ can take values $%
1\div 3.$ To generalize this result to so(4), it is convenient to begin with 
\begin{equation}
\left( \mathcal{A}_{1}\right) _{jk}=\left( 
\begin{array}{cccc}
0 & 0 & 0 & 0 \\ 
0 & 0 & -1 & 0 \\ 
0 & 1 & 0 & 0 \\ 
0 & 0 & 0 & 0%
\end{array}%
\right) ,\left( \mathcal{A}_{2}\right) _{jk}=\left( 
\begin{array}{cccc}
0 & 0 & 1 & 0 \\ 
0 & 0 & 0 & 0 \\ 
-1 & 0 & 0 & 0 \\ 
0 & 0 & 0 & 0%
\end{array}%
\right) ,\left( \mathcal{A}_{3}\right) _{jk}=\left( 
\begin{array}{cccc}
0 & -1 & 0 & 0 \\ 
1 & 0 & 0 & 0 \\ 
0 & 0 & 0 & 0 \\ 
0 & 0 & 0 & 0%
\end{array}%
\right)  \tag{E.7}
\end{equation}%
These matrices obey the so(3) commutation relations [$\mathcal{A}_{i}$,$%
\mathcal{A}_{j}$]=$\epsilon _{ijk}\mathcal{A}_{k}$ . Using just defined
matrices, \ we can now define the $\mathcal{B}$-matrices also obeying the
same commutator algebra [$\mathcal{B}_{i}$,$\mathcal{B}_{j}$]=$\epsilon
_{ijk}\mathcal{B}_{k}.$ Thus, we obtain so(4)$\simeq $so(3)$\oplus $so(3).
This is the Lie algebra of hydrogen atom [3$],[87].$ Explicitly, the $%
\mathcal{B}$ matrices are given by 
\begin{equation}
\left( \mathcal{B}_{1}\right) _{jk}=\left( 
\begin{array}{cccc}
0 & 0 & 0 & -1 \\ 
0 & 0 & 0 & 0 \\ 
0 & 0 & 0 & 0 \\ 
1 & 0 & 0 & 0%
\end{array}%
\right) ,\left( \mathcal{B}_{2}\right) _{jk}=\left( 
\begin{array}{cccc}
0 & 0 & 0 & 0 \\ 
0 & 0 & 0 & -1 \\ 
0 & 0 & 0 & 0 \\ 
0 & 1 & 0 & 0%
\end{array}%
\right) ,\left( \mathcal{B}_{3}\right) _{jk}=\left( 
\begin{array}{cccc}
0 & 0 & 0 & 0 \\ 
0 & 0 & 0 & 0 \\ 
0 & 0 & 0 & -1 \\ 
0 & 0 & 1 & 0%
\end{array}%
\right) .  \tag{E.8}
\end{equation}%
By looking at Eq.(E.3) we\ consider now the designing of a specific line-
the X axis. It can be constructed using 2 reference points 
\begin{equation*}
\left[ 
\begin{array}{c}
0 \\ 
0 \\ 
0 \\ 
1%
\end{array}%
\right] =\mathbf{w}\text{\textbf{\ }and\textbf{\ }}\left[ 
\begin{array}{c}
1 \\ 
0 \\ 
0 \\ 
0%
\end{array}%
\right] =\mathbf{v}
\end{equation*}%
representing the origin of X-axis and the point at infinity for the X axis.
Using Eq.(E.3), we obtain now for the X-line the following result: 
\begin{equation}
\mathbf{L}=\left[ 
\begin{array}{c}
0 \\ 
0 \\ 
0 \\ 
1%
\end{array}%
\right] \left[ 
\begin{array}{cccc}
1 & 0 & 0 & 0%
\end{array}%
\right] -\left[ 
\begin{array}{c}
1 \\ 
0 \\ 
0 \\ 
0%
\end{array}%
\right] \left[ 
\begin{array}{cccc}
0 & 0 & 0 & 1%
\end{array}%
\right] =\left( 
\begin{array}{cccc}
0 & 0 & 0 & -1 \\ 
0 & 0 & 0 & 0 \\ 
0 & 0 & 0 & 0 \\ 
1 & 0 & 0 & 0%
\end{array}%
\right) =\left( \mathcal{B}_{1}\right) _{jk}.  \tag{E.9}
\end{equation}%
Just obtained result allows us to rewrite Eq.(E.6) in 4 dimensional form%
\begin{equation}
\left( 
\begin{array}{cccc}
0 & p_{23} & -p_{13} & p_{12} \\ 
-p_{23} & 0 & p_{03} & -p_{02} \\ 
p_{13} & -p_{03} & 0 & p_{01} \\ 
-p_{12} & p_{02} & -p_{01} & 0%
\end{array}%
\right) \cdot \left( 
\begin{array}{c}
q_{0} \\ 
q_{1} \\ 
q_{2} \\ 
q_{3}%
\end{array}%
\right) =0  \tag{E.10}
\end{equation}%
The correctness of this result is easily checked following [140], page 201.
From \ this reference we find%
\begin{equation}
\mathbf{x}\times \mathbf{y}=\left( 
\begin{array}{ccc}
0 & -x_{3} & x_{2} \\ 
x_{3} & 0 & -x_{1} \\ 
-x_{2} & x_{1} & 0%
\end{array}%
\right) \cdot \left( 
\begin{array}{c}
y_{1} \\ 
y_{2} \\ 
y_{3}%
\end{array}%
\right) .  \tag{E.11}
\end{equation}%
The 4 dimensional extension of this result is straightforward. Evidently,
the result, Eq.(E.10), coincides with the constraint, Eq.(4.29). However
from Eq.(4.26b) it follows that there are 4 momenta while we just got 6
vectors since every so(4) \ matrix is decomposable in terms of the 6
skew-symmetric matrices $\mathcal{A}_{i}$ and $\mathcal{B}_{i}.$ To clarify
this mismatch, we follow [3$]$ and [139$].$ In \textbf{R}$^{4}$ we are
having \{\textbf{e}$_{0},...,\mathbf{e}_{3}\}$ as the basis. Using this
basis we can construct the set of exterior products. To label these products
we need to define the subsets $J_{p}=\{i_{1},...,i_{p}\},p\leq n.$ There are
exactly n!/(n-p)!p! of such subsets. Change of the basis: \{\textbf{e}$%
_{0},...,\mathbf{e}_{n}\}$ to \{$\mathbf{e}_{0}^{\prime },...,\mathbf{e}%
_{n}^{\prime }\}$ leads to the projective equivalence, e.g. to $\mathbf{e}%
_{i_{1}}^{\prime }\wedge \cdot \cdot \cdot \wedge \mathbf{e}_{i_{p}}^{\prime
}$ =det(A$_{ij})\mathbf{e}_{i_{1}}^{{}}\wedge \cdot \cdot \cdot \wedge 
\mathbf{e}_{i_{p}}.$ The Grassmannian Gr$_{p}$(\textbf{R}$^{n})$ is made out
of \ all of just defined $p$ -dimensional subspaces of \textbf{R}$^{n}$. Let
n=4 and consider all 2 dimensional subspaces of Gr$_{2}$(\textbf{R}$^{4})$.
These are made of 2- planes, e.g. \textbf{e}$_{i}\wedge \mathbf{e}_{j}$
There are 6=4!/2!2! planes which replace the 4 basis vectors \{\textbf{e}$%
_{0},...,\mathbf{e}_{3}\}.$ Because of the projective equivalence of such
planes, they are Pl\"{u}cker embedded into 5 dimensional projective space 
\textbf{P}$^{5}$. From Eq.s(E.4a-c) it is possible to establish 1-1
correspondence between the 6 planes set \textbf{e}$_{i}\wedge \mathbf{e}_{j}$
and the 6 planes set $p_{ij}$ . But $p_{ij}$ are the momenta! The first of
them, the \textbf{p}$_{0}$, is made out of planes \textbf{e}$_{0}\wedge 
\mathbf{e}_{1},$ \textbf{e}$_{0}\wedge \mathbf{e}_{2},$ \textbf{e}$%
_{0}\wedge \mathbf{e}_{3},$ There are 3 so(4) matrices associated with these
planes. The second- the \textbf{p}$_{1}$, is made of planes \textbf{e}$%
_{1}\wedge \mathbf{e}_{2}$ and \textbf{e}$_{1}\wedge \mathbf{e}_{3}$ and the
associated with them so(4) matrices and, the 3rd, the \textbf{p}$_{2},$ is
made out of plane \textbf{e}$_{2}\wedge \mathbf{e}_{3}$ and, the associated
with it, the so(4) matrix. These arguments explain the mismatch. The 6
infinitesimal generators obtained with help of 6 so(4) matrices are
obtainable in exactly the same way as 3 generators for the 3 dimensional
rigid rotator. Their explicit form is given in Appendix C of [87$].\bigskip $

\textbf{\ Appendix F. \ \ \ \ Essentials of the Sturmian problem and of the
conformal }\ \textbf{St\"{a}ckel transform\bigskip }

In our exposition of this topic we follow [69$],$[90$],[138]$ and $[139].$
In particular, following $[138]$ we begin with brief discussion of the
Sturmian problem. The hydrogen orbitals satisfy the 1-electron \ stationary
Schr\"{o}dinger equation (in the appropriately chosen system of units)%
\begin{equation}
\lbrack -\frac{1}{2}\nabla ^{2}-\frac{Z}{r}+\frac{Z^{2}}{2n}]\psi _{nlm}(%
\mathbf{x})=0  \tag{F.1a}
\end{equation}%
Denote $k=\dfrac{Z}{n},$ and rewrite Eq.(F.1a) \ using just introduced
notation:%
\begin{equation}
\lbrack -\frac{1}{2}\nabla ^{2}-\frac{nk}{r}+\frac{k^{2}}{2}]\tilde{\psi}%
_{nlm}(\mathbf{x})=0.  \tag{F.1b}
\end{equation}%
The eigenfunctions $\tilde{\psi}_{nlm}$ obey the \textsl{isoenergetic}
Sturmian (Schr\"{o}dinger) equation:%
\begin{equation}
\lbrack -\frac{1}{2}\nabla ^{2}+\frac{k^{2}}{2}]\tilde{\psi}_{nlm}(\mathbf{x}%
)=\beta _{n}\frac{1}{r}\tilde{\psi}_{nlm}(\mathbf{x}).  \tag{F.1.c}
\end{equation}%
In this equation, the energy $E=-\dfrac{k^{2}}{2}$ is fixed and the
eigenvalue $\beta _{n}$ is $\beta _{n}=nk.$ If the eigenfunctions $\psi
_{nlm}(\mathbf{x})$ are orthogonal, i.e. 
\begin{equation}
\dint d^{3}x\psi _{n^{\prime }l^{\prime }m^{\prime }}^{\ast }(\mathbf{x}%
)\psi _{nlm}(\mathbf{x})=\delta _{n^{\prime }n}\delta _{l^{\prime }l}\delta
_{m^{\prime }m}  \tag{F.2a}
\end{equation}%
, the eigenfunctions $\tilde{\psi}_{nlm}(\mathbf{x})$ are weighted
orthogonal, i.e.%
\begin{equation}
\dint d^{3}x\tilde{\psi}_{n^{\prime }l^{\prime }m^{\prime }}^{\ast }(\mathbf{%
x})\left( \frac{n}{kr}\right) \tilde{\psi}_{nlm}(\mathbf{x})=\delta
_{n^{\prime }n}\delta _{l^{\prime }l}\delta _{m^{\prime }m}.  \tag{F.2b}
\end{equation}%
The conformal St\"{a}ckel transform following [69$],$[90$]$ can be defined
as follows. Begin with the manifestly conformal (Laplace-type) equation%
\begin{equation}
\mathcal{H}\Psi =[-\frac{1}{\lambda (\mathbf{x})}\nabla ^{2}+V(\mathbf{x}%
)]\Psi =0.  \tag{F.3}
\end{equation}%
In this equation let $V(\mathbf{x})=W(\mathbf{x})-U(\mathbf{x})E.$ In accord
with Eq.(F.1b) \ we shall consider the parameter $E$ as fixed. The potential 
$U(\mathbf{x})$ defines the\textsl{\ conformal St\"{a}ckel transform}: from
manifestly conformal Eq.(F.3) (e.g. read section 3) to the Helmholtz-type
equation%
\begin{equation}
\mathcal{\tilde{H}}\tilde{\Psi}=E\tilde{\Psi},  \tag{F.4}
\end{equation}%
where 
\begin{equation}
\mathcal{\tilde{H}=-}\frac{1}{\tilde{\lambda}(\mathbf{x})}\nabla ^{2}+\tilde{%
V}(\mathbf{x}).  \tag{F.5}
\end{equation}%
Here $\tilde{\lambda}(\mathbf{x})=\lambda U,\tilde{V}$ $=\dfrac{W(\mathbf{x})%
}{U(\mathbf{x})}.$ The choice of the potential $U(\mathbf{x})$ is
nontrivial. It is determined by the Bertrand-Darboux system of partial
differential equations.\bigskip

\textbf{Theorem F.1.}\textit{\ Any second order conformal (Laplace-type)
superintegrable system admitting nonconstant potential U(\textbf{x}) can be
St\"{a}ckel-transformed to a Helmholtz superintegrable system. This
operation is invertible. The inverse mapping is taking true symmetries \ to
conformal symmetries.\bigskip }

\textbf{Remark F.2.} The nontriviality of the potential $U$(\textbf{x}) is
coming from the fact that it is obtainable as a solution of the system \ of
second order (conformal) Bertrand-Darboux (B-D) partial differential
equations (PDE).\bigskip

\textbf{Remark F.3.} Solutions of the B-D system of PDE's resulting \ in
obtaining of $U(\mathbf{x})$ is linked with solution of the Bertrand problem
in curved spacetimes (e.g. read section 5). D-O by passed \ the problem of
finding the U(\textbf{x}) by solving the B-D system of PDE's. They simply
guessed correctly the potential.\bigskip

\textbf{Remark F.4. }Eq.(F.3) is exactly of the same type as
Eq.s(3.14),(3.35). Therefore, methods described in section 3, also involving
solutions of PDE's, should be considered as complementary to those in [69$%
],[138],[139].\bigskip $

\textbf{Remark F.5. }There are three types of Laplace-type conformally
superintegrable equations. These are:

a) Recasting a Helmholtz superintegrable system $H\Psi =E\Psi $ into the
Laplace-type form $H^{\prime }\Psi =0,$ $H^{\prime }=H-E.$ Here $E$ is
absorbed into potential as a parameter. Clearly, this is the case of
Sturmian Eq.(F1.b).

b) \ Replacing a Helmholtz superintegrable system $H\Psi =E\Psi $ by the
Laplace-type $H^{\prime }\Psi =0$ in which $H^{\prime }=H-E_{0}$ with $E_{0}$
\ fixed once and for all eigenvalue. This is again the Sturmian-type problem.

c) Begin with a Helmholtz-type equation $H\Psi =E\Psi $ which is not
superintegrable and convert it into the Laplace-type. The resulting equation
will have the induced conformal symmetries.\bigskip

\textbf{Remark F.6. }Just described classification justifies the transition
from\ Eq.(1.2) to Eq.(1.3).\textbf{\bigskip }

\textbf{Appendix G.} \ \ \textbf{Thomas-Fermi treatment of multielectron
atoms. }\ \textbf{Contributions of Fermi, Tietz\bigskip\ and Klechkovski
into the proof of Madelung's rule\medskip \medskip }

This appendix is designed for chemists and atomic physicists. Mathematically
inclined researchers can skip it. Although the Thomas-Fermi (T-F) method is
only asymptotical (works well in the limit of high electron densities),
because of its simplicity and reasonably good accuracy, we provide some
facts about this method helpful for practical applications of the formalism
of the main text to multielectron atoms. The T-F method and contributions by
Tietz to this method used in section 4.5 are concisely summarized in the
book by Flugge [24]. Belokolos [39] used the Tietz potential
(Eq.s(G.8),(G.9) below) for the WKB-assisted "proof" of the Madelung rule.
His results are neither exact not conclusive even though he is citing the
paper by Wong [140$]$ in which the Madelung rule is seemingly derived. In
his derivation Wong used the Tietz potential. Wong's results are solely
based on seminal works by Fermi [121$],$Tietz [24] and Klechkowski [141$].$
It is impossible to develop theory of the Madelung rule without \
acknowledging contributions of these authors. The strongest and
chronologically the earliest contribution to this topic was made by Enrico
Fermi. Following publication \ of his ground breaking\ paper on the T-F
method in 1927, in 1928 Fermi came up with the paper [121$]$ \ describing
applications of the T-F method to the periodic system of elements. In it he
posed and solved the following problem:

Find analytically a relationship between the atomic number $Z$ and the
number of electrons $N_{l}$ with azimulal number $l.$

Knowledge of such a relationship allows then to solve the related problem of
major significance:

Find a sequence of $Z\prime $s for which \ for the first time electrons with
azimulal numbers $s,p,d,f,.$.emerge.

Fermi found a sequence : 1, 5, 21, 58. These are exactly the atomic numbers
initiating sequential irregularities in filing of the periodic system.
Subsequently, Tietz [142$]$ simplified \ Fermi calculations with help of the
potential bearing his name. The whole book by Klechkowski on "proving" the
Madelung rule is effectively a discussion around \ these works by Tietz .
The paper by Wong is a condensed summary of the results by Klechkowski
summarized in his book [141$]$. As such, it suffers from the same drawbacks
as Klechkowski's book. \ Specifically, in his book Klechkowski discusses
from various perspectives the result by Tietz for $N_{l}$:%
\begin{equation}
N_{l}=2(2l+1)(6Z)^{\frac{1}{3}}-2(2l+1)^{2}.  \tag{G.1}
\end{equation}%
Eq.(G.1) coincides with Eq.(14) of Wong's paper [120$]$ where it was derived
independently based directly on ideas of Enrico Fermi. \ Next, following
Klechkowski's philosophy, Wong runs into myriad of the same kinds of
problems. They began with an assumption 
\begin{equation}
\frac{N_{l}}{2(2l+1)}\simeq n-l-1=n_{r}.  \tag{G.2}
\end{equation}%
Here $n$ and $n_{r}$ are respectively \ the principal and radial quantum
numbers. A comparison between Eq.s(G.1) and (G.2) then produces:%
\begin{equation}
(6Z)^{\frac{1}{3}}-(2l+1)=n-l-1.  \tag{G.3a}
\end{equation}%
However, already from Eq.(G.1) it is evident that on the l.h.s the $N_{l}$
should be a nonnegative integer while on the r.h.s. \ the term $(6Z)^{\frac{1%
}{3}}$ is making the r.h.s. noninteger for general $Z^{\prime }s.$ Since
Eq.(G.3a) is incorrect for arbitrary nonnegative Z's, the same is true for
its corollary%
\begin{equation}
n+l=(6Z)^{\frac{1}{3}}.  \tag{G.3b}
\end{equation}%
Wong claims, however, that this result is mathematically equivalent to the
Madelung rule. Since $(6Z)^{\frac{1}{3}}$ is not always \ an integer, the
rest of Klechkowski's book [141$]$ is devoted to finding some corrections to
Eq.(G.1) making $n+l$ an integer. Wong missed this issue (studied by
Klechkowski) completely. His paper entitled "Theoretical justification of
the Madelung rule" culminates in Eq.(G.3b). None of the corrections
discussed in Klechkowski book are rigorous, systematic or convincing.
Surprisingly, he made no attempts to use directly the results by Fermi.
Instead, he made a few comments on the paper by Ivanenko and Larin [143$]$
which, in our opinion, deserves more \ attention than that given to it in
Klechkowski's book. The authors of [143$]$ used the Thomas-Fermi-Dirac
method to recalculate the results by Fermi in a direct and simple way. The
final result is expressible in compact form:%
\begin{equation}
Z_{l}=\gamma (Z_{l})(2l+1)^{3}.  \tag{G.4}
\end{equation}

The $Z_{l}-$ dependent factor $\gamma (Z_{l})$ is calculated iteratively for
each $Z_{l}$. Fig.1 of [143$]$ indicates that this variable factor can be
safely replaced by the constant factor: $\gamma (Z_{l})\simeq $ 0.169. In
such a case use of Eq.(G.4) reproduces (by rounding the numbers in Eq.(G.4))
the Fermi sequence of $\ Z^{\prime }s$: 1($l=0$), 5($l=1$), 21($l=2$) and 58(%
$l=3$) \footnote{%
Evidently, Fermi also used some rounding.}. \ Strictly speaking, the
minimization procedure \ involved \ in obtaining Eq.(G.4) is not extendable
to the values of $Z$ \ significantly different from those in the Fermi
sequence. \ In his book Klechkovski \ missed this restriction completely. If
however, in Klechkowski style, we do use Eq.(G.4) with unrestricted $%
Z^{\prime }s$ in Eq.(G.3b) in order to obtain%
\begin{equation}
n+l=(6Z)^{\frac{1}{3}}=(6\cdot 0.169)^{\frac{1}{3}}(2l+1)=1.004645(2l+1)%
\simeq 2l+1  \tag{G.5}
\end{equation}%
then, since $n=n_{r}+1+1,$ Eq.(G.5) reduces to the identity, if $n_{r}=0.$
This requirement makes perfect physical sense as explained in section 5.

Now we are in the position to make a number of comments about the Tietz
potential. In the book by Flugge [24] in\ sections devoted to the T-F theory
our readers will find a nice discussion about the Tietz potential and how it
fits the T-F theory. \ Typically, readers with standard physics background
are left with an impression that the T-F theory is reasonably well
describing miltielectron atoms \ only when the number of electrons tend to
infinity and is useless for description of, say, a hydrogen or helium atom.
\ In the lecture notes by Roi Baer [144$]$ it is demonstrated that,
surprisingly, the T-F theory provides reasonable results even for the
hydrogen atom. Additional details on T-F theory (without uses of the Tietz
potential) are beautifully summarized in the book by March [27$].$

To develop our intuition the scaling analysis by Feynman, Metropolis and
Teller [145$]$ is the most helpful. Following these authors, we consider a
similarity transformations for an atom. They are consisting of increasing
the nuclear charge of an atom while simultaneously increasing the number of
\ electrons so that the atom remains neutral. Evidently, such a
transformation is describing \ a\ relation between solutions for various
nuclear charges. It is assumed that the De Broglie length $\lambda $ remains
unchanged under scaling. Transformations consists of the following
infinitesimal changes: $Z\rightarrow Z(1+\zeta ),r\rightarrow r$(1+$\rho ),%
\bar{E}\rightarrow \bar{E}$(1+$\eta ).$ Here $\bar{E}$ represents any form
of the energy (classical or quantum) per electron. At the same time, the
electron density $q$ is changed by $q(1+\nu ).$ The charge balance leads to
the relation: $qr^{3}\rightarrow $ $q(1+\nu )r^{3}$(1+$\rho )^{3}=Z(1+\zeta
) $ implying $(1+\nu )$(1+$\rho )^{3}=1+\zeta $ or $\nu +3\rho =\zeta .$
Since the potential energy of an electron changes as $(1+\zeta )/$(1+$\rho
)= $(1+$\eta ),$ this leads to: $\zeta -\rho =\eta .$ In the T-F theory [24] 
$\bar{E}=\varkappa q^{\frac{2}{3}}$ where $\varkappa $ is known constant.
So, whenever this theory makes sense, we \ should have: $(1+\nu )^{\frac{2}{3%
}}=$(1+$\eta ),$ implying $\frac{2}{3}\nu =\eta .$ Elementary calculation
based on these results produces: $\rho =-\frac{1}{3}\zeta $ and $\eta =\frac{%
4}{3}\zeta .$ Thus, the energy $\bar{E}$ per electron changes as $\bar{E}$(1+%
$\frac{4}{3}\zeta )$ and for neutral atom initially made of $Z$ electrons it
changes as $\bar{E}$(1+$\frac{4}{3}\zeta )Z(1+\zeta )=\bar{E}Z(1+\frac{7}{3}%
\zeta ).$ Since in these arguments we assumed the validity of T-F
relationship $\bar{E}=\varkappa q^{\frac{2}{3}},$the dimensionless T-F
equation%
\begin{equation}
\frac{d^{2}\varphi }{dx^{2}}=\frac{\varphi ^{\frac{3}{2}}}{\sqrt{x}} 
\tag{G.6}
\end{equation}%
connected with this relationship automatically establishes the
characteristic length scale parameter%
\begin{equation}
a=0.8853a_{0}Z^{-\frac{1}{3}},  \tag{G.7}
\end{equation}%
where $a_{0}$ is the Bohr radius, $a_{0}=\frac{\hslash ^{2}}{me^{2}}.$ From
here emerges the dimensionless length scale: $x=\frac{r}{a}.$ As before, we
shall use the system of units in which $\hslash =m=1.$ In view of just
presented results, the potential energy of electron scales as $Z/a\simeq Z^{%
\frac{4}{3}}$ so that for $Z$ \ atomic electrons the total energy scales as $%
Z\cdot $ $Z^{\frac{4}{3}}=Z^{\frac{7}{3}}.$ This result is consistent with
the previously obtained scaling result $\bar{E}Z(1+\frac{7}{3}\zeta )$ for
the total energy. The result $Z^{\frac{7}{3}}$ was first derived by Milne [27%
$].$ Schwinger [146$]$ obtained a correction to just obtained scaling
result. With it, such a correction gives good results for all Z's. This
fact, is of importance if one is interested in comparing much more
technically cumbersome Hartree-Fock and T-F theories. Another example of
this kind is related to the discussion related to the Tietz potential. The
origins of this potential are connected with study of \ solutions of the T-F
Eq.(G.6). This nonlinear equation does not admit exact analytical solution
and, in addition, it requires the assignment of boundary conditions. For
neutral atoms these are simple: $\varphi (0)=$ 1 and $\varphi (\infty )=0.$
Because of their simplicity, Tietz suggested the analytical form for $%
\varphi :$%
\begin{equation}
\varphi (x)=\frac{1}{\left( 1+\alpha x\right) ^{2}},\text{ }\alpha =0.53625.
\tag{G.8}
\end{equation}%
The constant $\alpha $ is not a fitting parameter. \ It is determined by the
electroneutrality normalization condition [24]. The Tietz potential \ $V(x)$
is obtained with help of $\varphi (x)$\ as follows 
\begin{equation}
V(r)=-\frac{Ze}{r}\varphi (\frac{r}{a})  \tag{G.9}
\end{equation}%
For $r\rightarrow 0,$ $V(r)$ tend to the usual attractive Coulombic
potential while for larger r's it represents $V_{eff}(r)$ introduced in
Eq.(1.2). This fact was reported in [25$]$ on p.664, Fig.10. From this
figure it follows that the plot of $\varphi (x)$ vs. $x$ obtained with help
of the T-F method practically coincides with that obtained numerically with
help of the Hartree-Fock method.\bigskip \bigskip \newpage

\bigskip \bigskip

\textbf{References\bigskip \bigskip }

[1$]$ \ \ P.Thyssen, A. Ceulemans, Shattered Symmetry\textit{,}

\ \ \ \ \ \ Oxford University Press, New York, 2017.\ \ \ \ \ 

[2] \ \ H. Bethe, R. Jackiw,\ Intermediate Quantum Mechanics\textit{,}

\ \ \ \ \ \ CRC Press, New York, 2018.

[3$]$ \ \ A. Kholodenko,\ L. Kauffman, Huygens triviality of the
time-independent

\ \ \ \ \ \ Schr\"{o}dinger equation. Applications to atomic and high energy
physics,

\ \ \ \ \ \ Ann.Phys. 390 (2018) 1-59.

[4$]$ \ \ P-O. L\"{o}wdin, Some comments on periodic system of elements,

\ \ \ \ \ \ Int. J.Quant.Chemistry 3s, (1969) 331-334.

[5] \ \ L.Allen,E. Knight, The L\"{o}wdin challenge. Origin of the $n+l,n$
(Madelung)

\ \ \ \ \ \ rule\ for filling the orbital configurations,

\ \ \ \ \ \textit{\ }Int.J.Quantum Chemistry 90\textbf{\ }(2002) 80-88.

[6] \ \ E.\ Scerri, G. $\func{Re}$strepo, Mendeleev to Oganesson,

\ \ \ \ \ \ Oxford University Press, New York, 2018.

[7$]$ \ \ Y. Demkov, V. Ostrovsky, Internal symmetry of the Maxwell
"fish-eye"

\ \ \ \ \ \ problem and the Fock group for hydrogen atom. \textit{\ }

\ \ \ \ \ \ Sov.Phys. JETP 13 (1971)\textbf{\ }1083-1087.

[8$]$ \ \ Y. Kitagawara,A. Barut, Period doubling in the $n+l$ filling rule
and dynamical

\ \ \ \ \ \ symmetry of the Demkov-Ostrovsky atomic model, J.Phys.B 16
(1983) 3305-3322.

[9$]$ \ \ Y.Kitagawara, A. Barut, On the dynamical symmetry of the periodic
table:

\ \ \ \ \ \ II. Modified Demkov-Ostrovsky atomic model, J.Phys B 17 (1984)%
\textbf{\ }4251-4259.

[10] \ H.Goldstein, Ch. Poole, J. Safko, Classical Mechanics,

\ \ \ \ \ \ \ Pearson Education Ltd, London, 2014.

[11] \ \ M.Akila, D.Waltner, B. Gutkin, P. Braun,Th. Guhr,\ \ Semiclassical
identification

\ \ \ \ \ \ \ \ of periodic orbits in a quantum many-body system, PRL 118%
\textbf{,} (2017) 164101.

[12] \ \ M.\ Gutzwiller, Chaos in Classical and Quantum Mechanic,

\ \ \ \ \ \ \ \ Springer-Verlag, Berlin, 1990.

[13] \ \ P.Cvitanovic, Chaos: Classical and Quantum\textit{,} chaosbook.org

[14] \ \ J.Wheeler, From Mendeleev's atom to the collapsing star, \ Atti del
Convegno

\ \ \ \ \ \ \ \ Mendeleeviano, pp 189-233, 1971.

[15] \ \ R.Powers, Frequencies of radial oscillation and revolution as
affected by features

\ \ \ \ \ \ \ \ of a central potential, \ Atti del Convegno Mendeleeviano,
pp 235-242, 1971.

[16] \ \ J.Wheeler, Semi-classical analysis illuminates the connection \
between

\ \ \ \ \ \ \ \ potential and bound states and scattering, in Studies in
Mathematical Physics\textit{,}

\ \ \ \ \ \ \ \ pp 351-422. Princeton University Press, Princeton, 1976.

[17]\ \ \ V.Ostrovsky, Dynamic symmetry of atomic potential.

\ \ \ \ \ \ \ J.Phys. B 14\textbf{\ }(1981) 4425-4439.

[18] \ \ S.Kuru, \ J. Negro, O. Ragnisco, The Perlick system type I: From
the algebra

\ \ \ \ \ \ \ of symmetries to the geometry of the trajectories,

\ \ \ \ \ \ \ Phys.Lett. A 381(2017) 3355-3363.

[19] \ J. Little, Nondegenerate homotopies of curves on the unit 2-sphere.

\ \ \ \ \ \ \ J.Diff.Geom.\ 4 (1970)\textbf{\ }339-34.

[20] \ A.Ballesteros, A. Encio, F. Herranz, Hamiltonian systems admitting a

\ \ \ \ \ \ \ Runge--Lenz vector and an optimal extension of Bertrand's
theorem to curved

\ \ \ \ \ \ \ manifolds, Comm.Math.Phys\textit{.} 290 (2009) 1033-1049.

[21] \ R.\ Luneburg, Mathematical Theory of Optics\textit{,} University of
California Press,

\ \ \ \ \ \ \ Los Angeles, 1966.

[22] \ C.Caratheodory, Optics, Julius Springer, Berlin, 1937.

[23] \ V.Perlick, Bertrand spacetimes, Class.Quantum Grav\textit{. }9 (1992)%
\textbf{\ }1009-1021.

[24] \ S. Flugge, Practical Quantum Mechanics, Springer-Verlag, Berlin, 1999.

[25] \ D. Kirzhnitz,Y. Lozovik, G. Shpatkovskaya, Statistical model of
matter.

\ \ \ \ \ \ \ Sov.Phys. Uspekhi 18 (1976)\textbf{\ }649-672.

[26] \ R. Latter, Atomic energy levels for the Thomas-Fermi and
Thomas-Fermi-Dirac

\ \ \ \ \ \ \ potential, Phys.Rev\textit{. }99 (1955) 510-519.

[27] \ N.March, Self-Consistent Fields in Atoms, Pergamon Press Ltd.,
London, 1975.

[28] \ T.Tietz, Atomic energy levels for the approximate Thomas-Fermi
Potential,

\ \ \ \ \ \ \ J.Chem.Phys. 25 (1956) 789-790.

[29] \ T.Tietz, Contribution to the Thomas--Fermi Theory.

\ \ \ \ \ \ \ J.Chem.Phys. 49 (1968) 4391-4393.

[30] \ A.Kholodenko, How the modified Bertrand theorem explains regularities

\ \ \ \ \ \ \ and anomalies of the periodic system of elements,
arXiv.:2002.12128.

[31] \ U. Leonhardt, Th. Philbin, Geometry and Light. Dover Publications,\ 

\ \ \ \ \ \ \ New York, 2010.

[32] \ H. Stephani, General Relativity, Cambridge University Press,
Cambidge, 1990.

[33] \ J.Milnor, On the Geometry of the Kepler Problem,

\ \ \ \ \ \ \ Am.Math.Monthly 90\textbf{\ }(1983) 353-365.

[34] \ B. Dubrovin, S. Novikov, A. Fomenko, Modern Geometry-Methods and

\ \ \ \ \ \ \ Applications\textit{,} Vol.1, Springer-Verlag, Berlin, 1984.

[35] J.\ Moser, E. Zehnder, Notes on Dynamical Systems,\textit{.}

\ \ \ \ \ \ \ AMS Publishers, Providence, 2005.

[36] \ J. Moser, Regularization of Kepler's problem and the averaging method
on

\ \ \ \ \ \ \ a manifold, \textit{\ }Comm.Pure Appl.Math. 23(1970)\textbf{\ }%
609-636.

[37] $\ S.\func{Si}$nger, Linearity, Symmetry and Prediction in the Hydrogen
Atom,

\ \ \ \ \ \ \ Springer, Berlin, 2005.

[38] \ A. Makowski, K.Gorska, Phys.Rev\textit{. }A 79 (2009) 052116

[39] \ E. Belokolos, Mendeleev table: a proof of Madelung rule

\ \ \ \ \ \ \ and atomic Tietz potential, \textit{\ }SIGMA 13 (2017) 038

[40] \ P. Dirac, Lectures on Quantum Mechanics\textit{,} Yeshiva University,
New York, 1964.

[41] \ V. Arnol'd, Mathematical Methods of Classical Mechanics\textit{,}

\ \ \ \ \ \ \ Springer -Verlag, Berlin, 1989.

[42] \ O. Jones, Analytical Mechanics for Relativity and Quantum Mechanics,

\ \ \ \ \ \ \ Oxford University Press, Oxford, 2011.

[43] \ J. Struckmeier, Hamiltonian dynamics on the symplectic extended phase
space

\ \ \ \ \ \ \ for autonomous and non-autonomous systems, J.Phys.A 38 (2005)%
\textbf{\ }1257-1278.

[44] \ L. Prokhorov, S. Shabanov, Hamiltonian Mechanics of Gauge Fields%
\textit{,}

\ \ \ \ \ \ \ Cambridge University Press, Cambridge, 2011.

[45] \ E. Anderson, The problem of Time, Springer

\ \ \ \ \ \ \ International Publishing AG, Berlin, 2017.

[46] \ K.Kuchar, Time and interpretations of quantum gravity,

\ \ \ \ \ \ \ Int.J.Mod.Phys.D\textbf{\ }20 (2011)\textbf{\ }3-86.

[47] \ K. Kuchar, Gravitation, geometry, and nonrelativistic quantum theory,

\ \ \ \ \ \ \ Phys.Rev.D\textit{\ }22 (1980) 1285-1299.

[48] \ A. Makowski, P. Peplowski, Zero energy wave packets which follow

\ \ \ \ \ \ \ classical orbits, Phys.Rev. A 86 (2012) 042117.

[49] \ H.\ Rosu, M. $\func{Re}$yers, K. Wolf, O.Obregon, Supersymmetry of

\ \ \ \ \ \ \ Demkov-Ostrovsky effective potentials in the R$_{0}=0$ sector,

\ \ \ \ \ \ \ Phys.Lett. A 208 (1995)\textbf{\ }33\textbf{-}39.

[50] \ A. Keane, R. Barrett, J. Simmons, The classical Kepler problem and

\ \ \ \ \ \ \ geodesic motion\ on spaces of constant curvature, JMP 41
(2000) 8108-8116.

[51] \ \ S. Chanda, G.Gibbons, P. Guha, Jacobi--Maupertuis metric and Kepler

\ \ \ \ \ \ \ \ equation, Int. J.of Geom.Methods in Mod.Physics 14 (2017)
1730002.

[52] \ \ A.\ Petrov, Einstein Spaces, Pergamon Press, London, 1969.

[53] \ \ A. Makowski, Family of fish-eye-related models and their
supersymmetric

\ \ \ \ \ \ \ \ partners. Phys. Rev. A 81(2010) 052109.

[54] \ \ \ R.Wald, General Relativity, The University of Chicago Press,
Chicago, 1984.

[55] \ \ \ M.\ Tsamparlis, A. Paliathanasis, Conformally related metrics and
Lagrangians

\ \ \ \ \ \ \ \ and\ their physical interpretation\ in cosmology,

\ \ \ \ \ \ \ \ Gen.Rel. Grav.\textbf{\ }45\textbf{\ }(2013) 2003-2022.

[56] \ \ A.Kholodenko, E. Ballard, From Ginzburg Landau to Hilbert Einstein
via

\ \ \ \ \ \ \ Yamabe, Physica A 380 (2007) 115-162.

[57] \ \ A.Kholodenko, Towards physically motivated proofs of the Poincar%
\'{e} and

\ \ \ \ \ \ \ geometrization conjectures, Journal of Geometry and Physics%
\textit{\ }58\textbf{(}2008\textbf{)} 259.

[58] \ \ H.Urakawa, Geometry of Laplace-Beltrami operator on a complete
Riemannian

\ \ \ \ \ \ \ manifold, Adv.Stud. Pure Math. 22 (1993) 347-406

[59] \ \ L.Ovsyannikov, Group Analysis of Differential Equation\textit{,}

\ \ \ \ \ \ \ Academic Press, New York,1982.

[60] \ \ N.Ibragimov, Transformation Groups Applied to Mathematical Physics%
\textit{,}

\ \ \ \ \ \ \ D.Reidel Publ.Co, Boston, 1985.

[61] \ \ K.Yano, The Theory of Lie Derivatives and its Applications\textit{,}

\ \ \ \ \ \ \ \ North -Holland, Amsterdam, 1957.

[62] \ \ M.Obata, Conformal transformations of Riemannian manifolds,

\ \ \ \ \ \ \ \ J.Diff.Geom.\textit{\ }4\textbf{\ }(1970) 311-333.

[63] \ \ H.Yamabe, On a deformation of Riemannian structures on compact
manifolds,

\ \ \ \ \ \ \ \ Osaka Math.Journal 12 (1960)\textbf{\ }21-37.

[64] \ \ T.Aubin, Some Nonlinear Problems in Riemannian Geometry\textit{,}

\ \ \ \ \ \ \ \ Springer-Verlag, Berlin, 1998.

[65] $\ \ $L.$\func{Ei}$senhart, Riemannian Geometry, Princeton University
Press, Princeton, 1926.

[66] \ \ R.\ Courant, D. Hilbert, Methods of Mathematical Physics\textit{,}

\ \ \ \ \ \ \ \ Vol.2, Interscience Publishers, Amsterdam, 1966.

[67] \ \ \ V. Ostrovsky, Group theory applied to the periodic table of the
elements, in

\ \ \ \ \ \ \ \ The Mathematics of Periodic Table. Nova Science Publishers,
New York,\ 

\ \ \ \ \ \ \ \ pp. 265-311, 2006.

[68] \ \ \ C.\ Boyer, E. Ka$\ln $ins, W. Miller, Symmetry and separation of
variables for the

\ \ \ \ \ \ \ \ \ Helmholtz and Laplace equations,\textit{\ }Nagoya Math. J.
60\textbf{\ }(1976) 35-80.

[69] \ \ \ E. Kalnins, J. Kress, W. Miller, Separation of Variables and
Superintegrability.

\ \ \ \ \ \ \ \ IOP Publishing, Bristol, 2018.

[70] \ \ M. Chanachowicz, M. Chanu, R. McLenaghan, R-separation of variables
for

\ \ \ \ \ \ \ \ the conformally invariant Laplace equation, JGP 59 (2009)
876-884

[71] \ \ N.Evans, Superintegrability in classical mechanics,

\ \ \ \ \ \ \ \ Phys.Rev.A 41 (1990)\textbf{\ }5666-5676.

[72] \ \ T. Tuc, A. Danner, Absolute optical instruments, classical
superintegrability,

\ \ \ \ \ \ \ \ and separability of the Hamilton-Jacobi equation, Phys.Rev.A
96 (2017) 053838.

[73] \ \ A. Linner, Periodic geodesics generator. Exp.Math 13 (2004) 199-206.

[74] \ \ E.Kudryavtseva, D. Fedoseev, Superintegrable mechanical systems of

\ \ \ \ \ \ \ \ Bertrand type, Modern Math and Its Applications\ 148\textbf{%
, }(2018) 37-57 (in Russian).

[75] \ \ M.Cariglia, F. Alves, The Eisenhart lift: a didactical introduction
of modern

\ \ \ \ \ \ \ \ \ geometrical concepts from Hamiltonian dynamics, Eur. J.
Phys.\textbf{\ }36 (2015) 025018

[76] \ \ J.Carinena, F. Herranz, M. Ranada, Superintegrable systems on
3-dimensional

\ \ \ \ \ \ \ \ curved spaces: Eisenhart formalism and separability,

\ \ \ \ \ \ \ \ J. Math.Phys. 58\textbf{\ }(2017) 02270.

[77] \ \ M. Cariglia, Null lifts and projective dynamics, Ann.Phys\textit{. }%
362 (2015) 642-658.

[78]\ \ \ C.Boyer, E. Kalnins, W. Miller, St\"{a}ckel-equivalent integrable
Hamiltonian

\ \ \ \ \ \ \ \ systems, SIAM J.Math.Anal. 17 (1986) 778-797.

[79] \ \ W.Miller, Multiseparability and superintegrability for classical
and quantum

\ \ \ \ \ \ \ \ systems, CRM Proc. Lecture Notes 26 (2000)\textbf{,} 129-156.

[80] \ \ A. Tsiganov, Transformation of the St\"{a}ckel matrices preserving

\ \ \ \ \ \ \ \ superintegrability, JMP 60 (2019)\textbf{,} 042701.

[81] \ \ \ M.Berger, A Panoramic View of Riemannian Geometry\textit{,}

\ \ \ \ \ \ \ \ \ Springer-Verlag, Berlin, 2003.

[82] \ \ \ A.Fordy, A. Galajinsky, Eisenhart lift of 2-dimensional mechanics.

\ \ \ \ \ \ \ \ \ Eur.Phys.J. C 79 (2019)\textbf{\ }301.

[83] \ \ \ L.Landau, E. Lifshitz,Quantum Mechanics. Non-Relativistic Theory%
\textit{,}

\ \ \ \ \ \ \ \ \ Butterworth-Heineman, London, 2003.

[84] \ \ \ E. Kalnins, W. Miller, P. Winternitz, The group O(4), separation
of variables

\ \ \ \ \ \ \ \ \ and the hydrogen atom, SIAM J.Appl.Math. 30 (1976) 630-664.

[85] \ \ \ B.Adams, J. Cizek, L. Paldus, Representation Theory of so(4,2)
for the

\ \ \ \ \ \ \ \ \ perturbation treatment of hydrogenic type hamiltonians by
algebraic method,

\ \ \ \ \ \ \ \ \ Int. J. Quantum Chem\textit{.} 21 (1982) 153-171.

[86] \ \ \ M.\ \ Gadella, J. Negro, L. Nieto, G. Pronko, M. Santander,
Spectrum generating

\ \ \ \ \ \ \ \ \ algebras for the free motion in $S^{3}$, J.Math.Phys%
\textit{.}52 (2011) 063509.

[87] \ \ \ M.Englefield, Group Theory and the Coulomb Problem\textit{,}

\ \ \ \ \ \ \ \ \ Wiley-Interscience, New York, 1972.

[88]~\ \ \ A. Stone, Some Properties of Wigner coefficients and
hyperspherical harmonics,

$\ \ \ \ \ \ \ \ \ \Pr $oc.Camb.Phil.Soc. 52 (1956) 424-430.

[89] \ \ \ B. Bransden, C. Joahain, Physics of Atoms and Molecules\textit{,}

\ \ \ \ \ \ \ \ \ Logman, Ltd., London, 1983.

[90] \ \ \ \ E. Kalnins, J. Kress, W. Miller, S. Post, Laplace-type
equations as conformal

\ \ \ \ \ \ \ \ \ \ superintegrable systems, Adv.Appl.Math. 46\textbf{\ }%
(2011) 396-416.

[91] \ \ \ \ B. Kursunoglu, \textit{\ }Modern Quantum Theory,W.H.Freeman and
Co.,New York, 1962.

[92] \ \ \ H. Bateman, A. Erdelyi, Higher Transcendental Functions\textit{. }%
Vol.2\textit{.,}

\ \ \ \ \ \ \ \ \ McGraw Hill Inc., New York, 1953.

[93] \ \ \ \ E. Schr\"{o}dinger, A method of determining quantum-mechanical
eigenvalues and

\ \ \ \ \ \ \ \ \ \ eigenfunctions, Proc. Irish Acad, Section A 46 (1940)
9-16.

[94] \ \ \ \ A. Kholodenko, Applications of Contact Geometry and Topology

\ \ \ \ \ \ \ \ \ \ in Physics\textit{, } World Scientific, Singapore, 2013.

[95] \ \ \ \ P.Kustaanheimo, E.\ Steifel, Perturbation theory of Kepler
motion based

\ \ \ \ \ \ \ \ \ \ on spinor regularization, J. Reine Ang.Math\textit{.}218
(1965) 204-219.

[96] \ \ \ \ \ D. Parfitt, D. Portnoi, The two-dimensional hydrogen atom
revisited,

\ \ \ \ \ \ \ \ \ \ J.Math.Phys\textit{. }43(2002) 4681-4691.

[97] \ \ \ \ \ B. George, S. Doman, The Classical Orthogonal Polynomials,

\ \ \ \ \ \ \ \ \ \ World Scientific, Singapore, 2016.

[98] \ \ \ \ \ X.Yang, X. Guo, F. Chan, K.Wong, W. Ching, Analytic solution
of a

\ \ \ \ \ \ \ \ \ \ two-dimensional \ hydrogen atom. I. Nonrelativistic
theory.

\ \ \ \ \ \ \ \ \ \ Phys.Rev.A\ 43 (2001) 1186-1196.

[99] \ \ \ \ Y.\ Demkov,V. Ostrovsky, $n+l$ filling rule \ in the periodic
system and

\ \ \ \ \ \ \ \ \ \ focusing\ potentials, \ Sov. Phys. JETP 35(1972) 66-69.

[100] \ \ \ V.\ Ostrovsky, What and how physics contributes to understanding
the periodic

\ \ \ \ \ \ \ \ \ \ law, Found.Chemistry 3(2001) 145-181.

[101] \ \ \ J. Slater, Analytic Atomic Wave Functions, Phys.Rev.43\textbf{\ }%
(1932) 33-43.

[102] \ \ \ E. Kerner, The solution of the Schr\"{o}dinger equation for an
approximate atomic

\ \ \ \ \ \ \ \ \ \ field. Phys. Rev\textit{.} 83, (1951) 71-75.

[103] \ \ \ A.\ Fet, Group Theory of the Chemical Elements\textit{,}

\ \ \ \ \ \ \ \ \ \ Walter de\ Gr\"{u}yter, Boston, 2016.

[104] \ \ \ D.Dey, K. Bhattaharya, T. Sarkar, Astrophysics of Bertrand
space-times.

\ \ \ \ \ \ \ \ \ \ Phys.Rev.D\textit{\ }88 (2013) 083532.

[105] \ \ \ A.\ Kholodenko, Newtonian limit of Einsteinian gravity: from
dynamics of Solar

\ \ \ \ \ \ \ \ \ \ system to dynamics of stars in spiral galaxies,
arXiv:1006.4650 v3

[106] \ \ \ D.Riglioni, Quantum Bertand Systems\textit{, }

\ \ \ \ \ \ \ \ \ \ PhD Thesis. Roma III University, 2012.

[107] \ \ \ D.\ Latini, Algebraic and Coalgebraic Aspects of Classical and

\ \ \ \ \ \ \ \ \ \ Quantum Hamiltonian Systems\textit{,} PhD Thesis. Roma
III University, 2017.

[108] \ \ \ P.Woit, Quantum Theory, Groups and Representations, Springer,
Berlin, 2017.

[109] \ \ \ G.Naber, Foundations of Quantum Mechanics\textit{,}

\ \ \ \ \ \ \ \ \ \ \ http://inspirehep.net/record/1411865?ln=en

[110] \ \ \ \ M. Zworski, \ Semiclassical Analysis, AMS Publishers,
Providence, 2012.

[111] \ \ \ \ E.\ Heller, The Semiclassical Way\textit{,} Princeton
University Press, Princeton, 2018.

[112] \ \ \ \ S.\ Benenti, Separation of variables in the geodesic
Hamilton-Jacobi equation.

\ \ \ \ \ \ \ \ \ \ \ http://www.sergiobenenti.it/cp/52.pdf.

[113] \ \ \ \ A.\ Bruce, On the solution of the Hamilton-Jacobi equation by
the method

\ \ \ \ \ \ \ \ \ \ \ \ \ of separation of variables, PhD Thesis, University
of Waterloo, 2000.

[114] \ \ \ \ A.\ Ballesteros, A.Enciso, F.Herranz, O. Ragnisco, Bertrand
spacetimes as

\ \ \ \ \ \ \ \ \ \ \ Kepler/oscillator potentials,\ Class.Quantum Grav%
\textit{.} 25\textbf{, (}2008\textbf{) }165005.

[115] \ \ \ \ E. Kalnins, W, Miller, G, Pogosyan, Coulomb-oscillator duality
in spaces of

\ \ \ \ \ \ \ \ \ \ \ constant curvature, J.Math.Phys. 41(2000) 2629-2657.

[116] \ \ \ \ \ E. Kalnins, J. Kress, W. Miller, Second-order
superintegrable systems in

\ \ \ \ \ \ \ \ \ \ \ \ conformally flat spaces. V. Two- and
three-dimensional quantum systems,

\ \ \ \ \ \ \ \ \ \ \ J.Math.Phys. 47 (2006) 093501.

[117] \ \ \ \ E.Kalnins, J. Kress,W. Miller, Second order superintegrable
systems

\ \ \ \ \ \ \ \ \ \ \ in conformally flat spaces. IV. The classical 3D St%
\"{a}ckel transform and 3D

\ \ \ \ \ \ \ \ \ \ \ classification theory, J.Math.Phys.\textbf{\ }47
(2006) 043514.

[118] \ \ \ \ P.\ Petersen, \textit{Warped Products.}

\ \ \ \ \ \ \ \ \ \ \ https://www.math.ucla.edu/\symbol{126}%
petersen/warpedproducts.pdf.\ \ \ \ \ \ \ \ 

[119] \ \ \ \ \ B-Y. Chen, Differential Geometry of Warped Product Manifolds

\ \ \ \ \ \ \ \ \ \ \ \ and Submanifolds\textit{, }World Scientific,
Singapore, 2017.

[120] \ \ \ \ \ D. Kirzhnitz, Y. Lozovik, Plasma oscillations of the
electron shell of the atom,

\ \ \ \ \ \ \ \ \ \ \ \ Sov.Phys.Uspekhi 9\textbf{,}(1966) 340-345.

[121] \ \ \ \ \ E. Fermi, Sulla deduzione statistica di alcune proprieta
dell'atomo.

\ \ \ \ \ \ \ \ \ \ \ \ Applicazione alla theoria del sistema periodico
degli elementi,

\ \ \ \ \ \ \ \ \ \ \ \ Rend.Lincei\textit{\ }7(1928) 342-346.

[122] \ \ \ \ \ M.Bander, C.Itzykson, Group theory and the hydrogen atom (I).

\ \ \ \ \ \ \ \ \ \ \ \ Rev.Mod.Phys.38(1966) 330-345.

[123] \ \ \ \ \ L.\ Kauffman, On Knots, Princeton University Press,
Princeton, 1988.

[124] \ \ \ \ \ R.\ Albert, Computer-calculated explicit forms for
representations

\ \ \ \ \ \ \ \ \ \ \ \ of the three-dimensional pure rotation group,

\ \ \ \ \ \ \ \ \ \ \ \ Proc.Camb.Phil.Soc. 65 (1969) 107-110.

[125] \ \ \ \ \ A.\ Besse, Manifolds all of Whose Geodesics are Closed,

\ \ \ \ \ \ \ \ \ \ \ \ Springer-Verlag, Berlin,1978.

[126] \ \ \ \ \ B.\ McKay, A summary of progress on the Blaschke conjecture.

\ \ \ \ \ \ \ \ \ \ \ \ Notices of the International Congress of Chinese
Mathematics

\ \ \ \ \ \ \ \ \ \ \ \ 3(2015) 33--45.

[127] \ \ \ \ \ L.\ Xu , X. Wang, T.Tyc, C. Sheng, S. Shu, H. Liu, H. Chen,

\ \ \ \ \ \ \ \ \ \ \ \ Light rays and waves on geodesic lenses.
arXiv:1801.10438

[128] \ \ \ \ \ M.\ Sarbort, Non$-$Euclidean geometry in optics. PhD Thesis.

\ \ \ \ \ \ \ \ \ \ \ \ Masaryk University, (2013)

\ \ \ \ \ \ \ \ \
www.researchgate.net/publication/287993246\_Non-Euclidean\_Geometry\_in%
\_Optics

[129] \ \ \ \ L.\ \ Kuznia, E. Lundberg, Fixed points of conjugated Blaschke
products with

\ \ \ \ \ \ \ \ \ \ \ \ applications to gravitational lensing, Computational
methods and Function

\ \ \ \ \ \ \ \ \ \ \ \ Theory 9 (2009) 435-442.

[130] \ \ \ \ A.Shoom, A.: Metamorphoses of a photon sphere,

\ \ \ \ \ \ \ \ \ \ \ Phys.$\func{Re}$v. D 96 (2017) 084056.

[131] \ \ \ \ R.Nemiroff, Visual distortions near a neutron star and black
hole

\ \ \ \ \ \ \ \ \ \ \ Am.J.Phys\textit{.}61(1993) 619-632.

[132] \ \ \ A.Celletti, From Order to Disorder in Gravitational\ 

\ \ \ \ \ \ \ \ \ \ \ N-Body Dynamical Systems\textit{,} Chaotic Worlds%
\textit{, } pp. 203-226,

\ \ \ \ \ \ \ \ \ \ \ Springer, Dordrecht, 2006.

[133] \ \ \ \ S.Chanda, G. Gibbons, P.Guha, Jacobi-Maupertuis metric and
Kepler

\ \ \ \ \ \ \ \ \ \ \ equation, Int.J.of Geom.Methods in Mod.Physics
14(2016) 1730002.

[134] \ \ \ H.\ Geiges, An Introduction to Contact Topology\textit{,}

\ \ \ \ \ \ \ \ \ \ \ Cambridge University Press, Cambridge, 2008.\ \ \ 

[135] \ \ \ A.\ Kholodenko, Optical knots and contact geometry II. From
Ranada dyons to

\ \ \ \ \ \ \ \ \ \ \ transverse and cosmetic knots, Ann.Phys\textit{.}
371(2016) 77-124.

[136] \ \ \ W.Goldmann, Lines in \textbf{P}$^{3},$
http://terpconnect.umd.edu/\symbol{126}wmg/plucker.pdf

[137] \ \ \ W.F\"{o}rstner,B.Wrobel, Photogrammetric Computer Vision\textit{,%
}

\ \ \ \ \ \ \ \ \ \ \ Springer, Berlin, 2016.

[138] \ \ \ \ C. Coletti, D. Calderini, V. Aquilanti, D-dimensional
Kepler-Coulomb

\ \ \ \ \ \ \ \ \ \ \ sturmians and hyperspherical harmonics as complete
orthonormal atomic and

\ \ \ \ \ \ \ \ \ \ \ molecular orbitals, Adv.Quantum Chemistry 67 (2013)%
\textbf{\ }73-127.

[139] \ \ \ J.\ Avery, J. Avery,Jr, Hyperspherical Harmonics and Their
Physical

\ \ \ \ \ \ \ \ \ \ Applications,World Scientific, Singapore, 2018.

[140] \ \ \ D.\ \ Wong, Theoretical justification of Madelung's rule,

\ \ \ \ \ \ \ \ \ \ J.Chem.Educ. 56 (1979) 714-717.

[141] \ \ \ V.Klechkowski, Distribution of Atomic Electrons \ and the Rule
of Sequential

\ \ \ \ \ \ \ \ \ \ Filling of the (n+l) Groups, Atomizdat, 1968, (in
Russian)

[142] \ \ \ T.\ Tietz, Uber eine Approximation der Fermischen
Verteilungsfunktion,

\ \ \ \ \ \ \ \ \ \ \ Ann.der Phys\textit{.} 5 (1960) 237-240.

[143] \ \ \ D.\ \ Ivanenko, S. Larin, Remarks about the periodic system of
elements,

\ \ \ \ \ \ \ \ \ \ Sov. Physics Doklady 88 (1953)\textbf{\ }45-47.

[144] \ \ \ R.Baer, Electron Density Functional Theory\textit{,}

\ \ \ \ \ \ \ \ \ \ The Fritz Haber Center for Molecular Dynamics, 2009.

[145] \ \ \ R.\ Feynman, N. Metropolis, E. Teller, Equations of state of
elements based

\ \ \ \ \ \ \ \ \ \ on the generalized Fermi-Thomas theory, Phys.Rev\textit{.%
}75 (1949) 1561-1673.

[146] \ \ \ J.Schwinger,Thomas-Fermi model: The leading correction,

\ \ \ \ \ \ \ \ \ \ Phys.$\func{Re}$v. A 22 (1980) 1827-1832.

\bigskip \bigskip

\bigskip

\bigskip

\bigskip

\bigskip

\bigskip

\end{document}